\documentclass[12pt,twoside]{mitthesis}

\pdfoutput=1

\usepackage{lgrind}
\usepackage{cite}

\usepackage{amsmath,graphicx}
\usepackage{amsthm,amsfonts}

\newcommand{\braket}[2]{\langle#1\vert#2\rangle}
\newcommand{\ketsub}[2]{\vert#1\rangle_{#2}}
\newcommand{\brasub}[2]{\!\,_{#2}\!\langle#1\vert}
\newcommand{\kett}[1]{\vert#1\rangle\!\rangle}
\newcommand{\braa}[1]{\langle\!\langle#1\vert}
\newcommand{\braakett}[2]{\langle\!\langle#1\vert#2\rangle\!\rangle}
\newcommand{\tr}{\textrm{tr}}
\newcommand{\E}{\mathcal{E}}
\newcommand{\R}{\mathcal{R}}
\newcommand{\LL}{\mathcal{L}}
\newcommand{\HH}{\mathcal{H}}
\newcommand{\KK}{\mathcal{K}}
\newcommand{\SSS}{\mathcal{S}}
\newcommand{\AAA}{\mathcal{A}}

\newcommand{\CC}{C_{E,\E}}
\newcommand{\ol}[1]{\overline{#1}}
\newcommand{\Xo}{X^\star}
\newcommand{\Yo}{Y^\star}
\newcommand{\Ro}{R^\star}
\newcommand{\Yi}{Y^{(0)}}
\newcommand{\Zi}{Z^{(0)}}

\newtheorem{Pauli_Channel}{Theorem}
\newtheorem{close_isom}[Pauli_Channel]{Lemma}

\newtheorem{Gersgorin}[Pauli_Channel]{Theorem}

\input{Qcircuit}

\begin{document}

%
%
%
%
%
%
%
\title{Channel-Adapted Quantum Error Correction}

\author{Andrew Stephen Fletcher}
\department{Department of Electrical Engineering and Computer Science}
\degree{Doctor of Philosophy}
\degreemonth{June}
\degreeyear{2007}
\thesisdate{May 22, 2007}


\supervisor{Peter W. Shor}{Morss Professor of Applied Mathematics}
\supervisor{Moe Z. Win}{Associate Professor}

\chairman{Arthur C. Smith}{Chairman, Department Committee on Graduate Students}

\maketitle



\cleardoublepage
\setcounter{savepage}{\thepage}
\begin{abstractpage}
%
%
%
\vspace{-8pt}
Quantum error correction (QEC) is an essential concept for any quantum information processing device.  Typically, QEC is designed with minimal assumptions about the noise process; this generic assumption exacts a high cost in efficiency and performance.  We examine QEC methods that are adapted to the physical noise model.  In physical systems, errors are not likely to be arbitrary; rather we will have reasonable models for the structure of quantum decoherence.  We may choose quantum error correcting codes and recovery operations that specifically target the most likely errors.  This can increase QEC performance and also reduce the required overhead.

We present a convex optimization method to determine the optimal (in terms of average entanglement fidelity) recovery operation for a given channel, encoding, and information source. This is solvable via a semidefinite program (SDP).  We derive an analytic solution to the optimal recovery for the case of stabilizer codes, the completely mixed input source, and channels characterized by Pauli group errors.  We present computational algorithms to generate near-optimal recovery operations structured to begin with a projective syndrome measurement.  These structured operations are more computationally scalable than the SDP required for computing the optimal; we can thus numerically analyze longer codes.  Using Lagrange duality, we bound the performance of the structured recovery operations and show that they are nearly optimal in many relevant cases.

We present two classes of channel-adapted quantum error correcting codes specifically designed for the amplitude damping channel.  These have significantly higher rates with shorter block lengths than corresponding generic quantum error correcting codes.  Both classes are stabilizer codes, and have good fidelity performance with stabilizer recovery operations.  The encoding, syndrome measurement, and syndrome recovery operations can all be implemented with Clifford group operations.

\end{abstractpage}


\cleardoublepage

\section*{Acknowledgments}

I owe thanks to many for their guidance, support, and inspiration during the preparation of this dissertation.  I am grateful to Moe Win for first suggesting a topic in quantum computing.  We've come a long way from his original question, ``What role does diversity combining play in quantum communications?''  I am indebted to Peter Shor for many hours of technical conversation; he was patient when I was a novice in this field and gently guided me to greater understanding.

I am grateful to many at MIT Lincoln Laboratory.  The Lincoln Scholars Committee saw fit to fund these studies, and neither Bing Chang nor Dave Conrad, my group leaders, balked when my research departed so drastically from my original plans.

My grandfather, Robert C. Fletcher, preceded me in Ph.D.~studies here at MIT by nearly 60 years.  I have felt inspired by his legacy and feel a closer kinship with him as I've followed in his footsteps.  I'm also grateful to my parents, Bob and Gail Fletcher, who have encouraged me in my educational ambitions my entire life.

Finally, I cannot overstate my reliance on and gratitude for my wife, Mary Beth.  She embraced my dream of a Ph.D.~as her own, and would not let me abandon it.  She has been a pillar of support and resolve, uncomplaining through the challenges of graduate family life.  With love, I dedicate this thesis to her and our three beautiful daughters, Erin, Shannon, and Audrey.

\cleardoublepage
This work has been sponsored by the United States Air Force under AF Contract \#FA8721-05-C-0002.  Opinions, interpretations, recommendations and conclusions are those of the author and are not necessarily endorsed by the United States Government.

\pagestyle{plain}
\tableofcontents
\newpage
\listoffigures
\newpage
\listoftables

\chapter{Introduction}\label{chap:Intro}

\begin{center}
\emph{``Many authors have what appears to be a suspicious fondness for the depolarizing channel...''\\
\hspace{2in}-Michael Nielsen and Isaac Chuang in \cite{NieChu:B00}}
\end{center}

\section{Overview}

Quantum error correction (QEC) is an essential component of quantum information processing.  To realize its ground-breaking potential, a quantum computer must have a strategy to mitigate the effects of noise.  QEC protects information from noise by including redundancy in a manner analogous to classical error correction.  In this way, the effects of noise are reduced at the cost of extended overhead.

The noise suppression vs.~overhead tradeoff creates quite a conundrum as neither comes cheaply; these are two of the principal obstacles to a physical quantum computer.  Experimentalists have demonstrated several physical systems that exhibit the quantum effects necessary for quantum computing, but each suffers from decoherence and scalability issues.  It is one challenge to shield a quantum system from the environment and thus reduce noise.  It is yet another to construct an architecture which scales to process a large number of quantum bits (qubits).

Since overhead is so expensive, it behooves us to seek out the most efficient means of performing QEC.  To this end, we explore the concept of channel-adaptation.  QEC was developed with an intentionally generic model for the noise - indeed the early triumph of the Shor code was the demonstration of an encoding and decoding procedure which could correct for an \emph{arbitrary} error on a single qubit\cite{Sho:95}.  The subsequent development of CSS codes\cite{CalSho:96,Ste:96a} and the even more general stabilizer codes\cite{Got:96,CalRaiShoSlo:97,CalRaiShoSlo:98,Got:97} are all based on the concept of arbitrary qubit errors.  In essence, the only assumption is that errors would affect each qubit independently.  This assumption has aided greatly in connecting QEC to the mature field of classical error correcting codes.  Furthermore, the general applicability of QEC has enabled beautiful extensions to fault tolerant quantum computing\cite{Sho:96,Kit:97b,Kit:97c,DivSho:96,Got:98b,Got:97}.

The generic approach has its drawbacks, however.  Most notably, quantum codes impose a severe amount of overhead to correct for arbitrary errors.  As an example, the shortest block code that corrects an arbitrary qubit error embeds one qubit into five\cite{BenDivSmoWoo:96,LafMiqPazZur:96}.  The overhead tradeoff involved in QEC is steep when the code and recovery are designed for arbitrary errors.

QEC can be made more efficient if we no longer seek to correct arbitrary errors\cite{LeuNieChuYam:97}.  Any physical implementation of a quantum computer will interact with the environment in a specific way; this imposes a definite structure on the observed decoherence of the quantum state.  By designing the error correcting procedure to protect from such structured noise, we may improve efficiency and thus reduce the required overhead.  We will refer to this concept as \emph{channel-adapted quantum error correction}, the subject of this dissertation.

Channel-adapted QEC was introduced as `approximate' quantum error correction by Leung \emph{et.~al.}~in \cite{LeuNieChuYam:97}.  The name approximate was appropriate as the code did not perfectly satisfy the quantum error correcting conditions derived in \cite{BenDivSmoWoo:96,KniLaf:97}.  Instead, analogous approximate conditions were shown to apply without significant loss in performance.  The key criterion was the fidelity of the corrected state to the input - how well the encoding and recovery protect the information from the noise.
In the conclusion to \cite{LeuNieChuYam:97}, the authors state, ``It would be especially useful to develop a general framework for constructing codes based on approximate conditions, similar to the group-theoretic framework now used to construct codes that satisfy the exact conditions.''  Such results have been elusive.  Instead, channel-adapted QEC has recently found more traction when cast as an optimization problem\cite{FleShoWin:07,KosLid:06,ReiWer:05,YamHarTsu:05}.  Both encodings and recoveries can be designed by numerical methods that seek to maximize the overall fidelity.

While our research will be detailed in this dissertation, we feel compelled to note complementary work in channel-adapted QEC, particularly those focused on QEC via optimization methods.  In \cite{KosLid:06} and \cite{ReiWer:05}, encodings and decodings were iteratively improved using the performance criteria of ensemble average fidelity and entanglement fidelity, respectively.  A sub-optimal method for minimum fidelity, using a semi-definite program (SDP), was proposed in \cite{YamHarTsu:05}.  An analytical approach to channel-adapted recovery based on the pretty-good measurement and the average entanglement fidelity was derived in \cite{BarKni:02}.  (The various flavors of fidelity will be discussed in Sec.~\ref{sec:fidelity}).  The main point of each scheme was to improve error correction procedures by adapting to the physical noise process.

\section{Organization}

In the remainder of this chapter, we introduce some of the mathematical tools and notation to be used in the remainder of the dissertation.  We also lay out the channel models and quantum error correcting codes that will be used as examples in various subsequent sections.

Chapter \ref{chap:OptQER} explores channel-adaptation by considering a fixed encoding operation and computing the recovery operation that maximizes average entanglement fidelity.  In this form, the optimization problem turns out to be convex and has an efficient solution.  Several examples are given, which illustrate some of the performance gains available via channel-adaptation.  We derive the Lagrange dual of the optimum recovery operation and use the dual function to prove sufficient conditions for the generic QEC recovery operation to be optimal.

Chapter \ref{chap:NearOptQER} explores quantum error recovery operations where we have imposed additional constraints.  The recoveries have nearly optimal fidelity performance, but are structured in either physically simple or intuitively instructive forms.  The constraints also serve to enable the processing of higher dimensional channels, thus allowing channel-adaptation of longer quantum codes.  We present a general class of recovery operations that begin with projective error syndrome measurements as well as several specific algorithms that generate such recovery operations.

Chapter \ref{chap:DualBounds} uses the Lagrange dual to certify the near-optimality of the recovery operations from chapter \ref{chap:NearOptQER}.  We derive a numerical technique to generate dual feasible points given a structured recovery operation.  We show that the structured recovery operations are asymptotically optimal for the examples given.

Chapter \ref{chap:ThreeQubitCode} takes a closer look at channel-adapted QEC for the amplitude damping channel.  We begin with an analysis of the $[4,1]$ approximate code of \cite{LeuNieChuYam:97}.  We conclude that approximate is a bit of a misnomer, as in fact the code can perfectly correct a set of errors that approximate qubit dampings.  Furthermore, both the encoding and a good recovery operation can be understood in terms of the stabilizer formalism.  This discovery leads to two general classes of channel-adapted codes for the amplitude damping channel.

\section{Mathematical Notation and Background}

It is beyond the scope of this dissertation to provide an introduction to quantum computation or quantum information processing.  We presume familiarity with quantum states in both the \emph{bra-ket} and density matrix representations.  We refer readers who desire a more comprehensive introduction to the first two chapters of \cite{NieChu:B00}.  This section will, however, state succinctly some of the notation conventions used throughout the dissertation.  Furthermore, we will review the topics of quantum operations, channel fidelity metrics, and the classical optimization routine of semidefinite programming in more detail, as these will be of particular value throughout the remainder of the dissertation.

Pure quantum states will be denoted with the \emph{ket} notation $\ket{\cdot}$.  These are elements of a Hilbert space, which we will generally denote $\HH$ or $\KK$.  Bounded linear operators on this space are elements of $\LL(\HH)$.  A bounded linear operator that maps $\HH$ to $\KK$ is an element of $\LL(\HH,\KK)$.  Density matrices represent either a pure or a mixed quantum state; if the pure quantum state lives in $\HH$, then the density matrix is an element of $\LL(\HH)$.  We will generally refer to density matrices as $\rho$, or some operation acting on $\rho$ (\emph{i.e.}~$\AAA(\rho)$).

\subsection{Quantum operations}

A quantum operation must be a completely positive trace preserving (CPTP) linear map\cite{Kra:B83}.  This constraint arises as valid quantum states input to the operation must emerge as valid quantum states.  As either the input or the output of such an operation can be mixed, the map is defined as acting on density matrices and can be given (for example) as $\AAA:\LL(\HH)\mapsto\LL(\KK)$.

A map $\AAA:\LL(\HH)\mapsto\LL(\KK)$ is CPTP if and only if it can be represented by a set operators $\{A_k\}\in\LL(\HH,\KK)$ such that $\sum_k A_k^\dagger A_k=I$.  The input-output relation is given by $\AAA(\rho)=\sum_k A_k\rho A_k^\dagger$.  The operators $\{A_k\}$ are referred to equivalently as \emph{operator elements} or \emph{Kraus operators}.  The operator elements of a mapping are not a unique representation; any unitary recombination of the operator elements ($\{A_i'=\sum_j u_{ij}A_j\}$ where $\sum_k u_{ik}^*u_{kj}=\delta_{ij}$) yields an equivalent operation $\AAA$.

The Kraus operator representation of quantum operations is the most common, but its many-to-one nature will be inconvenient for some of our purposes.  In such cases, we will use an alternate description, in which a CPTP operation $\AAA:\mathcal{L}(\HH)\mapsto\mathcal{L}(\KK)$ is given in terms of a positive semidefinite (p.s.d.) operator $X_\AAA\in\mathcal{L}(\KK\otimes\HH^*)$\cite{Cho:75,DarLop:01,Hav:03,Cav:99,Dep:67}.  $X_\AAA$ is often called the \emph{Choi matrix}.

To derive the Choi matrix, we will make use of a convenient isomorphism in which bounded linear operators are represented by vectors and denoted with the symbol $\kett{\cdot}$.  While there are several choices for this isomorphism\cite{DarLop:01,Hav:03}, including most intuitively a ``stacking'' operation, we will follow the conventions of \cite{Tys:03} (also \cite{YamHarTsu:05}) which results in an isomorphism that is independent of the choice of basis.  For convenience, we will restate the relevant results here.

Let $A=\sum_{ij}a_{ij}\ket{i}\bra{j}$ be a bounded linear operator from $\HH$ to $\KK$ (\emph{i.e.}~$A\in\mathcal{L}(\HH,\KK)$, where $\{\ket{i}\}$ and $\{\ket{j}\}$ are bases for $\KK$ and $\HH$, respectively.  Let $\HH^*$ be the dual of $\HH$. This is also a Hilbert space, generally understood as the space of \emph{bras} $\bra{j}$.  If we relabel the elements as $\overline{\ket{j}}=\bra{j}$, then we represent $A$ as a vector in the space $\KK\otimes\HH^*$ as
\begin{equation}\label{eq:basis-free doubleket}
  \kett{A}=\sum_{ij}a_{ij}\ket{i}\overline{\ket{j}}.
\end{equation}

It is useful to note the following facts.  The inner product $\braakett{A}{B}$ is the Hilbert-Schmidt inner product $\tr A^\dagger B$.  Also, the partial trace over $\KK$ yields a useful operator on $\HH^*$:
\begin{equation}\label{eq:kett_trace}
\tr_{\KK}\kett{A}\braa{B}=\overline{AB^\dagger}.
\end{equation}
Finally, index manipulation yields the relation
\begin{equation}\label{eq:kett_triple_product}
A\otimes \overline{B}\kett{C}=\kett{ACB^\dagger},
\end{equation}
where $\overline{B}$ is the conjugate of $B$ such that $\overline{B\ket{\psi}}=\overline{B}\hspace{2pt}\overline{\ket{\psi}}$ for all $\ket{\psi}$.

The Choi matrix is calculated from the Kraus elements $\{A_k\}$ of $\AAA$ as
\begin{equation}\label{eq:Choi matrix}
  X_\AAA=\sum_k\kett{A_k}\braa{A_k}.
\end{equation}
(We will refer to $X_\AAA$ as the Choi matrix for $\AAA$, although most derivations do not use the basis-free free double-ket of (\ref{eq:basis-free doubleket}).)  The operation output is given by $\AAA(\rho)=\tr_\HH (I\otimes\overline{\rho}) X_\AAA$ and the CPTP constraint requires that $X_\AAA\geq 0$ and $\tr_\KK X_A=I$.

\subsection{Channel fidelity}\label{sec:fidelity}

In classical discrete communications, it is quite simple to describe the idea of `correct transmission' or, inversely, the probability of error.  As symbols are drawn from a discrete set, there is no fundamental barrier to observing what is sent and what is received and declaring success if the two match (and error if they do not).  The classical concept is essentially trivial.

Transmission metrics for quantum information are trickier.  The superposition principle for quantum states implies a continuum of states for a quantum system; if we defined an error for any output that did not \emph{exactly} match the input, then we must classify an infinitesimal rotation of $R(\epsilon)$ about some axis to be an error, despite the fact that the resulting state is essentially identical to the desired state.  Obviously, declaring errors in this manner is neither practical nor useful; we require an alternate metric for successful quantum communication analogous to `correct transmission.'

Standard QEC results provide one such metric, which essentially returns to the classical definition.  The triumph of QEC is the ability to \emph{perfectly} correct arbitrary errors on a single qubit.  The continuous errors are `discretized' by the syndrome measurement and the system is restored exactly to its initial quantum state.  We may declare the probability of successful transmission as the probability of observing a correctible error, \emph{i.e.}~an error on a single qubit.  For any channel model and a standard QEC operation, this probability is readily calculable.

Despite its simplicity, the standard QEC definition for the probability of error is too restrictive to enable channel-adaptivity.  As mentioned above, we intuitively understand that receiving $R(\epsilon)\ket{\psi}$ as the output when $\ket{\psi}$ is the input should be considered a successful transmission.  To account for this, we will rely upon the concept of the \emph{fidelity} of a quantum state.

For pure states $\ket{\psi_1}$ and $\ket{\psi_2}$, the fidelity has a perfectly natural form with a corresponding physical intuition:  $F(\ket{\psi_1},\ket{\psi_2})=|\braket{\psi_1}{\psi_2}|^2$.  (The fidelity is sometimes defined as the square root of this quantity.)  As this is the inner product squared of two unit length vectors, the fidelity is the cosine squared of the angle between $\ket{\psi_1}$ and $\ket{\psi_2}$.  If the second state is mixed, it is straightforward to see that this quantity becomes $F(\ket{\psi_1},\rho_2)=\bra{\psi_1}\rho_2\ket{\psi_1}$.  When both states are mixed, the fidelity has been generalized to be\cite{Joz:94}
\begin{equation}\label{eq:fidelity}
F(\rho_1,\rho_2)=(\tr\sqrt{\rho_1^{\frac{1}{2}}\rho_2\rho_1^{\frac{1}{2}}})^2.
\end{equation}
This quantity is consistent with the pure state definition of the fidelity, is symmetric in $\rho_1$ and $\rho_2$, takes values between 0 and 1, is equal to 1 if and only if $\rho_1=\rho_2$, and is invariant over unitary rotations of the state space.

While (\ref{eq:fidelity}) provides a measure of similarity between two states, what we really require is a \emph{channel fidelity} that will determine how well a noisy operation $\AAA:\HH\mapsto\HH$ preserves a quantum state.  For any given quantum state $\ket{\psi}$, the natural extension to (\ref{eq:fidelity}) is the quantity $F(\ket{\psi}\bra{\psi},\AAA(\ket{\psi}\bra{\psi}))$.  This input-output relation measures  how well the specific state $\ket{\psi}$ is preserved by $\AAA$.  While this may be sufficient, it is quite possible that $\AAA$ could successfully protect one quantum state from noise, while another is easily corrupted.  We would prefer a measure that more fully characterizes the behavior of $\AAA$.

We define the \emph{minimum fidelity} of $\AAA$ as the worst case scenario over all input states $\ket{\psi}$:\footnote{One might suppose we should have to minimize over all mixed states $\rho$.  In fact, it is sufficient to minimize over pure state inputs \cite{NieChu:B00}.}
\begin{equation}\label{eq:min_fidelity}
 F_{\min}(\AAA)=\min_{\ket{\psi}} F(\ket{\psi}\bra{\psi},\AAA(\ket{\psi}\bra{\psi})).
\end{equation}
By virtue of the minimization over $\ket{\psi}$, one need not assume anything about the input state.  This was the metric of choice in \cite{KniLaf:97} first establishing the theory of QEC, and translates nicely to the idea of perfectly correcting a set of errors.  The disadvantage arises through the complexity of the metric; indeed computation requires minimizing over all inputs.  This drawback makes minimum fidelity a difficult choice for optimization based channel-adaptation.  Efficient routines that have been developed for channel-adaptation using (\ref{eq:min_fidelity}) are sub-optimal\cite{YamHarTsu:05}.

\emph{Entanglement fidelity} and \emph{ensemble average fidelity} both provide more tractable metrics for $\AAA$.  To use them, we must make some assumption about the ensemble of input states.  We may define an ensemble $E$ consisting of states $\rho_i$ each with probability $p_i$.  The ensemble average fidelity is naturally defined as
\begin{equation}\label{eq:ensemble_ave_fidelity}
\bar{F}(E,\AAA)=\sum_ip_iF(\rho_i,A(\rho_i)).
\end{equation}
When $\rho_i$ are pure states, $\bar{F}$ is linear in $\AAA$.

Entanglement fidelity\cite{Sch:96} is defined for a mixed state $\rho$ in terms of a purification to a reference system.  Recall that $\rho$ can be understood as an ensemble of quantum states, $\rho=\sum_i p_i \rho_i$.  If $\ket{\psi}\in \HH_R\otimes\HH$ (where $\HH_R$ is a reference system) is a purification of $\rho$, then $\rho=\tr_{\HH_R}\ket{\psi}\bra{\psi}$.  The purification captures all of the information in $\rho$.  The entanglement fidelity is the measure of how well the channel $\AAA$ preserves the state $\ket{\psi}$, or in other words, how well $\AAA$ preserves the entanglement of the state with its reference system.  We write the entanglement fidelity as
\begin{equation}\label{eq:ent_fid}
F_e(\rho,\AAA)=\bra{\psi}\mathcal{I}\otimes\AAA(\ket{\psi}\bra{\psi})\ket{\psi},
\end{equation}
where $\mathcal{I}$ is the identity map on $\LL(\HH_R)$.  We have used the fact that $\ket{\psi}$ is pure to express (\ref{eq:ent_fid}) in a more convenient equation for the fidelity than the generic mixed state form of (\ref{eq:fidelity}).  The entanglement fidelity is linear in $\AAA$ for any input $\rho$, and is a lower bound to the ensemble average fidelity for any ensemble $E$ such that $\sum_ip_i\rho_i=\rho$.

The linearity of both ensemble average fidelity and entanglement fidelity in $\AAA$ is particularly useful for channel-adapted QEC.  It enables the use of the convex optimization problems called semidefinite programs, which will be summarized in the next section.  As all of the optimization problems in this dissertation could be performed using either metric, we will follow the lead of \cite{BarKni:02} and derive based on the \emph{average entanglement fidelity}, given by
\begin{equation}\label{eq:ave_ent_fidelity}
\bar{F}_e(E,\AAA)=\sum_ip_iF_e(\rho_i,\AAA).
\end{equation}
By so doing, all of the algorithms can be trivially converted to either entanglement fidelity or ensemble average fidelity with pure states, as both are special cases of average entanglement fidelity.

While the derivations will be in average entanglement fidelity, most examples will assume an ensemble $E$ of the completely mixed state $\rho=I/d_S$ with probability 1.  In essence, this will assume the minimum about the information source and apply the strictest fidelity condition.

The definition of entanglement fidelity given in  (\ref{eq:ent_fid}) is intuitively useful, but awkward for calculations.  An easier form arises when operator elements $\{A_i\}$ for $\AAA$ are given.  The entanglement fidelity is then
\begin{equation}\label{eq:ent_fid_kraus}
F_{e}(\rho,\AAA)=\sum_i|\tr(\rho A_i)|^2.
\end{equation}
From  (\ref{eq:ent_fid_kraus}), we may derive a calculation rule for the entanglement fidelity when the channel $\AAA$ is expressed via the Choi matrix.  Recalling the definition of the Hilbert-Schmidt inner product, we see that $\tr{A_i \rho}=\braakett{\rho}{A_i}$.  Inserting this into  (\ref{eq:ent_fid_kraus}), we obtain the entanglement fidelity in terms of $X_\AAA$:
\begin{eqnarray}
\nonumber F_{e}(\rho,\AAA)&=& \sum_i\braakett{\rho}{A_i}\braakett{A_i}{\rho}\\
&=& \braa{\rho}X_\AAA\kett{\rho}.
\end{eqnarray}
It is trivial to extend this expression to average entanglement fidelity given an ensemble $E$:
\begin{equation}
  \bar{F}_{e}(E,\AAA)=\sum_k p_k\braa{\rho_k}X_\AAA\kett{\rho_k}.
\end{equation}

\subsection{Semidefinite programming}\label{sec:SDP}

The choice of average entanglement fidelity provides a measure of performance that is linear in the operation $\AAA$.  The linearity is a particularly useful feature, as it enables many problems in channel-adapted QEC to be cast as a convex optimization problem called a semidefinite program (SDP).  Semidefinite programming is a useful construct for convex optimization problems; efficient routines have been developed to numerically evaluate SDP's.  The theory of SDP's is sufficiently mature that the numerical solution can be considered a `black-box routine' for the purposes of this dissertation.  We will here concisely state the definition of a SDP and refer the interested reader to the review article \cite{VanBoy:96} for a more extensive treatment.

A semidefinite program is defined as the minimization of a linear function of the variable $x\in\mathbf{R}^N$ subject to a matrix inequality constraint:
\begin{eqnarray}\label{eq:SDP}
\min_x c^Tx,\textrm{ such that } F(x)\geq 0,
\end{eqnarray}
where $F(x)=F_0+\sum_{n=1}^NxF_n$ for $F_n\in\mathbf{R}^{n\times n}$. The inequality $\geq$ in (\ref{eq:SDP}) is a matrix inequality that constrains $F(x)$ to be positive semidefinite.  The SDP is convex as both the objective function and the constraint are convex: for $F(x)\geq 0$ and $F(y)\geq 0$, we see that
\begin{equation}
  F(\lambda x+(1-\lambda) y)=\lambda F(x)+(1-\lambda)F(y)\geq 0,
\end{equation}
for all $\lambda\in[0,1]$.  Convex optimization is particularly valuable, as the problem is guaranteed to have a unique global minimum and is not troubled by the multiple local minima that often arise in non-convex optimization.

We will show in Chapter \ref{chap:OptQER} that the CPTP constraint for quantum operations can be understood as a semidefinite constraint, thus leading to the SDP.  (the Choi matrix representation of a CPTP map makes this particularly plain to see.)  SDP's have been applied to several quantum information topics including distillable entanglement \cite{Rai:01,DohParSpe:02,DohParSpe:05,BraVia:04}, quantum detection \cite{EldMegVer:03,Eld:03a,Eld:03b,EldStoHas:04,JezRehFiu:02}, optimizing completely positive maps (including channel-adapted QEC) \cite{AudDem:02,YamHarTsu:05,FleShoWin:07,KosLid:06}, and quantum algorithms for the ordered search problem\cite{ChiLanPar:07}.

\section{Channel Models}

We are interested in adapting an error correction scheme to a physical noise process.  To do so, we must choose relevant models to describe the form noise may take.  For an experimental procedure, the model for the noise will be governed by the observed decoherence process of the physical apparatus.  In such cases, the noise model will be chosen to best match the physical realities.  This dissertation is not tied to any specific physical process; we seek instead channel models that will illustrate the principles of channel-adapted QEC.\\

We prove in Sec.~\ref{sec:QECproof} that only some channels lead to effective channel-adaptation.  Specifically, in the case of a stabilizer code and channel operator elements that are members of the Pauli group, a maximum likelihood recovery after projecting onto code stabilizers is the optimal recovery operation.  Most of the time, this recovery is indeed the generic QEC recovery without any channel-adaptation.  We are therefore interested in quantum channel models whose operator elements cannot be written as scaled members of the Pauli group.

The remainder of this section will briefly describe the channel models of interest in the remainder of this dissertation.  The numerical techniques described throughout will be applied to each of these channels.  The results will be presented in the main body of the dissertation if they illustrate a particular principle; the remainder will be presented in App.~\ref{chap:App Figures}.

\subsection{Amplitude damping channel}\label{sec:AmpDampChannel}

The first channel for consideration is the amplitude damping channel, which we will denote $\E_a$.  Amplitude damping was the example used in \cite{FleShoWin:07} to illustrate optimal QER, as well as the example for channel-adapted code design of \cite{LeuNieChuYam:97}.  The channel is a commonly encountered model, where the parameter $\gamma$ indicates the probability of decaying from state $\ket{1}$ to $\ket{0}$ (\emph{i.e.}~the probability of losing a photon).  For a single qubit, $\E_a$ has operator elements
\begin{equation}\label{eq:ampdamp}
E_0=\left [ \begin{array}{ccc} 1 & 0 \\ 0 &\sqrt{1-\gamma} \end{array} \right ]\hspace{.5 cm} \textrm{and} \hspace{.5 cm}
E_1=\left [ \begin{array}{ccc} 0 & \sqrt{\gamma} \\ 0 & 0 \end{array} \right ].
\end{equation}

The amplitude damping channel is both physically relevant and conceptually simple.  In that way, it is perhaps the best choice for illustrating channel-adapted QEC.  We will often cite our results in terms of the amplitude damping channel, though it is important to point out that the numerical routines presented in this dissertation do not require such a simple channel model. Channel-adapted QEC for the amplitude damping channel will be examined quite closely in Chapter \ref{chap:ThreeQubitCode}.

\subsection{Pure states rotation channel}

We will next consider a qubit channel that is less familiar, though with a straightforward geometric description.  We will call this the `pure states rotation' channel and label it as $\E_{ps}$.  To describe the channel, we define a pure state by its angle in the $xz$-plane: $\ket{\theta}=\cos\theta\ket{0}+\sin\theta\ket{1}$.  The channel mapping is defined by its action on two pure states an angle $\theta$ apart, symmetric about the $z$-axis.  When $\ket{\pm\theta/2}$ is input to the channel, the result is $\ket{\pm(\theta-\phi)/2}$, also as a pure state.  Thus, these two states are rotated toward each other by $\phi$.  Any other state input to the channel will emerge mixed.  The operator elements for this channel can be written as
\begin{eqnarray}\label{eq:purestates channel}
\E_{ps}&\sim&\left\{
\alpha \begin{bmatrix}
  \cos\frac{\theta-\phi}{2}\sin\frac{\theta}{2} & \pm\cos\frac{\theta-\phi}{2}\cos\frac{\theta}{2}\\
  \pm\sin\frac{\theta-\phi}{2}\sin\frac{\theta}{2} &   \sin\frac{\theta-\phi}{2}\cos\frac{\theta}{2}
\end{bmatrix},
\beta \begin{bmatrix}
  \frac{\cos\frac{\theta-\phi}{2}}{\cos\frac{\theta}{2}} & 0\\
  0 & \frac{\sin\frac{\theta-\phi}{2}}{\sin\frac{\theta}{2}}
\end{bmatrix}\right\},
\end{eqnarray}
where $\alpha$ and $\beta$ are constants chosen to satisfy the CPTP constraint.

It is worth taking a closer look at the operators in (\ref{eq:purestates channel}).  The first two operators have the form $\ket{\pm(\theta-\phi)/2}\bra{\mp\theta+\pi/2}$.  If we think of $\ket{\pm\theta/2}$ as the states targeted for rotation by $\phi/2$, then $\ket{\mp\theta+\pi/2}$ are states orthogonal to the targets.  We understand the first operator as projecting onto all states orthogonal to $\ket{\theta/2}$ and mapping each to $\ket{-(\theta-\phi)/2}$.  The second operator performs the same function for $\ket{-\theta/2}$.  The third operator $E_3$ is constrained such that $E_3\ket{\pm\theta/2}\propto \ket{\pm(\theta-\phi)/2}$.

The pure states rotation channel has multiple parameters which characterize its behavior.  $\theta$ indicates the initial separation of the targeted states.  $\phi$, the amount of rotation, clearly parameterizes the `noise strength' as $\phi=0$ indicates no decoherence while $\phi=\theta$ is strong decoherence.  Furthermore, we have chosen the target states to be symmetric about the $z$-axis, but this is only for clarity in stating the channel; any alternate symmetry axis may be defined.  Furthermore, a similar channel with asymmetric rotations $\phi_1$ and $\phi_2$ may be defined.  This, however, corresponds to a symmetric channel followed by a unitary rotation.  While less physically motivated, the pure state rotation channel model provides an extended set of qubit channels which are not represented with Pauli group operator elements.  We will look at examples of this channel where $\theta= 5\pi/12$ and $\pi/4$.  There is no particular significance to these choices; they merely illustrate well the principles of channel-adapted QEC.

\section{Quantum Error Correcting Codes}

In many cases, we will choose to analyze channel-adapted QEC beginning with known and established quantum error correcting codes.  To that end, we will describe briefly each code of interest.  We will make use of the stabilizer formalism to describe each, which we will summarize.  A reader preferring a more detailed introduction to quantum error correction is referred to the good introductory article \cite{Got:00}.

In standard terminology, a quantum code is referred to as an $[n,k,d]$ code indicating that $k$ logical qubits of information are encoded into $n$ physical qubits.  (The third entry $d$ indicates the distance of the code, where $d\geq2t+1$ is necessary to correct $t$ arbitrary qubit errors.  We will not make much use of code distance in our discussions, and will often omit it when describing a code.)  We will refer to $\HH_S$ as the `source space,' or the space of logical qubits emerging from a source of information, which has dimension $d_S=2^k$.  After encoding, the quantum state lies in a subspace of $\HH_C$ which has dimension $d_C=2^n$.  The subscript $C$ is chosen to indicate `code,' but it should be noted the $\HH_C$ is the larger Hilbert space of $n$ qubits in which the encoded state lies, not the code subspace.

\subsection{The quantum error correction conditions}

Before discussing quantum codes, it is useful to understand the conditions that must be met for standard QEC\cite{KniLaf:97}.  We define $P_C\in\LL(\HH_C)$ as a projector onto the code subspace.  Let $\{E_i\}\in\LL(\HH_C)$ be a set of error operations.  There is a recovery operation $\R$ that perfectly corrects these errors if and only if
\begin{equation}\label{eq:quantum error correcting conditions}
  P_CE_i^\dagger E_j P_C = \alpha_{ij}P_C
\end{equation}
for some complex scalars $\alpha_{ij}$.  (\ref{eq:quantum error correcting conditions}) is known as the quantum error correction conditions.  The conditions are more easily understood by noting the following two facts.  First, if $\{E_i\}$ satisfy the error correcting conditions, then any linear combination of $\{E_i\}$ also satisfy the error correcting conditions.  Using this fact, we arrive at the second observation: for $\{E_i\}$ that satisfy (\ref{eq:quantum error correcting conditions}), we can always derive a set of operators $\{E'_i\}$ such that
\begin{equation}\label{eq:orthogonal qecc}
  P_CE_i'^\dagger E'_j P_C = \alpha'_{ii}\delta_{ij}P_C.
\end{equation}
(We can compute $\{E'_i\}$ by noting that $\alpha_{ij}$ is a Hermitian matrix and therefore unitarily diagonalizable.  See \cite{NieChu:B00} for details.)

We can gain an intuitive picture of how QEC works through (\ref{eq:orthogonal qecc}).  When the errors $\{E'_i\}$ act on a state in the quantum code, the state is rotated into an orthogonal subspace.  The rotation is uniform across the subspace.  Furthermore, each of the errors rotates into a distinct subspace.  The recovery operation may be constructed as a projection onto each of these error subspaces whose result specifies an error syndrome.  Depending on the syndrome measurement, we can rotate the resulting state back into the code subspace. In this way, the original state is perfectly preserved.

\begin{table}[t]
\begin{center}
\begin{tabular}{ccc}
  $X=\begin{bmatrix}
    0&1\\1&0
  \end{bmatrix}$
  &
  $Y=\begin{bmatrix}
    0&-i\\i&0
  \end{bmatrix}$
  &
  $Z=\begin{bmatrix}
    1&0\\0&-1
  \end{bmatrix}$
\end{tabular}
\end{center}
\caption{The Pauli matrices.}\label{tab:Pauli}
\end{table}

It is useful to consider the case when the errors $\{E_i\}$ are given by the Pauli matrices of Table \ref{tab:Pauli}.  It is not hard to see that the Pauli matrices, together with the identity operator $I$, form a basis for $\LL(\HH_2)$, the linear operators on a single qubit.  Imagine now the set $\{I^{\otimes n}, X_i,Y_i,Z_i\}\in\LL(\HH_2^{\otimes n})$, which are operators on $n$ qubits.  The subscript $i$ indicates that the Pauli operator acts on the $i^{th}$ qubit and the identity acts on the others.  If $\{I^{\otimes n}, X_i,Y_i,Z_i\}$ satisfy (\ref{eq:quantum error correcting conditions}), then an arbitrary operation restricted to a single qubit is also correctible.  In this way, we can design quantum codes that can correct for an arbitrary error on a single qubit.

\subsection{The stabilizer formalism}\label{sec:stabilizers}

We will make use of the stabilizer formalism \cite{Got:96,CalRaiShoSlo:97,CalRaiShoSlo:98,Got:97} to describe quantum error correcting codes, and their generic recovery operations.  The Pauli group on 1 qubit is given by $\mathcal{G}_1=\{\pm I, \pm i I, \pm X, \pm i X, \pm Y, \pm iY, \pm Z,\pm iZ\}$, where $X$, $Y$, and $Z$ are the Pauli matrices.  The multiplicative constants $\pm 1 $ and $\pm i$ are included so that $\mathcal{G}_1$ is closed under multiplication, and thus a proper group.  The stabilizer formalism for an $[n,k]$ code works with the Pauli group over $n$ qubits $\mathcal{G}_n$, the $n$-fold tensor product of $\mathcal{G}_1$.  We will use two equivalent notations for an element of $\mathcal{G}_n$: $Z_1X_2$ refers to a $Z$ on qubit 1 and an $X$ on qubit 2; $ZXIIIII$ indicates the same operation, where it is evident that we refer to an element of $\mathcal{G}_7$.  It is worth noting that any two elements $g_i,g_j\in\mathcal{G}_n$ either commute or anti-commute (\emph{i.e.}~$[g_i,g_j]=0$, or $\{g_i,g_j\}=0$).

A group $G$ can be specified in a compact form by defining its generator set $\langle g_1,\ldots,g_l \rangle$ where $[g_i,g_j]=0$.  Then any $g\in G$ can be written as a product of the generators (in any order, since they commute).  We connect back to error correction by noting that a group $G\subset \mathcal{G}_n$ can specify a subspace $C(G)$ on $\HH_2^{\otimes n}$, the space of $n$ qubits.  A state $\ket{\psi}$ is in the subspace if and only if $\ket{\psi}=g\ket{\psi}$ for all $g\in G$.  We note two useful facts: to show that $\ket{\psi}\in C(G)$, we need only check the generators of $G$.  Also, if $-I\in G$, then the subspace is trivial, with only the multiplicative identity $0\in C(G)$.

We may specify an $[n,k]$ quantum code by providing a set of $n-k$ generators in $\mathcal{G}_n$, which in turn determine the $2^k$ dimensional code subspace.  It is also very useful to create a structured set of operators to characterize the states of the code subspace.  To do so, we define $\bar{Z}_i$ and $\bar{X}_i$, $i=1,\ldots,k$ which act as logical Pauli operators.  These are elements of the normalizer of $G$, denoted $N(G)$ which means that they commute with all of the generators of $g$.  Thus, they transform states within the code subspace.  To fulfill the function of logical Pauli operation, we require the following five properties:
\begin{eqnarray}
  \left[ \bar{Z}_i,g\right ]=\left [\bar{X}_i,g\right ]&=& 0\textrm{ for all }g\in G,\\
  \left [\bar{Z}_i, \bar{Z}_j\right ]&=&0,\\
  \left [\bar{X}_i, \bar{X}_j\right ]&=&0,\\
  \left [\bar{Z}_i, \bar{X}_j\right ]&=&0,\textrm{ for }i\neq j,\\
  \textrm{and }\{\bar{Z}_i,\bar{X}_i\}&=&0.
\end{eqnarray}
We can then define the logical codeword with the logical Pauli matrices as
\begin{equation}
\ket{i_1\cdots i_k}=\bar{Z}_1^{i_1}\cdots\bar{Z}_k^{i_k}\ket{i_1\cdots i_k}.
\end{equation}

The syndrome measurement typically associated with a stabilizer code is to measure each generator $g_i$.  As the generators commute, this can be done in any order.  This is a projective measurement onto the $+1$ and $-1$ eigen-space of each generator.  If the state $\ket{\psi}$ is in the code subspace $C(G)$, then each generator will measure 1.  Suppose instead the state was corrupted by an error $E$ that anti-commutes with a generator, say $g_1$.  Then we see that the state $g_1E\ket{\psi}=-Eg_1\ket{\psi}=-E\ket{\psi}$ lies in the $-1$ eigen-space of $g_1$ and the measurement will so indicate.  If this is the only error, the syndrome measurement will detect it, and we will be able to apply $E^\dagger$ to recover the state.

Problems arise when an error $E$ commutes with all of the generators but is not itself in the group $G$.  This will corrupt the state in a way that cannot be detected.  In this case $E$ is in the normalizer of $G$, $N(G)$.  Furthermore, if two distinct errors $E_1$ and $E_2$ both yield the same error syndrome, the recovery operation will not be able to correct both.  In that case $E_1^\dagger E_2\in N(G)-G$.  In fact, this is the error-correcting condition for stabilizer codes: a set of errors $\{E_i\}\subset \mathcal{G}_n$ are correctible if and only if $E_i^\dagger E_j\in N(G)-G$ for all $i,j$.  Furthermore, any error that is a linear combination of $\{E_i\}$ is also correctible - the syndrome measurement will `discretize' the error by projecting onto one of the syndromes.

The generic QEC recovery operation for a stabilizer code consists of the error syndrome measurement of measuring each generator, followed by the appropriate recovery operation.  By appropriate, we mean the most likely element of the Pauli group $\mathcal{G}_n$ that returns the observed syndrome to the code subspace $C(G)$.  In general, it is assumed that the most likely correction will be the minimum weight (\emph{i.e.}~smallest number of non-Identity terms) element of $\mathcal{G}_n$.

We will now state several of the quantum error correcting codes, each in terms of the stabilizer formalism.  We will refer to each of these codes at various times in the dissertation.

\subsection{Shor code}

\begin{table}[tb]\label{table:codes}
\begin{center}
\begin{tabular}{c|c|c}

\begin{tabular}{c|c}
$[9,1]$&Shor code\\
\hline
\hline
Name & Operator\\
\hline
\begin{tabular}{c}
$g_1$\\
$g_2$\\
$g_3$\\
$g_4$\\
$g_5$\\
$g_6$\\
$g_7$\\
$g_8$\\
$\bar{Z}$\\
$\bar{X}$
\end{tabular}&
\begin{tabular}{c@{}c@{}c@{}c@{}c@{}c@{}c@{}c@{}c@{}c}
$Z$&$Z$&$ I$&$ I$&$ I$&$ I$&$ I$&$ I$&$ I$\\
$I $&$Z$&$ Z$&$ I$&$ I$&$ I$&$ I$&$ I$&$ I$\\
$I$&$ I$&$ I$&$ Z$&$ Z$&$ I$&$ I$&$ I$&$ I$\\
$I$&$ I$&$ I$&$ I$&$ Z$&$ Z$&$ I$&$ I$&$ I$\\
$I$&$ I$&$ I$&$ I$&$ I$&$ I$&$ Z$&$ Z$&$ I$\\
$I$&$ I$&$ I$&$ I$&$ I$&$ I$&$ I$&$ Z$&$ Z$\\
$X$&$ X$&$ X$&$ X$&$ X$&$ X$&$ I$&$ I$&$ I$\\
$I$&$ I$&$ I$&$ X$&$ X$&$ X$&$ X$&$ X$&$ X$\\
$X$&$ X$&$ X$&$ X$&$ X$&$ X$&$ X$&$ X$&$ X$\\
$Z$&$ Z$&$ Z$&$ Z$&$ Z$&$ Z$&$ Z$&$ Z$&$ Z$
\end{tabular}\\
\hline
\hline
\end{tabular}
&
\begin{tabular}{c|c}
$[7,1]$&Steane Code\\
\hline
\hline
Name & Operator\\
\hline
\begin{tabular}{c}
$g_1$\\
$g_2$\\
$g_3$\\
$g_4$\\
$g_5$\\
$g_6$\\
$\bar{Z}$\\
$\bar{X}$
\end{tabular}&
\begin{tabular}{c@{}c@{}c@{}c@{}c@{}c@{}c@{}c}
$I$&$ I$&$ I$&$ X$&$ X$&$ X$&$ X$\\
$I$&$ X$&$ X$&$ I$&$ I$&$ X$&$ X$\\
$X$&$ I$&$ X$&$ I$&$ X$&$ I$&$ X$\\
$I$&$ I$&$ I$&$ Z$&$ Z$&$ Z$&$ Z$\\
$I$&$ Z$&$ Z$&$ I$&$ I$&$ Z$&$ Z$\\
$Z$&$ I$&$ Z$&$ I$&$ Z$&$ I$&$ Z$\\
$Z$&$ Z$&$ Z$&$ Z$&$ Z$&$ Z$&$ Z$\\
$X$&$ X$&$ X$&$ X$&$ X$&$ X$&$ X$
\end{tabular}\\
\hline
\hline
\end{tabular}
&
\begin{tabular}{c|c}
[5,1]& Code\\
\hline
\hline
Name & Operator\\
\hline
\begin{tabular}{c}
$g_1$\\
$g_2$\\
$g_3$\\
$g_4$\\
$\bar{Z}$\\
$\bar{X}$
\end{tabular}&
\begin{tabular}{c@{}c@{}c@{}c@{}c@{}c}
$X$&$ Z$&$ Z$&$ X$&$ I$\\
$I$&$ X$&$ Z$&$ Z$&$ X$\\
$X$&$ I$&$ X$&$ Z$&$ Z$\\
$Z$&$ X$&$ I$&$ X$&$ Z$\\
$Z$&$ Z$&$ Z$&$ Z$&$ Z$\\
$X$&$ X$&$ X$&$ X$&$ X$
\end{tabular}\\
\hline
\hline
\end{tabular}
\end{tabular}
\end{center}
\caption{Generators and logical operations of the Shor code, Steane code, and five qubit code.}
\end{table}

The Shor code\cite{Sho:95} was the first example of a quantum error correcting code.  It is a $[9,1]$ code that is the quantum equivalent to the repetition code.  In the classical case, a logical $0$ is represented by the three bit codeword $000$, and logical $1$ as $111$.  This protects from a single bit flip error when decoding is performed via majority voting.  The Shor code works in a similar manner, but in this case, one must protect from both bit flip (Pauli $X$) and phase flip (Pauli $Z$) errors.  The stabilizers for the quantum code are provided in Table \ref{table:codes}, but in this case the actual logical code words are also revealing:
\begin{eqnarray}
  \ketsub{0}{L}=\frac{1}{2\sqrt{2}}(\ket{000}+\ket{111})\otimes(\ket{000}+\ket{111})\otimes
  (\ket{000}+\ket{111})\\
  \ketsub{1}{L}=\frac{1}{2\sqrt{2}}(\ket{000}-\ket{111})\otimes(\ket{000}-\ket{111})\otimes
  (\ket{000}-\ket{111}).
\end{eqnarray}
It is instructive to talk through the stabilizer measurements, as this may provide further intuition on the recovery procedure.  We can see that the first three qubits have the form $\ket{000}\pm\ket{111}$ which protect against bit flips.  Consider measuring the stabilizer $Z_1Z_2$ (\emph{i.e.}~Pauli $Z$ on the first and second qubits).  This will yield a $+1$ if the state is in the code space, and a $-1$ if the first and second bit are not aligned (\emph{e.g.}~$\ket{010}\pm\ket{101}$).  A measurement of $Z_2Z_3$ tests the second and third bits.  If both of these measurements is $-1$, we know the middle bit was flipped.  If only one of the first or second measurements results in a $-1$, that will indicate a flip of the first or the third bit, respectively.

Consider now the three blocks of three qubits.  We notice that the sign of the $\ket{111}$ terms are aligned when the state is in the code subspace.  Thus, we can measure $X_1X_2X_3X_4X_5X_6$ and $X_4X_5X_6X_7X_8X_9$ and determine if the phases match.  In a manner equivalent to the bit flips discussed above, we can determine if one of the blocks needs to have a Pauli $Z$ applied.  Notice that $Z_1$, $Z_2$, or $Z_3$ will all transform to the same state.  This is because the Shor code is a \emph{degenerate} code, and each of these errors yields the same syndrome and can be corrected by the same recovery.

The final note is that the Shor code can correct for both a $X$ error and a $Z$ error, which if they occur on the same qubit yield a $\pm iY$ error.  The code can thus correct for any of the single qubit Pauli operators, and thus can correct an arbitrary qubit error as the Pauli's, together with the identity, form a basis for qubit operators.

\subsection{Steane code}

The Steane code is a $[7,1]$ code of the CSS class of codes\cite{CalSho:96,Ste:96a}.  CSS codes come from classical codes that are self-dual, which allows an elegant extension to quantum codes.  The Steane code is created from the classical [7,4] Hamming code, a linear, self-dual code.  The Hamming code has several nice properties, especially the ease with which decoding can be performed.  While CSS codes are interesting in themselves, they are a subclass of the stabilizer codes, and it will be sufficient for our purposes to give the stabilizers for the Steane code in Table \ref{table:codes}.  We will, however, note that CSS codes are particularly valuable in fault tolerant quantum computing, as encoding circuits and encoded logical operations have a simple form.  For this reason, the Steane code is a popular, though costly in overhead, choice of experimentalists.

\subsection{Five qubit stabilizer code}

The five-qubit code was independently discovered by \cite{BenDivSmoWoo:96} and \cite{LafMiqPazZur:96}.  We will here follow the treatment in \cite{NieChu:B00} and specify the code via the generators $\{g_1,g_2,g_3,g_4\}$ and the logical $\bar{Z}$ and $\bar{X}$ operations given in Table \ref{table:codes}.  The code subspace $\mathcal{C}$ is the two-dimensional subspace that is the $+1$ eigenspace of the generators $g_i$.  The logical states $\ketsub{0}{L}$ and $\ketsub{1}{L}$ are the $+1$ and $-1$ eigenkets of $\bar{Z}$ on $\mathcal{C}$.  The five qubit code is the shortest block code that can correct for an arbitrary error on a single qubit.

\subsection{Four qubit [4,1] `approximate' amplitude damping code}\label{sec:4 qubit code}

We turn now to a code developed in 1997 on principles of channel-adapted QEC.  In \cite{LeuNieChuYam:97}, a [4,1] code and recovery operation was presented which was adapted specifically for the amplitude damping channel.  By channel-adaptation, the [4,1] code can duplicate the performance of a generic block code while only utilizing four physical qubits.

Leung \emph{et.~al.}~label their code as an `approximate' code, as it does not exactly satisfy the quantum error correction conditions.  Instead, they derive a set of approximate error correcting conditions, and show that their code achieves them.  The code maintains a high minimum fidelity for small values of $\gamma$, and in fact approximates the performance of the five qubit stabilizer code.

\begin{figure}
\begin{center}
\begin{tabular}{ccc}

\Qcircuit @C=1.3em @R=.7em {
\lstick{n_1} & \ctrl{1} &  \qw   & \qw & \multigate{2}{W_k}\\
\lstick{n_2} & \targ    & \meter & M_2 & \\
\lstick{n_3} & \ctrl{1} &  \qw   & \qw &\ghost{W_k} \\
\lstick{n_4} & \targ    & \meter & M_4 &
}\vspace{.3em}

& \hspace{1em} &
\Qcircuit @C=1.3em @R=.7em {
\lstick{n_1} & \targ     & \qw           & \ctrl{1}            & \qw    & \ket{\psi_{\textrm{out}}}\\
\lstick{n_3} & \ctrl{-1} & \gate{\theta} & \gate{\pi/4-\theta} & \meter &
}\\
(A) & & (B)\\

\\
\Qcircuit @C=1.3em @R=.7em {
\lstick{n_1}   & \qw     &\qw           &\qw\\
\lstick{\ket{0}}& \qw     &\gate{\theta'}&\meter\\
\lstick{n_3}   & \gate{X}&\ctrl{-1}     &\qw&\ket{\psi_{\textrm{out}}}
}
& &
\Qcircuit @C=1.3em @R=.7em {
\lstick{n_1}   & \gate{X}&\ctrl{1}     &\qw &\ket{\psi_{\textrm{out}}}\\
\lstick{\ket{0}}& \qw     &\gate{\theta'}&\meter\\
\lstick{n_3}   & \qw     &\qw           &\qw
}
\\
(C) & & (D)
\end{tabular}
\end{center}
\caption[Leung \emph{et.al.}~recovery circuit for the four qubit approximate code.]{Leung \emph{et.al.}~recovery circuit for the four qubit approximate code.  (A) is the circuit for error syndrome detection.  Measurement results $(M_2,M_4)$ determine the recovery operation $W_k$ to be performed.  If the result $(M_2,M_4)$ is 00, 10, or 01, use circuits (B), (C), or (D), respectively.  The angles $\theta$ and $\theta'$ are given by $\tan\theta=(1-\gamma)^2$ and $\cos \theta'=1-\gamma$.  The rotation gate with angle $\tilde{\theta}$ is understood to perform the operation $\exp(i\tilde{\theta}Y)$.}\label{fig:Approx QEC Recovery}
\end{figure}
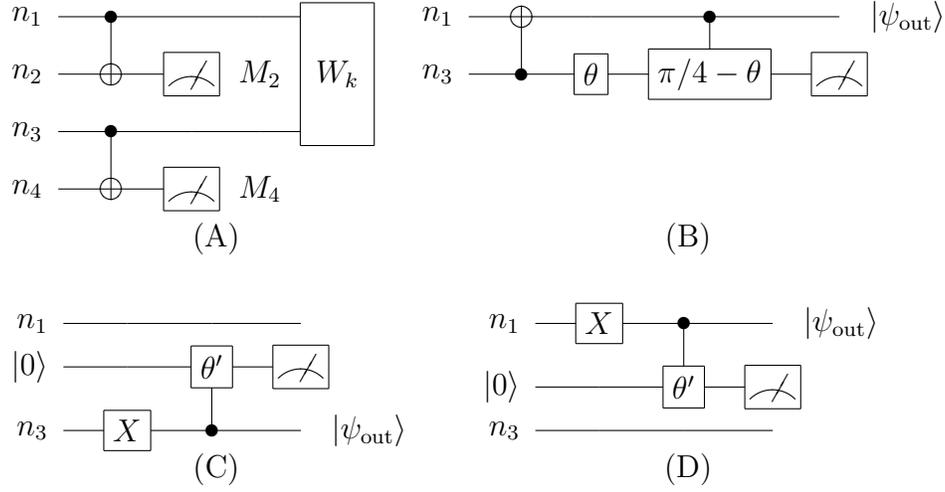

The logical states of the code are given by
\begin{eqnarray}\label{eq:4qubitcode}
\ketsub{0}{L}&=& \frac{1}{\sqrt{2}}(\ket{0000}+\ket{1111})\\
\ketsub{1}{L}&=& \frac{1}{\sqrt{2}}(\ket{0011}+\ket{1100}),
\end{eqnarray}
and the recovery operation is specified by the circuits in Fig.~\ref{fig:Approx QEC Recovery} which is a recreation of Fig.~2 of \cite{LeuNieChuYam:97}.  We note that the recovery operation depends explicitly on the parameter $\gamma$.  We revisit this recovery operation in Sec.~\ref{sec:qualitative analysis 4,1}.

\chapter{Optimum Channel-Adapted QEC}\label{chap:OptQER}

Standard quantum error correction is closely related to classical digital error correction.  Codes, errors, syndromes, and recovery operations are all derived from their classical counterparts.  Errors are either perfectly corrected or not corrected at all.  Standard QEC components are classical ideas extended to function in the Hilbert space of quantum systems.

In general, channel-adapted QEC is not so tidy.  The principles of error correction are unchanged, but we now explore more purely quantum effects.  To gain a full picture of the potential benefits of channel-adaptation, we depart from the classical analogues and explore the full space of quantum operations.

Rather than determining successful QEC through a binary question of corrected and uncorrected errors, we consider the fidelity of the composite operation.   More specifically, we consider the encoder, channel, and recovery operation as a single quantum channel and evaluate its performance in terms of the average entanglement fidelity.  In so doing, we utilize the computational power of semidefinite programming to establish the power of optimal channel-adapted QEC.

\section{Quantum Error Recovery (QER)}

\begin{figure}[tb]
  \begin{center}
    \includegraphics[width=\hsize]{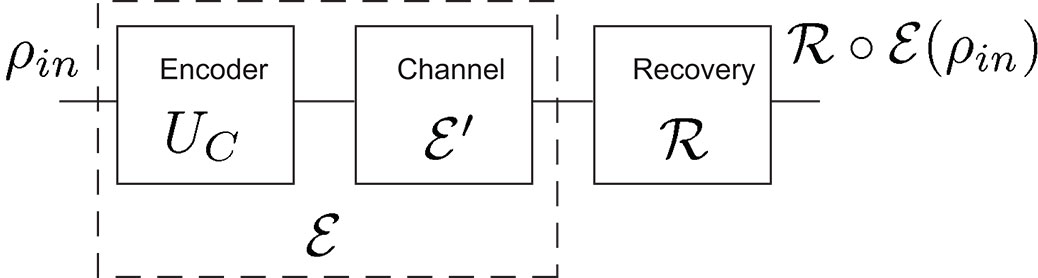}
  \end{center}
  \caption[Quantum error correction block diagram.]{Quantum error correction block diagram.  For channel-adapted recovery, the encoding isometry $U_C$ and the channel $\E'$ are considered as a fixed operation $\E$ and the recovery $\R$ is chosen according to the design criteria.}\label{fig:block diagram}
\end{figure}

The block diagram for quantum error correction is quite simple, as can be seen in Fig.~\ref{fig:block diagram}.  An isometry $U_C$ encodes the information in the quantum state $\rho$ into the Hilbert space $\HH_C$ of dimension $d_C$.  This encoded state is corrupted by the noisy channel $\E'$, after which the recovery operation $\R$ attempts to correct the state.

The design of a QEC system consists of the selection of the encoding isometry $U_C$ and the recovery operation $\R$.  The common practice is to select $U_C$ and $\R$ independent of the channel $\E'$, assuming only that errors are localized to individual qubits and occur independently.  Channel-adapted QEC selects $U_C$ and $\R$ based upon the structure of the channel $\E'$.

It is intuitive and correct to presume that channel-adaptation can be effective on both the choice of encoding and recovery operation.  However, we shall see that the optimization problem is greatly simplified when one of the two is held as fixed.  For most of this chapter, we assume a fixed choice of encoding isometry $U_C$ and optimize the choice of $\R$.  In this way, we will discover many of the principles of channel-adaptation and take advantage of the favorable optimization properties.  Thus, we define the channel $\E:\LL(\HH_S)\mapsto\LL(\HH_C)$ as the composition of the encoding isometry $U_C$ and the noisy operation $\E'$.  When the recovery operation is adapted for a fixed encoding and channel, we will declare the process channel-adapted \emph{Quantum Error Recovery} (QER).

\section{Optimum QER via Semidefinite Programming (SDP)}\label{sec:OptQERSDP}

To determine an appropriate recovery operation $\R$, we wish to maximize the fidelity of the input source to the the output of $\R\circ\E$.  We will make use of the average entanglement fidelity described in Sec.~\ref{sec:fidelity}, declaring the source to be an ensemble $E$ of states $\rho_k$ with probability $p_k$.  The optimization problem becomes
\begin{equation}\label{eq:ave_ent_fidelity_max}
\mathcal{R}^\star=\arg\max_{\{\mathcal{R}\}} \bar{F}_e(E,\mathcal{R}\circ\mathcal{E}),
\end{equation}
where $\{\R\}$ is the set of all CPTP maps from $\LL(\HH_C)\mapsto\LL(\HH_S)$ and $\arg$ refers to the element of $\{\mathcal{R}\}$ that achieves the maximum.

The problem given by  (\ref{eq:ave_ent_fidelity_max}) is a convex optimization problem, and we may approach it with sophisticated tools.  Particularly powerful is the semidefinite program (SDP), discussed in \ref{sec:SDP}, where the objective function is linear in an input constrained to a semidefinite cone.  Indeed, the power of the SDP is a primary motivation in choosing to maximize the average entanglement fidelity, which is linear in the quantum operation $\R$.

Using the expression for the average entanglement fidelity in (\ref{eq:ave_ent_fidelity}), we may now include the constraints in  (\ref{eq:ave_ent_fidelity_max}) to achieve an optimization problem readily seen to be a semidefinite program.  To do this, we must consider the form of the Choi matrix for the composite operation $\R\circ\E:\LL(\HH)\mapsto\LL(\HH)$.  If the operator elements for the recovery and channel are $\{R_i\}$ and $\{E_j\}$, respectively, then the operator $X_{\R\circ\E}$ is given by
\begin{equation}
X_{\R\circ\E}=\sum_{ij}\kett{R_iE_j}\braa{R_iE_j}.
\end{equation}
Applying  (\ref{eq:kett_triple_product}), this becomes
\begin{eqnarray}
\nonumber X_{\R\circ\E}&=&\sum_{ij}I\otimes \ol{E_j}\kett{R_i}\braa{R_i}I\otimes \ol{E_j}^\dagger\\
&=& \sum_j (I\otimes \ol{E_j})X_\R (I\otimes \ol{E_j}^\dagger).
\end{eqnarray}
The average entanglement fidelity is then
\begin{eqnarray}
\nonumber
\bar{F}_{e}(E,\R\circ\E)&=&\sum_{jk} p_k \braa{\rho_k}(I\otimes \ol{E_j})X_\R (I\otimes \ol{E_j}^\dagger)\kett{\rho_k}\\
\label{eq:ent_fid_Choi}&=& \tr{X_\R C_{E,\E}},
\end{eqnarray}
where
\begin{eqnarray}\nonumber
C_{E,\E}&=&\sum_{jk} p_k I\otimes \ol{E_j}\kett{\rho_k}\braa{\rho_k}I\otimes\ol{E_j}^\dagger\\
&=& \sum_{jk} p_k \kett{\rho_k E_j^\dagger}\braa{\rho_k E_j^\dagger}.
\end{eqnarray}

We may now express the optimization problem  (\ref{eq:ave_ent_fidelity_max}) in the simple form
\begin{eqnarray}\nonumber
X_{\R}^\star=\arg\max_X \tr ({X C_{E,\E}})\\ \textrm{such that }
X \geq 0, \hspace{10 pt} \tr_{\HH_S}{X}=I.\label{eq:ave_ent_fid_max_SDP}
\end{eqnarray}
This form illustrates plainly the linearity of the objective function and the semidefinite and equality structure of the constraints.  Indeed, this is the exact form of the optimization problem in \cite{AudDem:02}, which first pointed out the value of the SDP for optimizing quantum operations.

We should reiterate the motivation for using the average entanglement fidelity over an ensemble $E$.  The key attributes that lead to a semidefinite program are the CPTP constraint and the linearity of the objective function.  As both entanglement fidelity and ensemble average fidelity (when the states are pure) are linear in the choice of recovery operation, both can be solved via an SDP.  By deriving the SDP for average entanglement fidelity, it is trivial to convert to either entanglement fidelity or ensemble average fidelity.  In the former case, we simply define the ensemble $E$ as the state $\rho$ with probability 1.  For ensemble average fidelity, we define $E$ as a set of pure states $\{\ket{\psi_k}\}$ with probability $p_k$.

The value of an SDP for optimization is two-fold.  First, an SDP is a sub-class of convex optimization, and thus a local optimum is guaranteed to be a global optimum.  Second, there are efficient and well-understood algorithms for computing the optimum of a semidefinite program.  These algorithms are sufficiently mature to be widely available.  By expressing the optimum recovery channel as an SDP, we have explicit means to compute the solution for an arbitrary channel $\E$.  In essence, the numerical methods to optimize an SDP are sufficiently mature that we may consider them as a black box routine for the purposes of this dissertation.

\subsection{Optimal diversity combining}

Let us pause and consider the size of the optimization problem above.  We are optimizing over $X_\R$, which is an element of $\LL(\HH_S\otimes\HH_C^*)$ and thus has $d_C^2d_S^2=4^{n+k}$ matrix elements for an $[n,k]$ code.  It is not surprising that the size grows exponentially with $n$, since the Hilbert space $\HH_C$ has dimension $2^n$.  However, the fact that the growth goes as $4^n$ makes the SDP particularly challenging for longer codes.  This will be addressed in Chapter \ref{chap:NearOptQER}.

The dimensional analysis of the optimization problem motivates our choice of convention for $\R$.  Often, recovery operations are not written as decodings; instead of mapping $\LL(\HH_C)\mapsto\LL(\HH_S)$, it is common for $\R$ to be written $\LL(\HH_C)\mapsto\LL(\HH_C)$.  The structure of such an $\R$ is carefully chosen so that the output state lies in the code subspace.  This description of a non-decoding recovery operation is particularly valuable in the analysis of fault tolerant quantum computing, where the recovery operations restore the system to the code subspace but do not decode.  Were we to follow a similar convention, the number of optimization variables would grow as $16^n$.  Fortunately, this is not necessary.  We can calculate a decoding recovery operation and easily convert it back into a non-decoding operation by including the encoding isometry: $U_C\circ\R$.

The convention choice of $\E:\LL(\HH_S)\mapsto\LL(\HH_C)$ and $\R:\LL(\HH_C)\mapsto\LL(\HH_S)$ makes QER analogous to a common classical communications topic.  We may interpret $\E$ as a \emph{quantum spreading channel}, a channel in which the output dimension is greater than the input dimension.  The recovery operation is an attempt to combine the spread output  back into the input space, presumably with the intent to minimize information loss.  The recovery operation is then the quantum analog to the classical communications concept of diversity combining.

Classical diversity combining describes a broad class of problems in communications and radar systems.  In its most general form, we may consider any class of transmission problems in which the receiver observes multiple transmission channels.  These channels could arise due to multi-path scattering, frequency diversity (high bandwidth transmissions where channel response varies with frequency), spatial diversity (free-space propagation to multiple physically separated antennas),  time diversity, or some combination of the four.  Diversity combining is a catch-all phrase for the process of exploiting the multiple channel outputs to improve the quality of transmission (\emph{e.g.} by reducing error or increasing data throughput).

In a general description of classical diversity, the input signal is coupled through the channel to a receiver system of higher dimensionality.  Consider a communication signal with a single transmitter antenna and $N$ receiver antennae.  Often, the desired output is a signal of the same dimension as the input, a scalar in this case.  Diversity combining is then the process of extracting the maximum information from the $N$-dimensional received system.  In most communications systems, this combining is done at either the analog level (leading to beam-forming or multi-user detection) or digital level (making the diversity system a kind of repeater code).  Thus, the natural inclination is to equate diversity combining with either beam-forming or repeater codes.  The most general picture of diversity combining, however, is an operation that recombines the channel output into a signal of the same dimension as the input.  Thus, it is appropriate to consider a quantum spreading channel to be a quantum diversity channel, and the recovery operation to be a quantum diversity combiner.

Diversity combining provides extra intuition about the value of channel-adaptation.  Many routines to improve classical diversity combining begin with efforts to learn or estimate the channel response.  Channel knowledge greatly improves the efficacy of diversity combining techniques.  Analogously, information about the quantum noise process should allow more effective recovery operations.

\section{Examples}

\begin{figure}[tbh]
\begin{center}
\includegraphics[width=\columnwidth]{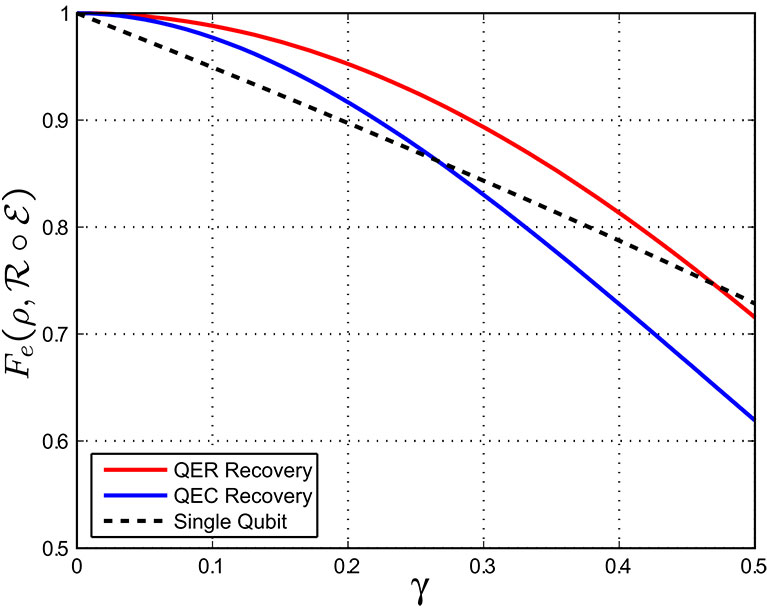}
\end{center}
\caption[Average entanglement fidelity vs.~$\gamma$ for the five qubit stabilizer code and the amplitude damping channel $\E_a^{\otimes 5}$.]{Average entanglement fidelity vs.~$\gamma$ for the five qubit stabilizer code and the amplitude damping channel $\E_a^{\otimes 5}$.  $\gamma$ refers to the damping parameter of the channel.  The performance of optimal channel-adapted QER is compared to non-adapted QEC.  Entanglement fidelity for a single qubit and no error correction (\emph{i.e.} $F_{e}(\rho,\E_a)$) is included and may be considered a baseline performance where no error correction is attempted.}
\label{fig:AmpDamp5_QER}
\end{figure}

\begin{figure}[tbh]
\begin{center}
\includegraphics[width=\columnwidth]{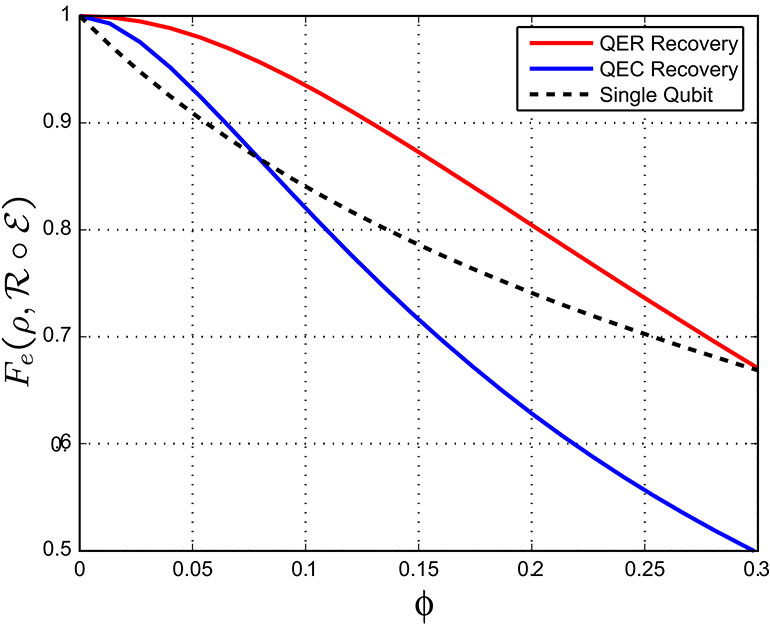}
\end{center}
\caption[Average entanglement fidelity vs.~$\phi$ for the five qubit stabilizer code and the pure state rotation channel with $\theta=5\pi/12$, $\E_{ps}^{\otimes 5}$.]{Average entanglement fidelity vs.~$\gamma$ for the five qubit stabilizer code and the pure state rotation channel with $\theta=5\pi/12$, $\E_{ps}^{\otimes 5}$.  $\phi$ refers to the amount by which the angle between pure states is reduced.  As $\phi$ increases, the channel may be considered noisier.  The performance of optimal channel-adapted QER is compared to non-adapted QEC.  Entanglement fidelity for a single qubit and no error correction (\emph{i.e.} $F_{e}(\rho,\E_{ps})$) is included and may be considered a baseline performance where no error correction is attempted.}
\label{fig:PureState5_QER}
\end{figure}

We illustrate the effects of channel-adapted QER by looking at the optimal recovery for the five qubit stabilizer code.  We consider the amplitude damping channel in Fig.~\ref{fig:AmpDamp5_QER} and the pure state rotation channel with $\theta=5\pi/12$ in Fig.~\ref{fig:PureState5_QER}.  We consider an ensemble $E$ that is in the completely mixed state $\rho=I/2$ with probability 1.  This simple ensemble is the minimal assumption that can be made about the source.  The optimal QER performance is compared to the non-adapted QEC performance.  We also include the average entanglement fidelity of a single qubit passed through the channel.  This indicates a baseline performance that is achieved when no error corrective procedure (encoding or recovery) is attempted.

Figures \ref{fig:AmpDamp5_QER} and \ref{fig:PureState5_QER} illustrate the potential gains of channel-adapted QEC.  We first note that the optimal recovery operation outperforms the non-adapted recovery by a non-trivial amount.  This confirms the intuition about the benefit of channel-adaptation and the inefficiency of non-adapted recovery.

To emphasize the benefits of channel-adaptation, consider respectively the high noise and low noise cases.  As the noise increases, moving to the right on the horizontal axis, we see the point where the recovery performance curve crosses the single qubit performance.  This threshold defines the highest noise for which the error correction scheme is useful; for noise levels beyond the threshold, the error correction procedure is doing more harm than good.  Notice that for a fixed encoding, channel-adaptation can significantly extend this threshold.  In the case of the amplitude damping channel, QEC performance dips below the baseline around $\gamma\approx 1/4$; optimal channel-adaptation crosses the baseline at nearly $\gamma\approx 1/2$.  The effect is even more pronounced for the pure state rotation channel; the $\phi$ where channel-adapted QER falls below the baseline is more than triple the cross-over threshold for non-adapted QEC.  (It is important to point out that this cross-over threshold is not directly related to the fault tolerant quantum computing (FTQC) threshold.  A much more extensive analysis is needed to approximate a channel-adapted FTQC threshold.  See Sec.~\ref{sec:FTQC}.)

Now consider the effect of channel-adapted QER as noise levels asymptotically approach 0.  This is particularly relevant as experimental methods for quantum computation improve the shielding of quantum systems from environmental coupling.  In both the amplitude damping and pure state rotation examples, the optimal channel-adapted performance is significantly greater than the non-adapted QEC.

We see this numerically by calculating the polynomial expansion of $F_{e}(\rho,\R\circ\E)$ as $\gamma$ goes to zero.  For the amplitude damping channel, the entanglement fidelity for the optimum QER has the form $F_{e}(\rho,\R\circ\E)\approx 1 -1.166 \gamma^2+\mathcal{O}(\gamma^3)$.  In contrast, the QEC recovery is $F_{e}(\rho,\R\circ\E)\approx 1 -2.5 \gamma^2+\mathcal{O}(\gamma^3)$.  For the pure state rotation channel with $\theta=5\pi/12$, the entanglement fidelity for the optimum QER has the form $F_{e}(\rho,\R\circ\E)\approx 1 -.13\phi -5.28 \phi^2+\mathcal{O}(\phi^3)$.  In contrast, the QEC recovery is $F_{e}(\rho,\R\circ\E)\approx 1 -1.24\phi -6.02 \phi^2+\mathcal{O}(\phi^3)$.

\section{QER Robustness}

Channel-adapted QEC is only useful if the model used for adaptation is a close match to the actual physical noise process.  This is an intuitively obvious statement - channel-adapting to the wrong noise process will be detrimental to performance.  If we are quite uncertain as to the form of the noise, a reasonable strategy is to stay with the generic QEC.  Consider instead a small amount of uncertainty; perhaps we know the form of the noise but are uncertain as to the strength.  How robust is the optimal QER operation to such an error?
\begin{figure}[tbh]
\begin{center}
\begin{tabular}{c}
\includegraphics[width=.7\columnwidth]{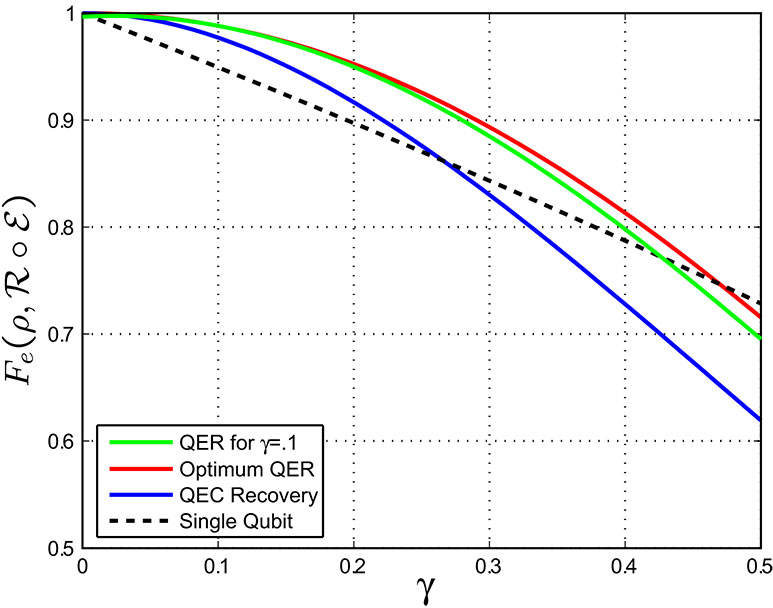}\\

(A)
\vspace{.5cm}\\
\includegraphics[width=.7\columnwidth]{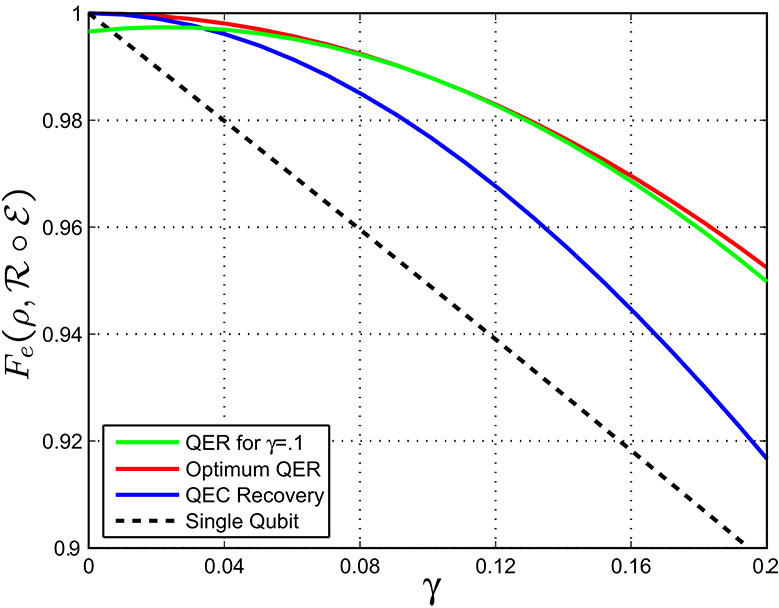}\\

(B)
\end{tabular}
\end{center}
\caption[Robustness of QER to $\gamma$ for the amplitude damping channel and five qubit code.]{Robustness of QER to $\gamma$ for the amplitude damping channel and five qubit code.  The optimal QER operation is computed for $\gamma=.1$.  This recovery operation is then applied for channels $\E$ for $0\leq\gamma\leq.5$ in (A) and $0\leq\gamma\leq.2$ in (B).  For comparison purposes, we include the optimal QER, standard QEC, and single qubit performance for each $\gamma$.}\label{fig:QER_Robustness}
\end{figure}

We can answer this question anecdotally with the example of the amplitude damping channel.  We channel adapt to the amplitude damping channel with $\gamma=.1$.  Figure \ref{fig:QER_Robustness} shows the entanglement fidelity performance of this recovery operation for other values of $\gamma$.  While the performance degrades as the actual parameter departs from $\gamma=.1$, we see that the degradation is not too severe unless the parameter is badly underestimated.  Even in those circumstances, the channel-adapted recovery operation outperforms the generic QEC.\\

We note in  Fig.~\ref{fig:QER_Robustness} (B), that when we have significantly overestimated $\gamma$, the channel-adapted recovery can performance worse than the generic QEC.  As $\gamma$ approaches 0 (as the probability of error goes to 0), channel-adapting to $\gamma=.1$ results in errors.  We conclude from this, that the optimal channel-adapted recovery does not have an operator element that simply projects onto the code subspace.  (We discuss this phenomenon in more detail in Sec.~\ref{sec:4qubit no damping}.

The formulation of the SDP can be adjusted to account for uncertainty in the channel.  Consider a channel $\E_\Lambda$ that can be parameterized by a random variable $\Lambda$ with density $f_\Lambda(\lambda)$.  We can write the output state (to be corrected by $\R$) as
\begin{equation}
  \E_\Lambda(\rho)=\int d\lambda f_\Lambda(\lambda)\E_\lambda(\rho)
\end{equation}
due to the linearity of quantum operations.  The linearity carries through the entire problem treatment and we can write the same optimization problem of (\ref{eq:ave_ent_fid_max_SDP}) as
\begin{eqnarray}\nonumber
X_{\R}^\star=\arg\max_X \tr ({X C_{E,\E,\Lambda}})\\ \textrm{such that }
X \geq 0, \hspace{10 pt} \tr_\HH{X}=I,\label{eq:robuestQER_SDP}
\end{eqnarray}
where
\begin{eqnarray}
C_{E,\E,\Lambda}&=& \int d\lambda f_\Lambda(\lambda)\sum_{jk} p_k \kett{\rho_k E^{\lambda\dagger}_j}\braa{\rho_k E_j^{\lambda\dagger}}.
\end{eqnarray}

\clearpage

\section{Channel-Adapted Encoding}\label{sec:opt encoding}

We have focused so far on the channel-adapted behavior of recovery operations while holding the encoding operation fixed.  This was done to exhibit the benefits of channel-adaptation within the framework of convex optimization.  It is also intuitive to think of an alternate recovery for a known encoding, whereas the reverse is less intuitive.  It should be pointed out, however, that there is no mathematical barrier to optimizing the encoding operation while holding the recovery operation fixed.  In this case, a SDP can again be employed to solve the convex optimization.

We can derive the optimum encoding for a fixed recovery operation just as we did in Sec.~\ref{sec:OptQERSDP}.  Let $\mathcal{C}:\LL(\HH_S)\mapsto\LL(\HH_C)$ be the encoding operation given by elements $\{C_k\}$ and now define the operator elements of $\E'$ to be $\{E_j'\}$.  We can write the composite Choi matrix as
\begin{eqnarray}
X_{\R\circ\E'\circ\mathcal{C}}&=&\sum_{ijk}\kett{R_iE_j'C_k}\braa{R_iE_j'C_k}\\
&=&\sum_{ijk}R_iE'_j\otimes I\kett{C_k}\braa{C_k}E_j'^\dagger R_i^\dagger \otimes I\\
&=& \sum_{ij} (R_iE_j'\otimes I)X_\mathcal{C}(E_j'^\dagger R_i^\dagger \otimes I).
\end{eqnarray}
We now write the average entanglement fidelity as
\begin{eqnarray}
\bar{F}_e(E,\R\circ\E'\circ\mathcal{C})&=&
\sum_kp_k\braa{\rho_k}X_{\R\circ\E'\circ\mathcal{C}}\kett{\rho_k}\\
&=& \tr D_{E,\R,\E'}X_\mathcal{C},
\end{eqnarray}
where
\begin{eqnarray}
D_{E,\R,\E'}&=& \sum_{ijk}p_kE_j'^\dagger R_i^\dagger \otimes I\kett{\rho_k}\braa{\rho_k}R_iE'_j\otimes I\\
&=& \sum_{ijk}p_k\kett{E_j'^\dagger R_i^\dagger \rho_k}
\braa{E_j'^\dagger R_i^\dagger \rho_k}.
\end{eqnarray}

We write the optimization problem for the optimum encoding problem:
\begin{eqnarray}\nonumber
X_{\mathcal{C}}^\star=\arg\max_X \tr ({X D_{E,\R,\E'}})\\ \textrm{such that }\label{eq:opt_encoding}
X \geq 0, \hspace{10 pt} \tr_(\HH_C){X}=I.\label{eq:ent_fid_max}
\end{eqnarray}
We should point out that we are merely constraining the encoding operation $\mathcal{C}$ to be CPTP.  Intuitively, we know that the encoding will be an isometry $U_C$; the result of the SDP yields encodings of this form even without the constraint.

From (\ref{eq:opt_encoding}), a simple iterative algorithm is evident.  For a fixed encoding, we may determine via the SDP the optimum recovery.  Holding the recovery operation fixed, we may determine the optimum encoding.  The procedure is iterated until convergence to a local maximum is achieved.  We can only claim a local maximum as the overall optimization of both $\mathcal{C}$ and $\R$ is no longer convex.

Iterative optimization of error correcting codes has been suggested and applied by several authors in recent years.  The idea was suggested in the context of calculating channel capacities in \cite{Sho:03}, though without discussion of the form of the optimization problem.  An iterative procedure based on eigen-analysis was laid out in \cite{ReiWer:05}.  We derived the convex optimization of optimal QER and pointed out the equivalent problem of optimal encoding in \cite{FleShoWin:07}, and suggested an iterative procedure.  Independently, the same results were derived by \cite{ReiWerAud:06} and \cite{KosLid:06}.

\subsection{The [4,1] `approximate' amplitude damping code}\label{sec:OptQER_4qubit}

Channel-adapted encoding need not be limited to iteratively derived codes.  Consider the [4,1] code of \cite{LeuNieChuYam:97} described in Sec.~\ref{sec:4 qubit code}.  While the authors labelled their code an `approximate' code, we may easily interpret it as a channel-adapted code.

The code was designed specifically for the amplitude damping channel, and even the proposed recovery operation is dependent on the parameter $\gamma$.  The code maintains a high minimum fidelity for small values of $\gamma$, and in fact approximates the performance of the five qubit stabilizer code.  We illustrate the accuracy of this approximation and also demonstrate that by channel adapting the recovery operation from the one proposed, we may even duplicate the five qubit code's optimal QER performance.

\begin{figure}[tbh]
\begin{center}
{\includegraphics[width=\columnwidth]{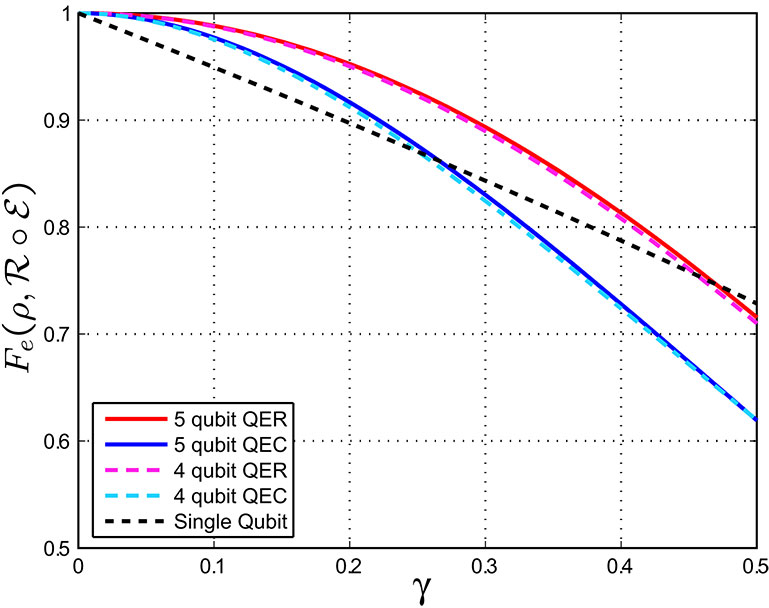}}
\end{center}
\caption[Entanglement fidelity vs.~$\gamma$ for the 4 qubit  code of Leung \emph{et.~al.}\cite{LeuNieChuYam:97}~and the amplitude damping channel $\E_a^{\otimes 4}$.]{Entanglement fidelity vs.~$\gamma$ for the 4 qubit  code of Leung \emph{et.~al.}\cite{LeuNieChuYam:97}~and the amplitude damping channel $\E_a^{\otimes 4}$.  The performance of both the channel-adapted optimum QER and the non-adapted QEC are compared with the equivalent performance of the five qubit stabilizer code.  Entanglement fidelity for a single qubit and no error correction (\emph{i.e.} $F_{e}(\rho,\E_a)$) is included as a performance baseline.}
\label{fig:Leung_comp_QER}
\end{figure}

  We compare the recovery of Leung \emph{et.~al.}~(which for consistency we will still call the QEC recovery) with the optimum QER computed according to  (\ref{eq:ave_ent_fidelity_max}), once again assuming the completely mixed input density $\rho=\frac{1}{2}\ketsub{0}{L}\brasub{0}{L}+\frac{1}{2}\ketsub{1}{L}\brasub{1}{L}$.  The numerical comparison for various values of $\gamma$ is provided in Fig.~\ref{fig:Leung_comp_QER}.  As $\gamma$ goes to zero, the entanglement fidelity for the optimum QER is numerically determined to have the form $F_{e}(\rho,\R\circ\E)\approx 1 -1.25 \gamma^2+\mathcal{O}(\gamma^3)$.  In contrast, the Leung \emph{et.~al.}~recovery is $F_{e}(\rho,\R\circ\E)\approx 1 -2.75 \gamma^2+\mathcal{O}(\gamma^3)$.

The approximate code and channel-adapted recovery illustrate the potential of channel-adaptation to improve QEC.  Consider that the approximate code reduces the overhead by 1 qubit (which halves the size of the Hilbert space $\HH_C$), and achieves essentially equivalent performance.  The equivalent performance continues when both codes are used together with channel-adapted recovery operations.  We will further explore the mechanism of the channel-adapted QER for this case in Chapter \ref{chap:ThreeQubitCode}.

\section{The Dual Function for Optimum QER}\label{sec:OptDual}

Every optimization problem has an associated dual problem\cite{BoyVan:B04}.  Derived from the objective function and constraints of the original optimization problem (known as the \emph{primal} problem), the dual problem optimizes over a set of dual variables often subject to a set of dual constraints.  The dual problem has several useful properties.  First of all, the dual problem is always convex.  In many cases, calculation of the dual function is a useful method for constructing optimization algorithms.  Most important for our purposes, the dual function provides a bound for the value of the primal function.  We define a \emph{dual feasible point} as any set of dual variables satisfying the dual constraint.  The dual function value for any dual feasible point is less than or equal to the primal function at any primal feasible point.  (We have implicitly assumed the primal function to be a minimization problem, which is the canonical form.)

We use the bounding feature of the dual problem in both this chapter and in Chapter \ref{chap:DualBounds}.  In this chapter, after deriving the dual function, we construct a proof of the optimal channel-adapted recovery for a class of codes and channels.  The dual function for channel-adapted recovery was derived in \cite{KosLid:06}; we will re-derive it here in a notation more convenient for our purposes.

The primal problem as given in (\ref{eq:ave_ent_fidelity_max}) can be stated succinctly as
\begin{equation}\label{eq:primal problem}
  \min_X -\tr X\CC,\textrm{ such that }X\geq0\textrm{ and }\tr_{\HH_S}X=I.
\end{equation}
The negative sign on the $\tr X\CC$ terms casts the primal problem as a minimization, which is the canonical form.
The Lagrangian is given by
\begin{equation}\label{eq:Lagrangian}
  L(X,Y,Z)=-\tr X\CC +\tr Y(\tr_{\HH_S}X-I)-\tr ZX,
\end{equation}
where $Y$ and $Z\geq 0$ are operators that serve as the lagrange multipliers for the equality and generalized inequality constraints, respectively.    The dual function is the (unconstrained) infimum  over $X$ of the Lagrangian:
\begin{eqnarray}\label{eq:dual_infimum}
  g(Y,Z)&=&\inf_X L(X,Y,Z)\\\label{eq:dual function}
  &=&\inf_{X} -\tr{X(\CC+Z-I\otimes Y)} - \tr Y,
\end{eqnarray}
where we have used the fact that $\tr (Y\tr_{\HH_S}X)=\tr (I\otimes Y) X$.  Since $X$ is unconstrained, note that $g(Y,Z)=-\infty$ unless $Z=I\otimes Y-\CC$ in which case the dual function becomes $g(Y,Z)=-\tr Y$.  $Y$ and $Z\geq 0$ are the dual variables, but we see that the dual function depends only on $Y$.  We can therefore remove $Z$ from the function as long as we remember the constraint implied by $Z=I\otimes Y-\CC$.  Since $Z$ is constrained to be positive semidefinite, this can be satisfied as long as $I\otimes Y-\CC\geq 0$.

We now have the bounding relation $-\tr X\CC\geq \tr -Y$ for all $X$ and $Y$ that are primal and dual feasible points, respectively.  If we now reverse the signs so that we have a more natural fidelity maximization, we write
\begin{eqnarray}\label{eq:primal dual inequality}
  \bar{F}_e(E_,\R\circ\E)=\tr X_\R \CC\leq \tr Y,
\end{eqnarray}
where $\R$ is CPTP and $I\otimes Y-\CC\geq 0$.  To find the best bounding point $Y$, we solve the dual optimization problem
\begin{equation}\label{eq:dual problem}
  \min_Y \tr Y,\textrm{ such that }I\otimes Y-\CC\geq 0.
\end{equation}
Notice that the constraint implies that $Y=Y^\dagger$.  Note also that $Y\in\LL(\HH_C^*)$.

\subsection{Optimality equations}\label{sec:optimality_eqs}

Semidefinite programming is a convex optimization routine, which provides several useful results relating the primal and dual problems.  As suggested by their names, the primal and dual problems are essentially the same optimization problem in different forms; a solution to one provides the solution to the other.  This fact is numerically exploited in the routines for a computational solution, but we will not concern ourselves with such details.  Instead, we will provide formulae that relate the optimal primal and dual points $\Xo$ and $\Yo$.

We utilize two characteristics of $\Xo$ and $\Yo$.  First, the optimal primal and dual function values are the same, so $\tr \Xo\CC=\tr \Yo$.  This condition is called \emph{strong duality} and it is true for most convex optimization problems.  Second, we have the \emph{complementary slackness} conditions which can be derived for optimization problems that are strongly dual, as is true in this case.  We derive complementary slackness for our context; a general derivation may be found in \cite{BoyVan:B04}.

We defined the dual function in (\ref{eq:dual_infimum}) as the infimum of the Lagrangian $L(X,Y,Z)$ over all $X$.  This implies an inequality when we include the optimal points $\Xo$ and $\Yo$ in the definition of the Lagrangian given in (\ref{eq:Lagrangian}):
\begin{eqnarray}\label{eq:opt_Lagrangian}
  g(\Yo,Z)\leq -\tr \Xo\CC +\tr \Yo(\tr_{\HH_S}\Xo-I)-\tr Z\Xo.
\end{eqnarray}
Since $\Xo$ is a primal feasible point, $\tr_{\HH_S}\Xo-I=0$ so $\tr \Yo(\tr_{\HH_S}\Xo-I)=0$. We also know that $\Xo\geq 0$ and $Z\geq 0$, so we can upper bound the right hand side of (\ref{eq:opt_Lagrangian}) with $-\tr \Xo\CC$.  On the left hand side of (\ref{eq:opt_Lagrangian}), we note that the dual function value at $\Yo$ is $-\tr \Yo=-\tr\Xo\CC$.  Thus,
\begin{equation}\label{eq:comp_slackness_ineq}
-\tr\Xo\CC\leq -\tr \Xo\CC -\tr Z\Xo \leq -\tr \Xo\CC
\end{equation}
which implies that $\tr Z\Xo=0$.  Furthermore, since $Z$ and $\Xo$ are positive semidefinite, $Z\Xo$ is positive semidefinite.  The only positive semidefinite matrix with trace 0 is the 0 operator, so $Z\Xo=0$.

Let's include the definition of $Z=I\otimes Y-\CC$ and state succinctly the two conditions that $\Xo$ and $\Yo$ satisfy due to strong duality:
\begin{eqnarray}
  \label{eq:strong_duality}
  \tr \Yo &=& \tr\Xo\CC\\
  \label{eq:comp_slackness}
  (I\otimes \Yo-\CC)\Xo &=& 0.
\end{eqnarray}

We use (\ref{eq:strong_duality}) and (\ref{eq:comp_slackness}) to provide a means of constructing $\Yo$ given $\Xo$.  (The reverse direction is given in \cite{KosLid:06}.)  We write (\ref{eq:comp_slackness}) as a set of equations in the eigenvectors $\{\kett{\Ro_k}\}$ of $\Xo$:
\begin{eqnarray}
  I\otimes \Yo\kett{\Ro_k}&=&\CC\kett{\Ro_k}\Leftrightarrow\\
  \kett{\Ro_k\ol{\Yo}}&=& \sum_{ij} p_i\kett{\rho_iE_j^\dagger}\braakett{\rho_iE_j^\dagger}{\Ro_k}\Leftrightarrow\\
\Ro_k\ol{\Yo}&=&\sum_{ij} p_i\rho_iE_j^\dagger \tr{E_j\rho_i\Ro_k}.
\end{eqnarray}
Recalling that $\sum_k R^{\star\dagger}_k\Ro_k=I$, we left multiply by $R^{\star\dagger}_k$ and sum over all $k$ to conclude
\begin{equation}\label{eq:YgivenX}
  \ol{\Yo}=\sum_{ijk} p_i R^{\star\dagger}_k \rho_iE_j^\dagger \tr{E_j\rho_i\Ro_k}.
\end{equation}

The form of (\ref{eq:YgivenX}) is interesting given what we know about dual feasible points $Y$. First of all, we know that $Y$ is Hermitian, which is not at all obvious from (\ref{eq:YgivenX}).  Inserting an arbitrary CPTP map specified by $\{R_k\}$ into the right hand side of (\ref{eq:YgivenX}) does not in fact always yield a Hermitian result.  Furthermore, it is not hard to see that the trace of the right hand side is always the average entanglement fidelity $\bar{F}_e(E,\R\circ\E)$ whether $\R$ is optimal or not.  But when $\R$ is the optimal recovery, the constructed $Y$ is not only Hermitian, but is a dual feasible point.  We concisely state this result as an optimality condition.  The operation given by operator elements $\{R_k\}$ is the optimal recovery if and only if
\begin{equation}
  \label{eq:optimality_equation}
  I\otimes \ol{\sum_{ijk} p_i R^{\star\dagger}_k \rho_iE_j^\dagger \tr{E_j\rho_i\Ro_k}}
  -\CC\geq 0.
\end{equation}

\section{Stabilizer Codes and Pauli Group Channels}\label{sec:QECproof}

We have shown several examples where channel-adapted QER has higher fidelity than the standard QEC recovery operation.  To further our understanding, we now present sufficient conditions for the non-adapted QEC to be the optimal QER recovery operation.  Strictly speaking, we analytically construct the optimal recovery for a class of codes, channels, and input ensembles; in most cases, this constructed recovery will be identical to the QEC recovery operation.  The cases where this is not the QEC recovery operation are intuitively clear by construction.  We prove optimality by constructing a dual feasible point where the dual function value equals the average entanglement fidelity.

We can construct the optimal recovery operation for a stabilizer code when the channel $\E'$ is characterized by Pauli group errors and the input ensemble is the completely mixed state.  That is, $E$ is given by $\rho=I/d_S$ with $p=1$ and the channel can be represented by Kraus operators $\{E_i\}$ where each $E_i$ is a scaled element of the Pauli group.  (Notice that this does not require every set of Kraus operators that characterize $\E'$ to be scaled elements of the Pauli group, since unitary combinations of Pauli group elements do not necessarily belong to the Pauli group.)

Let us pause for a moment to consider the interpretation Pauli group channels.  A Pauli group channel on $n$ qubits can be described as $\{\sqrt{p_i} e_i\}$ where $e_i\in\mathcal{G}_n$ and $\sum_i p_i=1$.  We can describe this channel as having the error $e_i$ occur with probability $p_i$.  The depolarizing channel
\begin{equation}
\E_{dp}(\rho)=(1-3p)\rho +p(X\rho X+Y\rho Y+Z\rho Z)
\end{equation}
is a Pauli group channel.  Another example is a channel in which bit flips and phase flips ($X$ and $Z$) occur independently on each qubit.  These are the two primary channels considered for standard QEC, since an ability to correct these errors for one qubit implies the ability to correct arbitrary errors on that qubit.

With a stabilizer code and Pauli group errors, the situation is essentially classical.  The information is initially embedded in the $+1$ eigenspace of the code generators $\langle g_1,\ldots,g_{n-k}\rangle$.
With probability $p_i$, the Pauli group operation $e_i$ is performed.  Since $e_i$ either commutes or anti-commutes with the generators $g_j$, the resulting state lies in a syndrome subspace of the code.  That is, $e_i$ rotates the state into the union of the $\pm 1$ eigenspaces of the generators $g_j$.

The output of the channel $\E'$ is an ensemble of states lying in the stabilizer syndrome subspaces.  It is thus intuitive that the first stage of the optimal recovery operation will be to perform the standard projective syndrome measurement.  The standard QEC recovery operation performs the minimum weight operation that transforms from the code subspace to the observed syndrome subspace.  For the optimal recovery, instead of the minimum weight Pauli operation, we choose the most likely error operation, given the observed syndrome.  In many cases, this will be the same as the minimum weight operator (which is the reason for the choice in standard QEC).

Let us establish this construction more formally.  To do so, we carefully define the syndrome measurement subspaces and the Pauli group operators that connect the subspaces.  We must do this in a way to consistently describe the normalizer operations of the code.  Consider an $[n,k]$ stabilizer code with generators $\langle g_1,\ldots,g_{n-k}\rangle$ and logical $\bar{Z}$ operators $\bar{Z}_1,\ldots\bar{Z}_k$ such that $\{g_1,\ldots,g_{n-k},\bar{Z}_1,\ldots\bar{Z}_k\}$ form an independent and commuting set.  Define logical $\bar{X}$ operators such that $[\bar{X}_i,g_j]=[\bar{X}_i,\bar{X}_j]=0$ $\forall$ $i,j$, $[\bar{X}_i,\bar{Z}_j]=0$ for $i\neq j$ and $\{\bar{X}_i,\bar{Z}_i\}=0$.

The syndrome subspaces correspond to the intersection of the $\pm 1$ eigenspaces of each generator.  Accordingly, we label each space $\SSS_q$ where $q=0,1,\ldots,2^{n-k}-1$, where $\SSS_0$ corresponds to the code subspace.  Let $P_q$ be the projection operator onto $\SSS_q$.  Let $\{\ketsub{i_1 i_2\cdots i_k}{q}\}$ form a basis for $\SSS_q$ such that
\begin{equation}
  \bar{Z}_1\bar{Z}_2\cdots\bar{Z}_k\ketsub{i_1 i_2\cdots i_k}{q}=(-1)^{i_1}(-1)^{i_2}\cdots(-1)^{i_k}\ketsub{i_1 i_2\cdots i_k}{q},
\end{equation}
where $i_j\in\{0,1\}$.  In this way, we have a standardized basis for each syndrome subspace which can also be written as $\{\ketsub{m}{q}\}$, $m=0,\ldots,2^k-1$.

Let us recall the effect of a unitary operator on a stabilizer state.  If $\ket{\psi}$ is stabilized by $\langle g_1,\ldots,g_{n-k}\rangle$, then $U\ket{\psi}$ is stabilized by $\langle Ug_1U^\dagger,\ldots,Ug_{n-k}U^\dagger\rangle$.  What happens if $U\in G_n$, the Pauli group on $n$ qubits?  In that case, since U either commutes or anti-commutes with each stabilizer, $U\ket{\psi}$ is stabilized by $\langle \pm g_1,\ldots,\pm g_{n-k}\rangle$ where the sign of each generator $g_i$ is determined by whether it commutes or anti-commutes with $U$.  Thus, a Pauli group operator acting on a state in the code subspace $\SSS_0$ will transform the state into one of the subspaces $\SSS_q$.

We have established that the Pauli group errors always rotate the code space onto one of the stabilizer subspaces, but this is not yet sufficient to determine the proper recovery.  Given that the system has be transformed to subspace $\SSS_q$, we must still characterize the error by what happened within the subspace.  That is to say, the error consists of a rotation to a syndrome subspace and a normalizer operation within that subspace.

Let us characterize these operations using the bases $\{\ketsub{m}{q}\}$.  Define $W_{qq'}\equiv\sum_m\ketsub{m}{q'}\brasub{m}{q}$ as the operator which transforms $\SSS_q\mapsto \SSS_{q'}$ while maintaining the ordering of the basis.  Define the encoding isometry $U_C\equiv\sum_m \ketsub{n}{0}\brasub{n}{S}$ where $\ketsub{n}{S}\in\HH_S$, the source space.  Further define $U_{cq}\equiv W_qU_C$, the isometry that encodes the $q^{th}$ syndrome subspace.  We will define the $4^k$ code normalizer operators as
\begin{equation}
  A_p\equiv \bar{X}_1^{i_1}\bar{X}_2^{i_2}\cdots\bar{X}_k^{i_k}
  \bar{Z}_1^{j_1}\bar{Z}_2^{j_2}\cdots\bar{Z}_k^{j_k}
\end{equation}
where $p$ is given in binary as $i_1i_2\cdots i_kj_1j_2\cdots j_k$.  Notice that if a similarly defined $A_p^S$ is an element of the Pauli group $\mathcal{G}_k\in\LL(\HH_S)$ with generators $\langle X_1^S,\ldots,X_k^S,Z_1^S,\ldots,Z_k^S\rangle$, we can conclude $A_pU_C=U_CA_p^S$.

The preceding definitions were chosen to illustrate the following facts.  First, we can see by the definitions that $[W_{qq'},A_p]=0$.  That is, $W_{qq'}$ characterizes a standard rotation from one syndrome subspace to another, and $A_p$ characterizes a normalizer operation within the subspace.  These have been defined so that they can occur in either order.  Second, let $\E'$ be a quantum channel represented by operator elements that are scaled members of the Pauli group $\mathcal{G}_n$.  Then the composite channel $\E$ which includes the encoding isometry $U_C$ can be represented by operator elements of the form
\begin{equation}\label{eq:Pauli_channel}
  \{E_{pq}=a_{pq}A_pW_qU_C=a_{pq}A_pU_{Cq}\},
\end{equation}
where the CPTP constraint requires $\sum_{pq}|a_{pq}|^2=1$.

We can understand the amplitudes $a_{pq}$ by noting that with probability $|a_{pq}|^2$, the channel $\E$ transforms the original state to $\SSS_q$ and applies the normalizer operation $A_p$.  To channel-adaptively recover, we project onto the stabilizer subspaces $\{\SSS_q\}$ and determine the most likely normalizer operation for each syndrome subspace $\SSS_q$.  Let $p_q=\arg\max_p|a_{pq}|^2$, and let $\tilde{a}_q\equiv a_{p_qq}$.  With these definitions in place, we can state the following theorem:
\begin{Pauli_Channel}\label{thm:PauliChannel}
Let $\E$ be a channel in the form of (\ref{eq:Pauli_channel}), i.e.~a stabilizer encoding and a channel with Pauli group error operators.  For a source in the completely mixed state $\rho=I/d_S$ the optimal channel-adapted recovery operation is given by $\R\sim\{U_{Cq}^\dagger A_{p_q}\}$, which is the stabilizer syndrome measurement followed by maximum likelihood  normalizer syndrome correction.
\end{Pauli_Channel}
\begin{proof}
We prove Theorem \ref{thm:PauliChannel} by constructing a dual feasible point $Y$ such that the dual function value $\tr Y$ is equal to the entanglement fidelity $F_e(\rho,\R\circ\E)$.

We begin by calculating $F_e(\rho,\R\circ\E)$.  For later convenience, we will do this in terms of the Choi matrix $\CC$ from (\ref{eq:ent_fid_Choi}):
\begin{equation}
  \CC=\sum_{pq}|a_{pq}|^2\kett{\rho U_{Cq}^\dagger A_p}\braa{\rho U_{Cq}^\dagger A_p}.
\end{equation}
Following (\ref{eq:ent_fid_Choi}), we write the entanglement fidelity in terms of the recovery operator elements $\kett{U_{Cq}^\dagger A_{p_q}}$:
\begin{eqnarray}
  F_e(\rho,\R\circ\E)&=& \tr X_\R \CC\\
  &=&\sum_{q'}\braa{U_{Cq'}^\dagger A_{p_{q'}}}\CC\kett{U_{Cq'}^\dagger A_{p_{q'}}}.\label{eq:proof_ent_fid}
\end{eqnarray}
To evaluate (\ref{eq:proof_ent_fid}), we note that
\begin{eqnarray}
  \label{eq:proof_trace1}
  \braakett{\rho U_{Cq}^\dagger A_{p}}{U_{Cq'}^\dagger A_{p_{q'}}}&=&
  \tr A_pU_{Cq}\rho U_{Cq'}^\dagger A_{p_{q'}}\\
  &=& \tr A_pW_qU_C\rho U_C^\dagger W_{q'}^\dagger A_{p_{q'}}\\\label{eq:proof_trace2}
  &=&\tr A_pW_{q'}^\dagger W_qU_C\rho U_C^\dagger A_{p_{q'}}\\\label{eq:proof_trace3}
  &=&\delta_{qq'}\tr A_pU_C\rho U_C^\dagger A_{p_{q'}}\\\label{eq:proof_trace4}
  &=&\delta_{qq'}\tr A_p^C\rho A_{p_{q'}}^C.
\end{eqnarray}
We have used the commutation relation $[W_{qq'},A_p]=0$ to arrive at (\ref{eq:proof_trace2}) and the facts that $W_{q'}^\dagger W_q=\delta_{qq'}P_0$ and $P_0U_C=U_C$ to conclude (\ref{eq:proof_trace3}).
Since $\rho=I/d_S$ and $\tr A_p^C A_{p_{q'}}^C=\delta_{pp_{q'}}d_S$, we see that $\tr A_p^C\rho A_{p_{q'}}^C=\delta_{pp_{q'}}$.  Thus,
\begin{equation}\label{eq:proof_braakett}
  \braakett{\rho U_{Cq}^\dagger A_{p}}{U_{Cq'}^\dagger A_{p_{q'}}P_{q'}}=
  \delta_{pp_{q'}}\delta_{qq'}.
\end{equation}
Using (\ref{eq:proof_braakett}), it is straightforward to evaluate (\ref{eq:proof_ent_fid}):
\begin{eqnarray}
  F_e(\rho,\R\circ\E)&=&\sum_{pqq'}
  |a_{pq}|^2|\braakett{\rho U_{Cq}^\dagger A_{p}}{U_{Cq'}^\dagger A_{p_{q'}}}|^2\\
  &=& \sum_{pqq'}|a_{pq}|^2\delta_{qq'}\delta_{pp_{q'}}\\
  &=& \sum_{q}|\tilde{a}_q|^2.
\end{eqnarray}

We now propose the dual point $Y=\sum_q |\tilde{a}_q|^2\ol{P_q}/d_S$.  Since
\begin{eqnarray}
\tr Y &=& \sum_q |\tilde{a}_q|^2\tr{\ol{P_q}}/d_S\\
&=& \sum_q |\tilde{a}_q|^2\\
&=& F_e(\rho,\R\circ\E),
\end{eqnarray}
we complete the proof by demonstrating that
\begin{equation}
  I\otimes Y-\CC\geq0,
\end{equation}
\emph{i.e.}~$Y$ is a dual feasible point.  We show this by demonstrating that $I\otimes Y$ and $\CC$ have the same eigenvectors, and that the associated eigenvalue is always greater for $I\otimes Y$.

By the same argument used for (\ref{eq:proof_braakett}), we note that
\begin{equation}
  \braakett{\rho U_{Cq}^\dagger A_{p}}{\rho U_{Cq'}^\dagger A_{p'}}
    =\delta_{pp'}\delta_{qq'}/d_S^2.
\end{equation}
This means that $\kett{\rho U_{Cq}^\dagger A_{p}}$ is an eigenvector of $\CC$ with eigenvalue $|a_{pq}|^2/d_S$.  We normalize the eigenvector to unit length and apply it to $I\otimes Y$:
\begin{eqnarray}
  I\otimes Y\kett{\rho U_{Cq}^\dagger A_{p}/d_S}&=&
  \sum_{q'} |\tilde{a}_{q'}|^2\ol{P_{q'}}/d_S\kett{\rho U_{Cq}^\dagger A_{p}/d_S}\\
  &=& \frac{1}{d_S}\sum_{q'}|\tilde{a}_{q'}|^2\kett{\rho U_{Cq}^\dagger A_{p}P_{q'}/d_S}\\
  &=& \frac{1}{d_S}|\tilde{a}_{q}|^2\kett{\rho U_{Cq}^\dagger A_{p}/d_S}.
\end{eqnarray}
Thus we see that $\kett{\rho U_{Cq}^\dagger A_{p}}$ is an eigenvector of $I\otimes Y$ with eigenvalue $|\tilde{a}_{q}|^2/d_S\geq |a_{pq}|^2/d_S$ $\forall$ $p$.  Thus $I\otimes Y-\CC\geq 0$ and $Y$ is a dual feasible point.
\end{proof}

As mentioned above, this theorem is an intuitive result.  Stabilizer codes, like virtually all quantum error correcting codes, are designed to correct arbitrary single qubit errors.  Since the Pauli matrices $X$, $Y$, and $Z$ together with $I$ constitute a basis for all qubit operations, the codes are designed to correct all of those errors.  Essentially, the code is optimally adapted to the channel where these errors occur with equal probability.  For a Pauli error channel, the QEC recovery only fails to be optimal if the relative probabilities become sufficiently unequal.  For example, if $X$ and $Z$ errors occur independently with $p_X=.01$ and $p_Z=.2$, we see that a term such as $Z_1Z_2$ is more likely than $X_1$ and the recovery operation should adapt accordingly.

We may conclude from this section that numerically obtained channel-adaptation is useful only when the channels are not characterized by Pauli errors.  This was asserted when we introduced our emphasis on channels such as the amplitude damping channel and pure state rotation channel.  When the channel is, in fact, a Pauli error channel, channel-adaptation is relatively trivial.  In most cases, the optimal recovery will be the standard QEC result of the minimum weight.  When this is not best, one should be able to quickly determine the appropriate alternative. 

\chapter{Near-Optimal Quantum Error Recovery}\label{chap:NearOptQER}

The optimal quantum error recovery results of Chapter \ref{chap:OptQER}  demonstrate the utility of channel-adaptivity.  Such efforts show that quantum error correction designed for   generic errors can be inefficient in the face of a particular noise process.  Since overhead in physical quantum computing devices is challenging, it is advantageous to maximize error recovery efficiency.

The optimal recovery operations generated through semidefinite programming suffer two significant drawbacks.  First, the dimensions of the optimization problem grow exponentially ($4^n$) with the length of the code, limiting the technique to short codes.  Second, the optimal operation, while physically legitimate, may be quite difficult to implement.  The optimization routine is constrained to the set of completely positive, trace preserving (CPTP) operations, but is not restricted to more easily implemented operations.  In addition to these fundamental drawbacks, the SDP provides little insight into the mechanism of channel adaptivity, as the resulting operation is challenging to interpret.

In this chapter, we describe efforts to determine near-optimal channel-adapted quantum error recovery procedures that overcome these drawbacks of optimal recovery.  While still numerical procedures, we have developed a class of algorithms that is less computationally intensive than the SDP and which yields recovery operations of an intuitive and potentially realizable form.  While the imposed structure moves us a small amount from the optimum, in many ways the resulting operations are of greater use both physically and intuitively.

\section{EigQER Algorithm}\label{sec:EigQER}

To achieve a near-optimal QER operation, an algorithm must have a methodology to approach optimality while still satisfying the CPTP constraints.  Furthermore, to ease implementation of such a recovery, we can impose structure to maintain relative simplicity.

Let us begin by considering the structure of a standard QEC recovery operation.  QEC begins by defining a set of correctable errors, \emph{i.e.}~errors that satisfy the quantum error correction conditions.  To correct this set, we construct the recovery operation by defining a projective syndrome measurement.  Based on the detected syndrome, the appropriate unitary rotation restores the information to the code space, thereby correcting the error.  This intuitive structure, projective measurement followed by unitary syndrome recovery, provides a simple geometric picture of error correction.  Furthermore, it is a  relatively straightforward task to  translate such a recovery operation into a quantum circuit representation.

Let us impose the same constraint on the channel-adapted recovery operation.  We construct an operation with operator elements that are a projective syndrome measurement followed by a classically controlled unitary operation.  Thus the operator elements can be written $\{R_k=U_kP_k\}$ where $P_k$ is a projection operator.  While we could merely constrain $U_k$ to be unitary, we will instead continue with the convention from Chapter \ref{chap:OptQER} that the recovery operation performs a decoding: $\R:\LL(\HH_C)\mapsto\LL(\HH_S).$  Under this convention, $U_k\in\LL(\HH_C,\HH_S)$ and $U_k^\dagger U_k=I$.  In words, both $U_k^\dagger$ and $R_k^\dagger$ are isometries.

The CPTP constraint
\begin{eqnarray}
  I&=&\sum_kR_k^\dagger R_k\\
  &=& \sum_k P_kU_k^\dagger U_k P_k\\
  &=& \sum_k P_k\label{eq:EigQER CPTP}
\end{eqnarray}
can be satisfied if and only if the projectors span $\HH_C$.  This provides a method to construct a recovery while satisfying the CPTP constraints.  $\{P_k\}$ partitions $\HH_C$ into orthogonal subspaces, each identified with a correction isometry\footnote{In fact, $U_k^\dagger$ is the isometry.  For ease of explication, we will refer to $U_k$ as an isometry as well.} $U_k$.

Since the $\{P_k\}$ project onto orthogonal subspaces, we see that $R_j^\dagger R_k=\delta_{jk}P_k$.  From this we conclude that $\{\kett{R_k}\}$ are an orthogonal set and thus are eigenvectors of the Choi matrix $X_\R$.  The  eigenvalue $\lambda_k$ associated with $\kett{R_k}$ is the rank of $P_k$ and is thus constrained to be an integer.  Furthermore, since $U_k$ restores the $k^{th}$ syndrome to $\HH_S$, $\lambda_k \leq d_S$.

We can conceive of a `greedy' algorithm to construct a recovery operation $\R$.  The average entanglement fidelity can be decomposed into the contributions of each individual operator element as $\braa{R_k}C_{E,\E}\kett{R_k}$.  We can construct $\R$ by successively choosing the syndrome subspace to maximize the fidelity contribution.  As long as each syndrome is orthogonal to the previously selected subspaces, the resulting operation will be CPTP and will satisfy our additional constraints.  In fact, this greediest algorithm has no immediate method for computation; the selection of the syndrome subspace to maximize the fidelity contribution has no simple form.  We propose instead a greedy algorithm to approximate this procedure.

We motivate our proposed algorithm in terms of eigen analysis.  Let us assume for the moment that the rank of each syndrome subspace is exactly $d_S$ which is the case for QEC recoveries for stabilizer codes.  By such an assumption, we know that there will be $d_C/d_S$ recovery operator elements.  Consider now the average entanglement fidelity, in terms of the eigenvectors of $X_\R$:
\begin{equation}
  F(\rho,\R\circ\E)=\sum_{k=1}^{d_C/d_S}\braa{R_k}C_{E,\E}\kett{R_k}.
\end{equation}
If we were to maximize the above expression with the only constraint being a fixed number of orthonormal vectors $\kett{R_k}$, the solution would be the eigenvectors associated with the $d_C/d_S$ largest eigenvalues of $C_{E,\E}$.  In fact, the actual constraint differs slightly from this simplification, as we further must constrain $R_k^\dagger$ to be an isometry (\emph{i.e.} $R_kR_k^\dagger=I$).  The analogy to eigen-analysis, however, suggests a computational algorithm which we dub `EigQER' (for eigen quantum error recovery).  We use the eigenvectors of $\CC$ to determine a syndrome subspace with a large fidelity contribution.

The algorithm proceeds as follows:
\begin{enumerate}
\item {Initialize $C_1=C_{E,\E}$.}\\For the $k^{th}$ iteration:
  \item Determine $\kett{X_k}$, the eigenvector associated with the largest eigenvalue of $C_k$.
    \item Calculate $R_k^\dagger$, the isometry `closest' to $X_k^\dagger$ via the singular value decomposition.  Call $R_k$ an operator element of $\R$.
    \item Determine $C_{k+1}$ by projecting out of $C_k$ the support of $R_k$.
    \item Return to step 2 until the recovery operation is complete.
\end{enumerate}
The EigQER algorithm is guaranteed to generate a CPTP recovery operation, and will satisfy the criterion that it can be implemented by a projective syndrome measurement followed by a syndrome dependent unitary operation.

Steps 2 and 3 in the above algorithm require further exposition.  Given an operator $X\in\LL(\HH_C,\HH_S)$, what is the closest isometry $R_k$?  A straightforward answer uses the norm derived from the Hilbert-Schmidt inner product where $\|A\|^2=\tr A^\dagger A$.  We will now allow the rank of $k^{th}$ subspace to be $d_k\leq d_S$.\footnote{Inclusion of reduced rank subspaces may seem unnecessary or even undesirable - after all, such a projection would collapse superpositions within the encoded information.  We allow the possibility since such operator elements are observed in the optimal recovery operations of Chapter \ref{chap:OptQER}.  We will discuss the phenomenon further in Chapter \ref{chap:ThreeQubitCode}.}  Thus $R_kR_k^\dagger=I_{d_k}$ where $I_{d_k}$ is a diagonal operator where the first $d_k$ diagonal matrix elements are 1 and the rest are 0.
We have the minimization problem
\begin{eqnarray}\label{eq:closest isometry}
\min_{R_k} \tr(X-R_k)^\dagger(X-R_k) \textrm{ such that } R_kR_k^\dagger=I_{d_k}.
\end{eqnarray}

We will state the solution as the following lemma.
\begin{close_isom}\label{thm:closest isometry}
  Let $X$ be an operator with singular value decomposition $X=U\Sigma V^\dagger$.  The rank $d$ isometry $R$ that minimizes the Hilbert-Schmidt norm difference $\|X-R\|$ is given by $R=UI_{d_I}V^\dagger$.
\end{close_isom}
\begin{proof}
Let $\mathcal{U}_d$ be the set of rank $d$ isometries;  that is $\mathcal{U}_d=\{U|U^\dagger U =I_d\}$.
We wish to find the $R^\dagger \in\mathcal{U}$ that minimizes $\tr(X-R)^\dagger(X-R)$.  Since  this can be written as
\begin{equation}
\tr(X-R)^\dagger(X-R)=\tr X^\dagger X + \tr R^\dagger R - \tr(X^\dagger R+R^\dagger X)
\end{equation}
and $\tr R^\dagger R=d$, an equivalent problem is
\begin{eqnarray}\label{eq:closest isometry2}
\max_{R\in\mathcal{U}} \tr(X^\dagger R+R^\dagger X)=\max_{R\in\mathcal{U}}\tr(V\Sigma U^\dagger R+R^\dagger U\Sigma V^\dagger),
\end{eqnarray}
where we have replaced $X$ with its singular value decomposition.

We can simplify the above expression by noting that $C^\dagger=U^\dagger R\in\mathcal{U}$.  We can thus equivalently maximize the following expression over $C^\dagger\in\mathcal{U}$:
\begin{eqnarray}
\tr(V\Sigma C^\dagger+C\Sigma V^\dagger)&=&\tr\Sigma (C^\dagger V+V^\dagger C)\\
\label{eq:closest_isom_proof2}
&=& \sum_{i=1}^d \sigma_i (c_i^\dagger v_i+v_i^\dagger c_i)\\
&=& 2\sum_{i=1}^d \sigma_i \textrm{Re}\{v_i^\dagger c_i\}\\
&\leq& 2\sum_{i=1}^d \sigma_i |v_i^\dagger c_i|\\
&\leq& 2\sum_{i=1}^d \sigma_i \|v_i\|\|c_i\|\\
&=& 2\sum_{i=1}^d \sigma_i.
\end{eqnarray}
In (\ref{eq:closest_isom_proof2}), $\sigma_i$ is the $i^{th}$ largest singular value of $X$ and $v_i$ and $c_i$ are the $i^{th}$ columns of $V$ and $C$, respectively.
We have used the fact that $\Sigma$ is a diagonal matrix of the singular values in descending order.  The inequality is saturated when $c_i=v_i$, which also implies that $C= V I_d\Rightarrow R=UI_dV^\dagger$.
\end{proof}

One item not mentioned above is the determination of the desired rank $d_k$.  In our implementation of EigQER, this is accomplished by setting a relatively high threshold on the singular values of $X$.  We only considered singular values such that $\sigma^2\geq.05$.  This \emph{ad hoc} value was chosen as it led to acceptable numerical results in the examples.

We turn now to step 3 of the EigQER algorithm.  Recall that the CPTP constraint as written in (\ref{eq:EigQER CPTP}) requires that the syndrome subspaces are mutually orthogonal.  Thus, the syndrome measurement for the $k^{th}$ iteration must be orthogonal to the first $k-1$ iterations: $P_{k}P_i=0$ for $i<k$.  We satisfy this constraint by updating the data matrix $C_{k-1}$.

To understand the update to $C_{k-1}$, recall that the first step of the $k^{th}$ iteration is the computation of the dominant eigenvector $\kett{X_k}$.  To satisfy the constraint, we require that
\begin{equation}
X_kP_i=0\Leftrightarrow\kett{X_kP_i}=I\otimes\overline{P_i}\kett{X_k}=0
\end{equation}
for $i<k$.  All $\kett{X}$ for which this is not satisfied should be in the nullspace of $C_k$.  Thus, after each iteration we update the data matrix as
\begin{equation}
  C_{k}=(I-I\otimes \overline{P_{k-1}})C_{k-1}(I-I\otimes \overline{P_{k-1}}).
\end{equation}

The algorithm terminates when the recovery operation is complete, \emph{i.e.} $\sum_k R_k^\dagger R_k=\sum_kP_k=I$.  Given the structure of the recovery operations, this can be determined with a simple counter that is increased by $d_k$ at each step $k$.  When the counter reaches $d_C$, the recovery is complete.\\

\begin{figure}
  \begin{center}
  \includegraphics[width=\columnwidth]{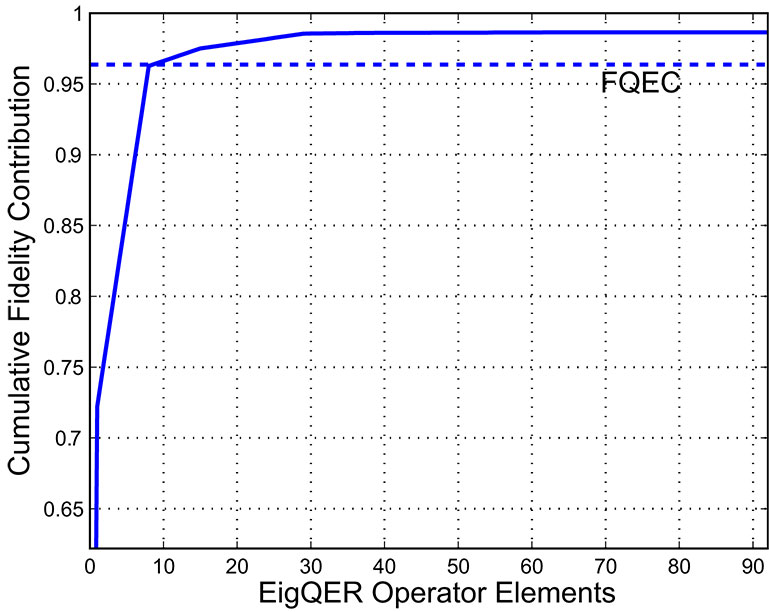}
    \caption[Fidelity contribution of EigQER recovery operators for the amplitude damping channel and the Shor code.]{Fidelity contribution of EigQER recovery operators for the amplitude damping channel and the Shor code.  Notice that the QEC performance is equaled with only 9 operator elements, and the relative benefit of additional operators goes nearly to zero after 30.}
    \label{fig:EigQER greedy}
  \end{center}
\end{figure}

In fact, the greedy nature of EigQER allows early termination of the above algorithm.  Each $R_k$ contributes $\braa{R_k}C_{E,\E}\kett{R_k}$ to the average entanglement fidelity.  Since the algorithm seeks to maximize its gain at each step, the performance return of each $R_k$ diminishes as $k$ grows.  This is illustrated in Fig.~\ref{fig:EigQER greedy}, where we show the cumulative contribution for each recovery operator element with the Shor code and the amplitude damping channel.  The greedy construction results in simplifications in both computation and implementation.  When the contribution $\braa{R_k}\CC\kett{R_k}$ passes below some selected threshold, the algorithm may terminate and thus reduce the computational burden.  This results in an under-complete recovery operation where $\sum_k R_k^\dagger R_k\leq I$.  An under-complete specification for the recovery operation may significantly reduce the difficulty in physically implementing the recovery operation.  In essence, an under-complete recovery operation will have syndrome subspaces whose occurrence is sufficiently rare that the recovery operation may be left as a `don't care.'

Before we consider examples of EigQER recovery performance, we should say a few words about the algorithm complexity when channel adapting an $[n,k]$ code.  Recall that the SDP of Chapter \ref{chap:OptQER} had $4^{n+k}$ complex optimization variables constrained to a semidefinite cone with a further $4^k$ equality constraints.  From \cite{BoyVan:B04}, an SDP with $n$ variables and a $p\times p$ semidefinite matrix constraint requires $\mathcal{O}(\max \{np^3,n^2p^2,n^3\})$ flops per iteration (with typically 10-100 iterations necessary).  For our case, this yields $\mathcal{O}(2^{5(n+k)})$ flops per iteration.

For the EigQER operation, the dominant computation is the calculation of $\kett{X_k}$, the eigenvector associated with the largest eigenvalue of $C_k$.  $C_k$ is a $2^{n+k}\times 2^{n+k}$ dimensional matrix, but the eigenvector has only $2^{n+k}$ dimensions.  Using the \emph{power method} for calculating the dominant eigenvector requires $\mathcal{O}(2^{2(n+k)})$ flops for each iteration of the power method.  While both problems grow exponentially with $n$, the reduced size of the eigenvector problem has a significant impact on the computational burden.

We should note that the eigenvector computation must be repeated for each operator element of $\R$.  If we were to compute all of them, not truncating early due to the diminishing returns of the greedy algorithm, this would require iterating the algorithm approximately $d_C/d_S=2^{n-k}$ times.  In fact, we have a further reduction as the algorithm iterates.  At the $j^{th}$ iteration we are calculating the dominant eigenvector of $C_j$ which lives on a $(d_C-jd_S)d_S=2^k(2^n-j2^k)$ dimensional subspace.  We can therefore reduce the size of the eigenvector problem at each iteration of EigQER.

\subsection{EigQER examples}

To demonstrate the value of the EigQER algorithm, we consider several channels and codes.  We look at the same channels as in Chapter \ref{chap:OptQER}, but can now consider channel-adapted QER for longer codes.  We compare the EigQER recovery performance to the optimal channel-adapted recovery performance for the 5 qubit stabilizer code\cite{BenDivSmoWoo:96,LafMiqPazZur:96}.  We also compare the EigQER performance for the 5 qubit code, the 7 qubit Steane code\cite{Ste:96a,CalSho:96}, and the 9 qubit Shor code\cite{Sho:95}.  All comparisons consider an ensemble $E$ of qubit states that are in the completely mixed state $\rho=I/2$.

\begin{figure}
\includegraphics[width=\columnwidth]{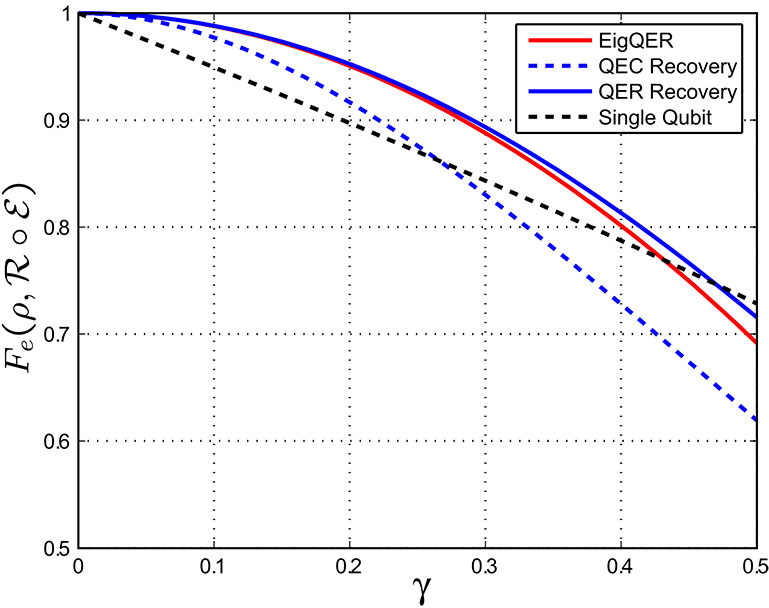}
  \caption[EigQER and Optimal QER for the amplitude damping channel and the five qubit stabilizer code.]{EigQER and Optimal QER for the amplitude damping channel and the five qubit stabilizer code.  EigQER nearly duplicates the optimal channel-adapted performance, especially for lower noise channels (small $\gamma$).}
\label{fig:AmpDamp5_eigQER}
\end{figure}

Figure \ref{fig:AmpDamp5_eigQER} compares the performance of the EigQER algorithm to the optimal QER recovery for the case of the five qubit stabilizer code and the amplitude damping channel.  Also included are the generic QEC recovery and the entanglement fidelity of a single qubit acted upon by $\E_a$ (\emph{i.e.}~no error correction performed).  From this example we observe that the EigQER performance nearly achieves the optimum, especially for the values of $\gamma$ below $.4$.  For higher $\gamma$, the EigQER performance begins to diverge, but this is less important as that region is one in which even the optimal QER lies below the fidelity of a single qubit obtainable with no error correction.

\begin{figure}
\includegraphics[width=\columnwidth]{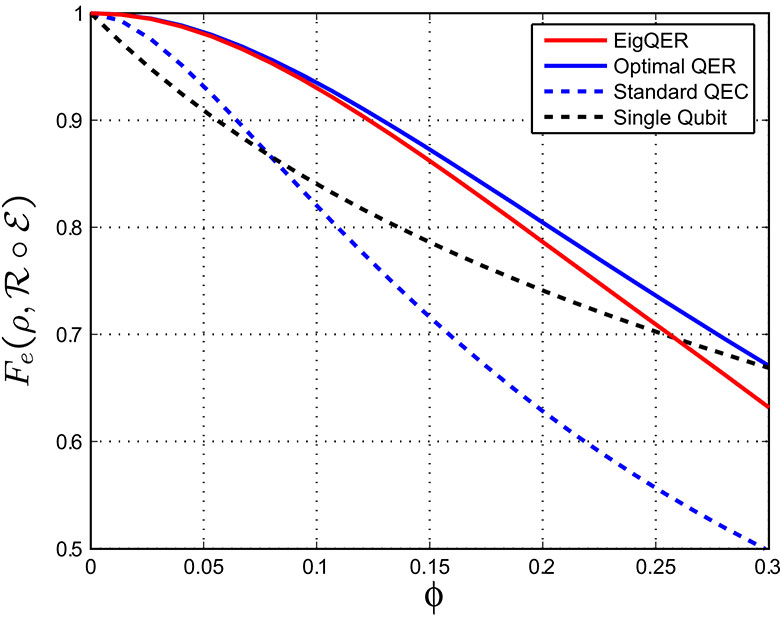}
  \caption[EigQER and Optimal QER for the pure state rotation channel with $\theta=5\pi/12$ and the five qubit stabilizer code.]{EigQER and Optimal QER for the pure state rotation channel with $\theta=5\pi/12$ and the five qubit stabilizer code.  EigQER nearly duplicates the optimal channel-adapted performance, especially for lower noise channels (small $\phi$).}
  \label{fig:PureStates5_EigQER}
\end{figure}

Figure \ref{fig:PureStates5_EigQER} compares EigQER and optimal QER for the five qubit stabilizer code and the pure state rotation channel with $\theta=5\pi/12$.  We see again that the EigQER algorithm achieves a recovery performance nearly equivalent to the optimum, especially as the noise level approaches $0$.

\begin{figure}
\includegraphics[width=\columnwidth]{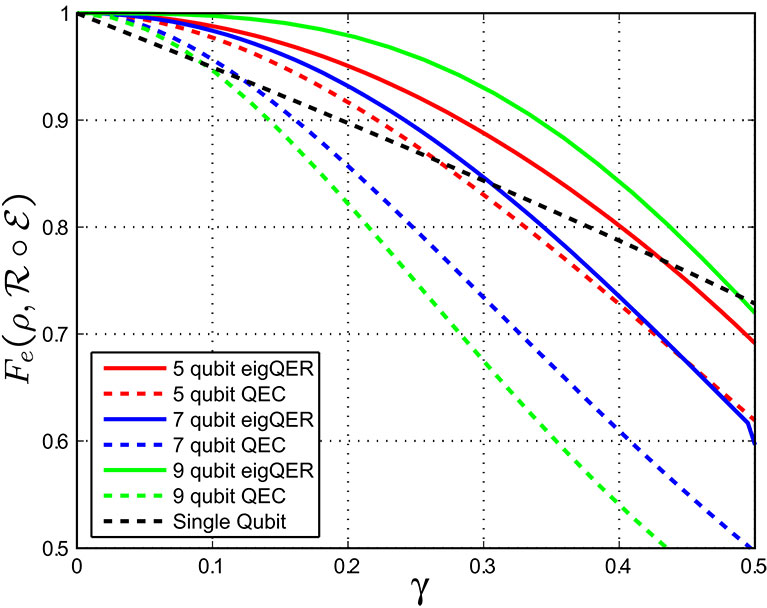}
  \caption[EigQER and standard QEC recovery performance for the five, seven, and nine qubit codes and the amplitude damping channel.]{EigQER and standard QEC recovery performance for the five, seven, and nine qubit codes and the amplitude damping channel.  Note that generic QEC performance decreases for longer codes, as multiple qubit errors become more likely.  While the EigQER performance for the nine qubit Shor code is excellent, the seven qubit Steane code shows only modest improvement, with performance similar to the generic five qubit QEC recovery.}
\label{fig:AmpDamp579_eigQER}
\end{figure}

Figure \ref{fig:AmpDamp579_eigQER} demonstrates the performance of several codes and the amplitude damping channel.  We compare the EigQER performance for the five, seven, and nine qubit codes, contrasting each with the generic QEC performance.  Notice first the pattern with the standard QEC recovery: the entanglement fidelity decreases with the length of the code.  The five qubit stabilizer code, the Steane code, and the Shor code are all designed to correct a single error on an arbitrary qubit, and fail only if multiple qubits are corrupted.  For a fixed $\gamma$, the probability of a multiple qubit error rises as the number of physical qubits $n$ increases.

The QEC performance degradation with code length is a further illustration of the value of channel adaptivity.  All three codes in Figure \ref{fig:AmpDamp579_eigQER} contain one qubit of information, so longer codes include more redundant qubits.  Intuitively, this should better protect the source from error.  When we channel adapt, this intuition is confirmed for the Shor code, but not for the Steane code. In fact, the EigQER entanglement fidelity for the Steane code is only slightly higher than the generic QEC recovery for the five qubit code.  From this example, it appears that the Steane code is not particularly well suited for adapting to amplitude damping errors.  We see that the choice of encoding significantly impacts channel-adapted recovery.  We will return to the channel-adapted performance of the Steane code in Chapter \ref{chap:ThreeQubitCode}.

\begin{figure}
\includegraphics[width=\columnwidth]{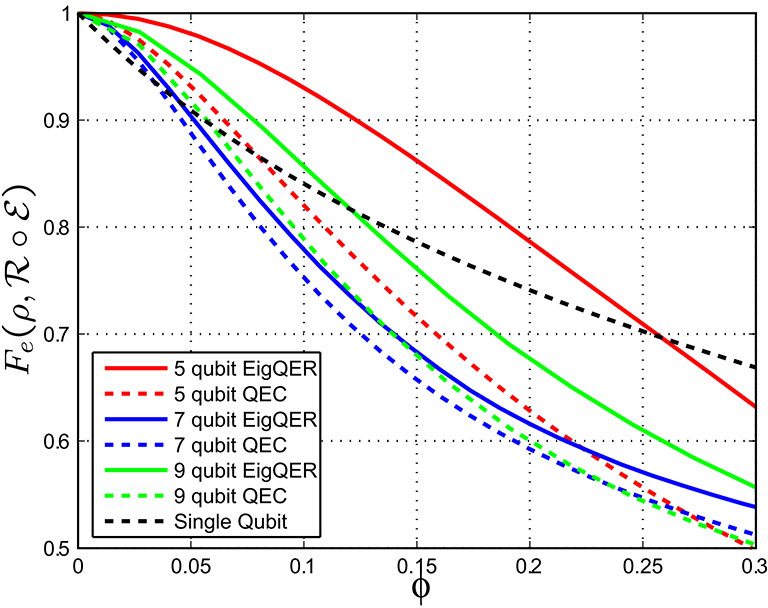}
  \caption[EigQER and standard QEC recovery performance for the five, seven, and nine qubit codes and the pure state rotation channel with $\theta=5\pi/12$.]{EigQER and standard QEC recovery performance for the five, seven, and nine qubit codes and the pure state rotation channel with $\theta=5\pi/12$.  Despite the least redundancy, the five qubit code has the best channel-adapted performance.  The Steane code appears particularly poor for this channel: both the generic QEC and the adapted recovery have lower fidelity than the other codes.}
\label{fig:PureStates579_eigQER}
\end{figure}

The effect is even more dramatically (and puzzlingly) illustrated in the pure state rotation channel.  Figure \ref{fig:PureStates579_eigQER} compares the EigQER recoveries for the five qubit, Steane, and Shor codes with $\theta=5\pi/12$.  It is interesting to see that the five qubit code outperforms each of the others, despite less redundancy to protect the information.  Furthermore, both the standard QEC and channel-adapted recoveries for the Steane code perform worse than the generic recovery of the Shor code!  This suggests that the five qubit code is particularly well suited to adapt to errors of this type, while the Steane code is particularly ill-suited.  (We suspect that the Shor code with QEC recovery outperforms the Steane due to its degenerate structure.)

\section{Block SDP QER}\label{sec:BlockSDP}

The recovery operation generated by the EigQER algorithm of the preceding section is one of a broader class of quantum error recoveries.  The class is characterized by an initial projective syndrome measurement, followed by a syndrome-specific recovery operation.  The projective measurement partitions $\HH_C$ and provides some knowledge about the observed noise process.

Projective syndrome measurements for quantum error correction are tricky to design.  We wish to learn as much as possible about the error while learning as little as possible about the input state, so as not to destroy quantum superposition.  The EigQER algorithm aggressively designs the syndrome measurement, as the $R_k=U_kP_k$ structure of the operator elements implies a finality about the syndrome selection.  The outcome of the syndrome measurement completely determines the correction term $U_k$.

\begin{figure}[tb]
\includegraphics[width=.8\columnwidth]{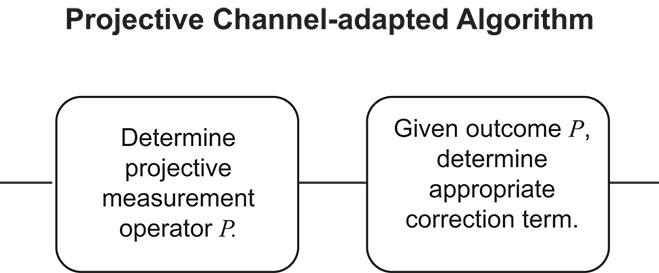}
\caption[Two stage diagram for design of a projective channel-adapted algorithm.]{Two stage diagram for design of a projective channel-adapted algorithm.  The first stage selects a projective syndrome operator $P_k$.  The second determines the corrective action necessitated by $P_k$.}\label{fig:Block algorithm}
\end{figure}

We can conceive of a less aggressive projective measurement.  If we projected onto larger subspaces of $\HH_C$, we would learn less about the noise but perhaps have less chance of destroying the superposition of the input state.  We could consider this an intermediate syndrome measurement, a preliminary step to further error correction.  To design a recovery operation of this type, we must have a strategy to select a projective measurement.  Given the outcome $P_k$, we must further design the syndrome recovery operation $\R_k$.  This general framework is illustrated in Fig.~\ref{fig:Block algorithm}.

Consider the projective syndrome measurement operator $P_k$.  For the EigQER algorithm, $P_k=R_k^\dagger R_k$ always projects onto a subspace of dimension less than or equal to the source space: $\textrm{rank}(P_k)\leq d_S$.  This is an aggressive condition that arises from constraining the subsequent syndrome recovery to be a unitary operator.  We will relax this constraint and allow an arbitrary syndrome recovery $\R_k$ for the $k^{th}$ syndrome measurement.  It turns out that we can determine the optimum such recovery $\R_k^{opt}$ via semidefinite programming, just as in Chapter \ref{chap:OptQER}.  The intermediate syndrome measurement $P_k$ reduces the dimension of the SDP, and thus the technique is still applicable to long codes where computing the global optimum recovery is impractical.

We will demonstrate how the optimum syndrome recovery $\R_k$ can be calculated via a semidefinite program.  Let $\{P_k\}_{k=1}^{K}$ be a set of projectors such that $\sum_k P_k = I\in\HH_C$ that constitute an error syndrome measurement.  Let $\SSS_k$ be the support of $P_k$ with dimension $d_k$; it is clear that $\SSS_1\oplus\SSS_2\oplus\cdots\oplus\SSS_K=\HH_C$.  Given the occurrence of syndrome $k$, we must now design a recovery operation $\R_k:\SSS_k\mapsto\HH_S$.  $\R_k$ is subject to the standard CPTP constraint on quantum operations, but only has support on $\SSS_k$.  We may calculate the recovery $\R_k$ that maximizes the average entanglement fidelity using the SDP in a structure identical to that of (\ref{eq:ave_ent_fid_max_SDP}) while accounting for the reduced input space:
\begin{eqnarray}\label{eq:reduced SDP}
X_{\R_k}=\arg \max_X \tr X (\CC)_k,\\
\nonumber \textrm{such that }X\geq 0,\textrm{ }\tr_{\HH_S}X=I\in\SSS_k.
\end{eqnarray}
Here, $(\CC)_k=I\otimes\overline{P_k}\CC I\otimes\overline{P_k}$ is the data matrix projected into the $k^{th}$ subspace.  Notice that $X_{\R_k}$ and $(\CC)_k$ are operators on $\HH_S\otimes\SSS_k^*$.  In contrast to $\CC$, which requires $d_S^2d_C^2$ matrix elements, $(\CC)_k$ is fully specified by $d_S^2d_k^2$ matrix elements.  By partitioning $\HH_C$ into subspaces $\{\SSS_k\}$ through a careful choice of a syndrome measurement $\{P_k\}$, we may apply semidefinite programming to high dimensional channels without incurring the full computational burden of computing the optimal recovery.  In the following sections we discuss two strategies for determining the syndrome measurement.

\subsection{Block EigQER}

The first step of an iteration of EigQER computes the dominant eigenvalue and corresponding eigenvector of $\CC$.  This eigenvector corresponds to the operator that maximizes the average entanglement fidelity gain at a single step.  While such an operator may violate the CPTP constraint for the recovery operation, it serves to identify an important subspace onto which we may project.  Indeed, the success of the EigQER algorithm rests on the successful identification of syndrome subspaces via eigen-analysis.

An intuitive extension of this concept is to use multiple eigenvectors to specify a higher-dimension subspace.  If $\{\kett{X_m}\}_{m=1}^{M}$ are the eigenvectors corresponding to the $M$ largest eigenvalues of $\CC$, then it is reasonable to define the subspace $\SSS_1$ as the union of the support of the operators $\{X_m\}$.  We define the corresponding projector $P_1$ and calculate the syndrome recovery $\R_1$ via the SDP of (\ref{eq:reduced SDP}).  As in the EigQER algorithm, we update the data matrix $C$ by projecting out the subspace $\SSS_1$, at which point we select another set of eigenvectors.  We will refer to this algorithm as BlockEigQER.

How many eigenvectors should be selected to define a block?  A simple solution is for a fixed block size, say $M$, to be processed until the recovery is complete.  For $M=1$, BlockEigQER is identical to EigQER.  For $M=d_Sd_C$, BlockEigQER computes the optimal recovery operation, as the syndrome measurement is simply the identity operator.  For values in between, one would expect to trade off performance for computational burden.  While there is no guarantee that performance will improve monotonically, we would anticipate improved performance as $M$ increases.

\begin{figure}
\begin{center}
\begin{tabular}{c}
\includegraphics[width=.7\columnwidth]{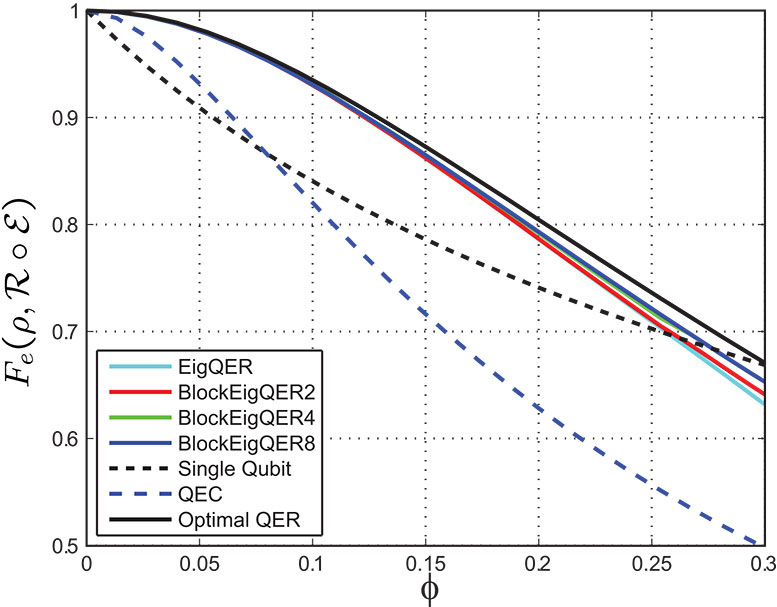}\\

(A)
\vspace{.5cm}\\
\includegraphics[width=.7\columnwidth]{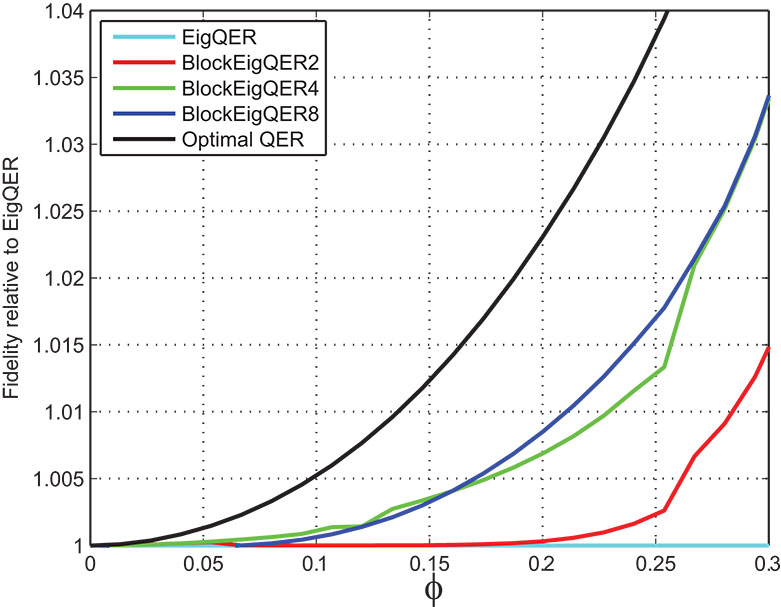}\\

(B)
\end{tabular}
\end{center}

  \caption[BlockEigQER performance for the five qubit code and the pure state rotation channel with $\theta=5\pi/12$.]{BlockEigQER performance for the five qubit code and the pure state rotation channel with $\theta=5\pi/12$.  BlockEigQER is computed with fixed block lengths of 2, 4, and 8.  In (A) we compare the entanglement fidelity to the EigQER recovery, standard QEC recovery and Single Qubit baseline.  The different block lengths have nearly indistinguishable performance from EigQER.  In (B), we compute the fidelity relative to the EigQER recovery and show that the fidelity improves by less than $4\%$ for the displayed region.  We can note, however, that longer block lengths tend to better performance.}\label{fig:BlockEigQER}
\end{figure}

We illustrate the performance for several choices of $M$ in fig.~\ref{fig:BlockEigQER}.  We use the pure state rotation channel ($\theta=5\pi/12$) and the five qubit code with block sizes of 2, 4, and 8.  The expected improvement as $M$ increases is evident, though the gain is quite modest for noise levels of interest (below the cross-over with the single qubit recovery) and is not strictly monotonic.  The variations in performance , including the non-monotonicity, are likely the result of syndrome measurements that collapse the input superpositions.  While the eigenvectors of $\CC$ that identify the syndrome subspace generally avoid collapsing the input state, the mechanism is imperfect.

\begin{figure}
  \begin{center}
    \includegraphics[width=\columnwidth]{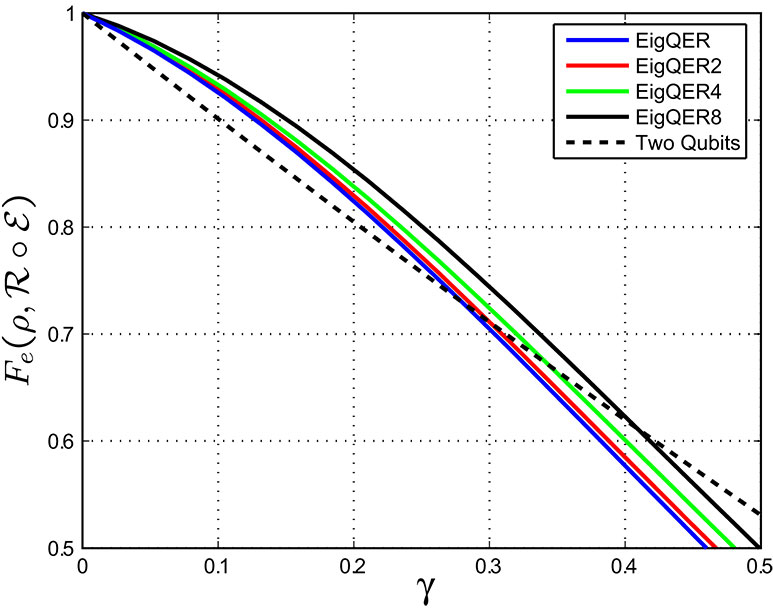}
    \caption[BlockEigQER for the amplitude damping channel and a random six qubit code.]{BlockEigQER for the amplitude damping channel and a random [6,2] code.  We compare the BlockEigQER algorithm for block sizes of 2,4, and 8 with EigQER algorithm.  We see significant performance improvement for larger block sizes, at the cost of computational and recovery complexity.  Baseline in this case is the entanglement fidelity for two qubits input to the channel without error correction.}\label{fig:Random_BlockEigQER}
  \end{center}
\end{figure}

While BlockEigQER outperforms EigQER in the $[5,1]$ code, we see in (B) of \ref{fig:BlockEigQER} that the improvement is less than $5\%$ within the $\phi$ of interest.  We see more significant gains when we encode multiple qubits.  Consider a random $[6,2]$ encoding for the amplitude damping channel, shown in Figure \ref{fig:Random_BlockEigQER}.  In this case we see a distinct performance gain as $M$ increases and the difference is non-trivial.

Fixing the block size $M$ ignores some of the inherent symmetries in the channel and encoding.  In particular, it is quite common for $\CC$ to have degenerate eigenvalues.  By fixing the number of eigenvectors to simultaneously consider, one may inadvertently partition such a degenerate subspace according to the numerical precision of the eigen-analysis software.  To avoid this unwanted circumstance, we may select a variable block size based on the magnitude of the eigenvalues.  This approach necessitates a strategy for parsing the eigenvalues into variable size blocks which can be a tricky procedure.  Due to the modest returns of such an attempt, we have not pursued such a strategy.

While BlockEigQER shows modest performance improvements when compared to EigQER, it has one significant drawback.  Unlike EigQER, the recovery operation from BlockEigQER is not constrained to a collection of isometries.  Once the initial projective syndrome measurement is performed, the subsequent correction terms are arbitrary CPTP maps.  This may complicate attempts to physically implement such an operation.  Furthermore, BlockEigQER does not provide much more intuition for recovery design than EigQER.  For this reason, we consider BlockEigQER a numerical tool whose principal value is its incremental improvement approaching optimality.  It also prove useful for the performance bounds derived in Chapter \ref{chap:DualBounds}.

\subsection{OrderQER}

We now consider a block QER algorithm that does provide intuition for error recovery design.  We are often interested in channels where each qubit is independently corrupted; thus the overall channel is the tensor product of single qubit channels.  We can use this structure to design an intuitive projective measurement.  We illustrate using the classical bit flip channel with probability of error $p$.  If a single bit of the codeword is flipped, we label this a `first order error' as the probability of such an error is $\mathcal{O}(p)$.  If two codeword bits are flipped, this is a `second order error', which occurs with probability $\mathcal{O}(p^2)$.

This intuition can easily yield a choice of syndrome subspaces $\{\SSS_k\}$.  Consider, for example, the amplitude damping channel given in (\ref{eq:ampdamp}).  Recognizing $E_1$ as the `error event,' we declare first order errors to be of the form $E^1_k=E_0\otimes\cdots E_1\otimes E_0\otimes\cdots$ where the error is on the $k^\textrm{th}$ qubit.  In this case we can declare the first order syndrome subspace to be
\begin{equation}\label{eq:first order subspace}
\SSS_1=\textrm{span}(\{\ket{E_0^{\otimes n}0_L},\ket{E_0^{\otimes n}1_L},\ket{E_1^10_L},\ket{E_1^11_L},\cdots\ket{E_n^11_L}\}),
\end{equation}
where $\ket{0_L}$ and $\ket{1_L}$ are the logical codewords for an $n$-length code.  We include the `no error' term as numerical experience suggests that the code projector $P_C$ is not always an optimal syndrome measurement.  By parallel construction, we can define the second order syndrome subspace $\SSS_2$.  While these two will probably not complete the space $\HH_C$, quite possibly we may neglect any higher orders.  Alternatively we can analyze the remaining subspace with either the SDP or the numerically simpler EigQER algorithm.  We will refer to this block SDP algorithm as OrderQER.

The SDP's for first and second order subspaces significantly reduce the dimension from the full optimal SDP, though the effect is not as dramatic as BlockEigQER.  Consider the case of the amplitude damping channel which has only two operator elements for the single qubit channel.  For an $[n,k]$ code, there is one `no error' operator and $n$ first order error operators.  This suggests that $\SSS_1$ has dimension $(n+1)d_S=(n+1)2^k$.  The SDP then has $(n+1)^22^{4k}$ optimization variables. Contrast this $n^2$ growth with the $4^n$ growth of the optimal SDP.  For second order errors, there are $\begin{pmatrix} n\\2\end{pmatrix}\approx \frac{n^2}{2}$ error operators.  The subspace $\SSS_2$ has approximate dimensions of $n^22^{k-1}$ and thus the SDP has $n^42^{4k-2}$ optimization variables.  For the $[7,1]$ Steane code, computing the full optimal SDP requires an impractical $4^7\cdot4=65536$ variables.  However, the first order SDP requires $8^22^4=1024$ variables and the actual second order SDP has $42^2\cdot4=7056$ optimization variables.  For contrast, the full SDP and the five qubit code requires 1024 optimization variables.  For the $[9,1]$ Shor code, the second order SDP has an impractical $72^2\cdot4=20736$ optimization variables.  We therefore do not use OrderQER for the Shor code.

\begin{figure}
  \begin{center}
    \includegraphics[width=\columnwidth]{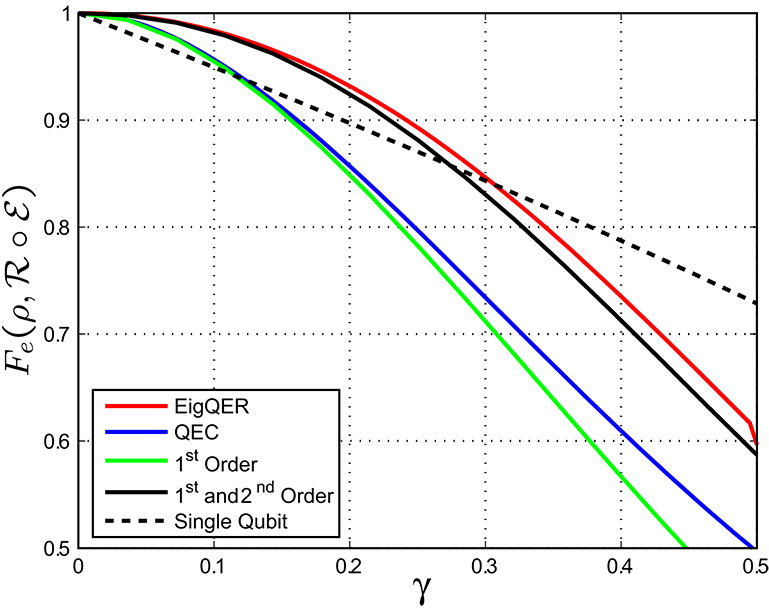}
    \caption[OrderQER recovery for the seven qubit Steane code and the amplitude damping channel.]{OrderQER recovery for the seven qubit Steane code and the amplitude damping channel.  We compare the recovery fidelity of the $1^{st}$ order error to the standard QEC performance.  The performance of the $1^{st}$ and $2^{nd}$ order recoveries together are comparable to the EigQER recovery, especially as $\gamma$ approaches 0.}\label{fig:OrderQER}
  \end{center}
\end{figure}

While the scaling of OrderQER grows quickly with $n$ making its use challenging for codes as long as nine qubits, OrderQER results provide significant insight into the mechanism of channel-adaptation.  Consider the $1^{st}$ and $2^{nd}$ order recovery performance for the Steane code and the amplitude damping channel from Figure \ref{fig:OrderQER}.  We note that the fidelity performance for the recovery from $\SSS_1$ is comparable to the performance of standard QEC, especially as $\gamma$ approaches 0.  This matches the intuition that standard QEC is correcting single qubit errors which are almost completely restricted to $\SSS_1$.  For small $\gamma$, the most likely syndrome measurement will be a Pauli $X$ or $Y$, as these characterize single qubit dampings.  These same errors are corrected by $1^{st}$ order OrderQER.   As $\gamma$ grows, the distortion from the `no error' term $E_0\otimes\cdots\otimes E_0$ becomes more pronounced and the QEC outperforms $1^{st}$ order OrderQER.

We see that $1^{st}$ and $2^{nd}$ order recovery performance is quite comparable to the EigQER performance.  Thus, the performance gains observed for channel adapted QER can be understood as corrections of higher order errors.  Since $\SSS_1$ has dimension significantly less than $d_C$ and yet approximates the QEC recovery performance, it is only reasonable that the remaining redundancy of the code can be exploited to protect from further error.  We will further explore the consequences of this insight for code and recovery design in Chapter \ref{chap:ThreeQubitCode}.

\section{Summary}
Channel-adapted quantum error recovery is possible even with structured recovery operations.  We showed with the EigQER algorithm that high performing channel-adapted QER can be achieved with projective syndrome measurements and syndrome-dependent unitary operations.  As this structure mirrors standard QEC recovery operations and has simple physical interpretation, we can conceivably implement such recovery operations.  Furthermore, the imposed structure of EigQER as well as the block SDP algorithms BlockEigQER and OrderQER allow numerical analysis of longer codes.  While all algorithms will necessarily scale exponentially with the code length $n$, our structured QER algorithms are more scalable than the optimal SDP of Chapter \ref{chap:OptQER}.

\chapter{QER Performance Upper Bounds}\label{chap:DualBounds}

In Chapter \ref{chap:NearOptQER}, we imposed constraints on the recovery operations to provide structure and aid computation.  While the resulting channel-adapted recoveries out perform the generic QEC recovery operation in all of the examples, the constraints essentially guarantee sub-optimality.  For the five qubit code (where computation of the optimal QER operation is practical), we observe that the proposed algorithms (EigQER, BlockEigQER, and OrderQER) closely approximate the optimal performance.  This anecdotal evidence, however, is hardly sufficient to justify the bold description in the chapter title of `near-optimal' channel-adapted QER.  In this chapter, we more fully justify the near-optimal label by deriving channel-adapted performance bounds.

We accomplish this by using the Lagrange dual function derived in Section \ref{sec:OptDual}.  Specifically, we will use the bounding properties of the dual function.  Recall that $Y\in \LL(\HH_C^*)$ is a \emph{dual feasible point} if and only if $I\otimes Y-\CC\geq 0$.  (As in preceding chapters, the inequality is a generalized matrix inequality indicating that the left hand side is positive semidefinite.) Recall from (\ref{eq:primal dual inequality}) that $\bar{F}_e(E_,\R\circ\E)\leq \tr Y$ for all $\R$ if $Y$ is dual feasible; $Y$ is thus a certificate of convergence for a recovery operation.

To provide a good performance bound, it is desirable to find a dual feasible point with a small dual function value.  Indeed, the best such bound is the solution to (\ref{eq:dual problem}), that is to find the dual feasible point with the smallest trace.  However, finding the optimal $Y$ is the equivalent of solving for the optimal recovery due to the strong duality of the SDP.  As this suffers the same computational burden as computing the optimal recovery, we require an alternate method for generating useful dual feasible points.  We will establish methods to convert the sub-optimal recovery operations of Chapter \ref{chap:NearOptQER} into dual feasible points.

We need to determine a good dual feasible point beginning with one of the sub-optimal recoveries computed by the EigQER, BlockEigQER, or OrderQER algorithms.  In Sec.~\ref{sec:optimality_eqs}, we established a method to construct the optimal dual $Y^\star$ given the optimal recovery $\R^\star$.  We might be tempted to apply the same construction using the sub-optimal recovery operations.  Unfortunately, the method suggested by (\ref{eq:YgivenX}) relies upon the fact $\R^\star$ is known to be optimal and thus $\tr X_{\R^\star} \CC=\tr Y^\star$.  Applying (\ref{eq:YgivenX}) will only yield a dual feasible point if the input recovery is optimal.

We instead utilize the structure of the sub-optimal recovery operations to generate a dual feasible point.  We present two methods that exploit the projective syndrome measurement to achieve performance bounds.  The first bound is motivated by the proof of Theorem \ref{thm:PauliChannel} where the optimal dual feasible point is constructed for Pauli group errors.  Beginning with this construction and the recovery generated by EigQER, we use the Ger\v{s}gorin disc theorem to generate a dual feasible point.  The resulting dual function we denote the Ger\v{s}gorin dual bound.  The second construction iteratively generates dual feasible points given an initial infeasible point.  While it is more computationally burdensome, it generates tighter bounds for the considered examples.  We begin with a trial dual variable that may or may not be feasible and iteratively extend this point until it is feasible.  We will call this construction the iterative dual bound.  We present several methods for providing an initial trial point.

Discussion of both bounding methods is facilitated by choosing an appropriate basis for $\HH_S\otimes \HH_C^*$.  Both methods begin with a recovery operation generated by one of the sub-optimal methods of Chapter \ref{chap:NearOptQER}.  As they all begin with a projective measurement, the recovery provides a partition of $\HH_C$ into subspaces $\SSS_q$ of dimension $d_q$ described by projection operators $\{P_q\}\in\LL(\HH_C)$.  We are interested in a basis $\{\ket{v_i}\}_{i=1}^{2^{n+k}}$ where the first block of $d_Sd_0$ basis vectors span $I\otimes\SSS_0^*$ and the $q^{th}$ block spans $I\otimes\SSS_q^*$.  Let us define
\begin{equation}
(\CC)_{qq'}\equiv I\otimes \ol{P_q}\CC I\otimes \ol{P_{q'}}
\end{equation}
as we did in (\ref{eq:reduced SDP}) and then write
\begin{equation}
  \CC=\begin{bmatrix}
    (\CC)_{00} & \cdots & (\CC)_{0q} & \cdots\\
    \vdots & \ddots &  \vdots & \\
    (\CC)_{q0} & \cdots& (\CC)_{qq} &\\
    \vdots & & & \ddots
  \end{bmatrix}
\end{equation}
in our defined basis.
This block structure delineates the relationship of the data operator $\CC$ on each of the subspaces $\SSS_q$ which will be useful when discussing dual feasible points.

\section{Ger\v{s}gorin Dual Bound}

The first method for constructing dual feasible points imposes a convenient structure on $Y$.  In the proof of Theorem \ref{thm:PauliChannel}, the optimal dual feasible point has the form
\begin{equation}\label{eq:dual feasible form}
  Y=\sum_qw_q \ol{P_q},
\end{equation}
where $w_q$ are a set of weights corresponding to the probability of the most likely error resulting in the $q^{th}$ syndrome measurement.  The form of (\ref{eq:dual feasible form}) is appealing due its simplicity, especially for the EigQER recovery operation where the rank $d_q$ of the $P_q$ is constrained to be $\leq d_S$ as is the case in Theorem \ref{thm:PauliChannel}.  While we cannot necessarily generate the optimal dual feasible point in this form, we can use similar methods to generate a reasonable performance bound.

Before we state the Ger\v{s}gorin dual bound, we take a second look at the optimal dual point of Theorem \ref{thm:PauliChannel}.  For an $[n,k]$ stabilizer code, recall that $\HH_C$ is partitioned into $2^{n-k}$ syndrome subspaces $\SSS_q$ and we establish a basis $\{\ketsub{m}{q}\}$ for each subspace.  We also determined that $\kett{U_{Cq}^\dagger A_p}$ is an eigenvector of $\CC$.  Note that $\{\kett{U_{Cq}^\dagger A_p}\}_{p=0}^{2^{2k}-1}$ span the space $I\otimes\ol{\SSS_q}$.

If we write out the operator $(\CC)_{qq}$ in this basis, we have
\begin{eqnarray}
  (\CC)_{qq}&=&\begin{bmatrix}
  a_{0q} &&\\
  & \ddots &\\
  & & a_{(2^{2k}-1)q}
  \end{bmatrix}
\end{eqnarray}
which is diagonal because $\{\ketsub{m}{q}\}$ are eigenvectors of $\CC$.  This also implies that all of the off-diagonal blocks $(\CC)_{qq'}$ where $q\neq q'$ are also 0.   We can now see that $Y=\sum_q \tilde{a}_q\ol{P_q}$ where $\tilde{a}_q=\max_p |a_{pq}|$ is a dual feasible point since
\begin{equation}
  I\otimes Y^\star = \begin{bmatrix} \tilde{a}_0 I & 0 & \cdots & 0\\ 0 &\tilde{a}_1I & \cdots& 0\\ \vdots & \vdots  & \ddots & \vdots\\
  0 & 0 & \cdots & \tilde{a}_{2^{n-k}-1}I\end{bmatrix}\\
\end{equation}
is diagonal in the chosen basis.

We return now to the general case.  Unlike in the case of a Pauli error channel and a stabilizer code, we cannot guarantee that $\CC$ will be either diagonal or block diagonal in this basis.  However, if our sub-optimal recovery $\R$ is generated from the EigQER algorithm, then the subspaces $\SSS_q$ are selected based on the eigenvectors of $\CC$ and we can expect $\CC$ to be approximately block diagonal when we partition according to the subspaces $I\otimes \SSS_q^*$.  We say that $\CC$ is approximately block diagonal in this basis if $\|(\CC)_{qq}\|\gg\|(\CC)_{qq'}\|$ for $q\neq q'$.

To generate a dual feasible point of the form $Y=\sum_q w_q \ol{P_q}$, we need to choose $w_q$ so that $I\otimes Y-\CC\geq 0$.  If $\CC$ were exactly block diagonal in this basis, we could accomplish this by setting $w_q=\lambda_{\max}((\CC)_{qq})$.  Since the block terms off the diagonal are not strictly 0, we must account for their contributions in the location of the eigenvalues of $\CC$.\\

We will make use of a linear algebra theorem known as the Ger\v{s}gorin disc theorem.  This theorem provides bounds on the location in the complex plane of the eigenvalues of an arbitrary matrix.  As will be evident, the theorem is most valuable when the matrix is dominated by its diagonal entries.  We state the theorem as it is given in \cite{HorJoh:B85} $\S$ 6.1:

\begin{Gersgorin}
  \label{thm:Gersgorin}
  Let $A=[a_{ij}]\in \mathbb{C}^{n\times n}$, and let
  \begin{equation}
    R_i'(A)\equiv\sum_{j=1, j\neq i}^n |a_{ij}|,\hspace{20pt}1\leq i\leq n
  \end{equation}
denote the \emph{deleted absolute row sums} of $A$.  Then all the eigenvalues of $A$ are located in the union of $n$ discs
\begin{equation}
  \bigcup_{i=1}^n \{ z\in\mathbb{C}:|z-a_{ii}|\leq R_i'(A)\}\equiv G(A).
\end{equation}
Furthermore, if a union of $k$ of these $n$ discs forms a connected region that is disjoint from all the remaining $n-k$ discs, then there are precisely $k$ eigenvalues of $A$ in this region.
\end{Gersgorin}

Theorem \ref{thm:Gersgorin} is particularly useful for proving the positivity of a matrix.  The $R_i'(A)$ are the radii of discs centered at the diagonal entries $a_{ii}$ and the eigenvalues are constrained to lie within the union of these discs.  If $A$ is a Hermitian matrix, then we can be certain it is positive semidefinite if $a_{ii}\geq R_i'(A)$ for all $i$ as all of the eigenvalues would be constrained to lie to the right of the origin (or on the origin) on the real line.

We can apply Theorem \ref{thm:Gersgorin} to generating a dual feasible point structured as (\ref{eq:dual feasible form}).  In this case we use the weights $w_q$ to ensure that the diagonal entries of $I\otimes Y-\CC$ are greater than the deleted absolute row sums.  Let $c_{ij}$ denote the matrix elements of $\CC$ in our defined basis and let the basis vector $\ket{v_i}$ lie in the subspace $\SSS_q$.  We then the have the $i^{th}$ diagonal element $[I\otimes Y-\CC]_{ii}=w_q-c_{ii}$ and the $i^{th}$ deleted absolute row sum is $\sum_{i\neq j} |c_{ij}|$.  We can assure non-negativity if
\begin{equation}
  w_q\geq \sum_j |c_{ij}|,\textrm{ for all }i\textrm{ such that }\ket{v_i}\in \SSS_q.
\end{equation}
Thus, we can guarantee a dual feasible point if $w_q$ is set to be the maximum absolute row sum for all rows $i$ such that $\ket{v_i}\in\SSS_q$.
We may express $w_q$ concisely in terms of the induced $\infty$-norm(\cite{HorJoh:B85} $\S$ 5.6.5), denoted $\|\cdot\|_\infty$:
\begin{eqnarray}
  w_q&=&\left\|\begin{bmatrix} (\CC)_{q0} & \cdots & (\CC)_{qq} & \cdots \end{bmatrix}\right\|_\infty\\
  &=&\|I\otimes\ol{P_q}\CC\|_\infty.
\end{eqnarray}

\begin{figure}
    \includegraphics[width=\columnwidth]{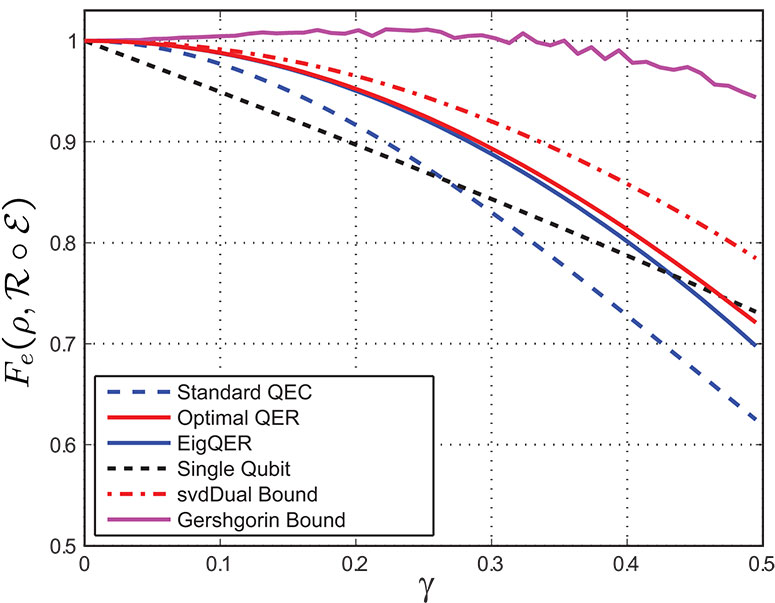}
  \caption[Ger\v{s}gorin and SVD dual bound for the amplitude damping channel and the 5 qubit stabilizer code.]{Ger\v{s}gorin and SVD dual bound for the amplitude damping channel and the 5 qubit stabilizer code.  The Ger\v{s}ogrin bound is clearly not very useful as in some cases it is greater than 1.  The SVD dual bound clearly tracks the optimal performance, although the departure from optimal of the bound exceeds the EigQER recovery.}\label{fig:EigDUAL 5}
\end{figure}

The Ger\v{s}gorin disc theorem is a computationally simple way to guarantee construction of a dual feasible point given a partition of $\HH_C$ into subspaces $\{\SSS_q\}$.  Unfortunately, the induced infinity norm does not provide a particularly useful performance bound as can be seen in Figure
\ref{fig:EigDUAL 5}.  When we compare to the optimal recovery performance for the five qubit code and the amplitude damping channel, we see that the dual bound is far from tight.  In fact, for many values of $\gamma$, the bound is greater than 1, which is truly useless for upper bounding fidelities.  While we have generated a dual point $Y$ that is guaranteed to be feasible, such a guarantee imposes too strict a cost to have a useful bounding property.

The Ger\v{s}gorin dual bound provides useful insight for a tighter dual construction.  If we replace the induced infinity norm with the induced 2-norm, we generate a dual point that is often dual feasible.  That is, choose
\begin{eqnarray}
  w_q&=&\|I\otimes\ol{P_q}\CC\|_2\\
  &=& \max_{\kett{x}} \braa{x}I\otimes\ol{P_q}\CC\kett{x}\\
  \label{eq:svd 2norm}
  &=& \sigma_{\max}(I\otimes\ol{P_q}\CC),
\end{eqnarray}
where $\sigma_{\max}(\cdot)$ in (\ref{eq:svd 2norm}) indicates the maximum singular value and is the computational method for the induced 2-norm.  We will refer to this construction as the SVD (for singular value decomposition) dual point.  The $Y$ generated in this way is not guaranteed to be dual feasible as was the case with the $\infty$-norm, but has proven to be dual feasible in all of the examples that we have tried.  If for some circumstance the SVD dual point is not feasible, it can be iteratively adjusted to become dual feasible in a manner we present in the following section.

\section{Iterative Dual Bound}\label{sec:IterativeDual}

We now present an iterative procedure to generate a dual feasible point given an initial dual point $Y^{(0)}$ that is presumably not dual feasible.  After presenting the algorithm, we will discuss choices for the initial dual point.

At the $k^{th}$ iteration, we update the dual point to produce $Y^{(k)}$ until we achieve feasibility.  For convenience we will define
\begin{equation}
Z^{(k)} \equiv I\otimes Y^{(k)}-\CC.
\end{equation}
Let $x$ and $\kett{x}$ be the smallest eigenvalue and associated eigenvector of $Z^{(k)}$.  If $x\geq 0$, we may stop, as $Y^{(k)}$ is already dual feasible.  If $x\leq 0$, we wish to update $Y^{(k)}$ a small amount to ensure that $\braa{x}Z^{(k+1)}\kett{x}\geq 0$.  Essentially, we are replacing a negative eigenvalue with a 0 eigenvalue.  Given no constraints on the update, we could accomplish this as $Z^{(k+1)}=Z^{(k)}+x\kett{x}\braa{x}$ but we must instead update $Y^{(k)}$ with the tensor product structure implicit.

We determine the properly constrained update by means of the Schmidt decomposition of the eigenvector:
\begin{equation}
\kett{x}=\sum_i \lambda_i\ketsub{\hat{x}_i}{\HH_S}\ketsub{\tilde{x}_i}{\HH_C^*}.
\end{equation}
As we can only perturb $Z^{(k)}$ in the $\HH_C^*$ slot, we choose the smallest perturbation guaranteed to achieve $\braa{x}Z^{(k+1)}\kett{x}\geq 0$.
Let
\begin{equation}
Y^{(k+1)}=Y^{(k)}+\frac{|x|}{|\lambda_1|^2}\ket{\tilde{x}_1}\bra{\tilde{x}_1}.
\end{equation}
Then
\begin{eqnarray}
  \braa{x}Z^{(k+1)}\kett{x}&=& x+\frac{|x|}{|\lambda_1|^2}\braa{x}(I\otimes \ket{\tilde{x}_1}\bra{\tilde{x}_1})\kett{x}\\
  &=& x+\frac{|x|}{|\lambda_1|^2}|\lambda_1|^2\\
  &=& 0,
\end{eqnarray}
since $x<0$.
While we have not yet guaranteed that $Z^{(k+1)}\geq 0$, $\kett{x}$ is no longer associated with a negative eigenvalue.  By repeatedly perturbing $Y^{(k)}$ in this manner, we iteratively approach a dual feasible point while adding as little as possible to the dual function value $\tr Y^{(k)}$.

As a final point, we demonstrate that the iterative procedure will converge to a dual feasible point.  Let's consider the effect of the $k^{th}$ iteration on the space orthogonal to $\kett{x}$.  Let $\kett{y}\in\HH_S\otimes\HH_C^*$ be orthogonal to $\kett{x}$.  Then, for $Z^{(k+1)}$ we see that
\begin{eqnarray}
  \braa{y}Z^{(k+1)}\kett{y}=\braa{y}Z^{(k)}\kett{y}+\frac{|x|}{|\lambda_1|^2}\braa{y}(I\otimes \ket{\tilde{x}_1}\bra{\tilde{x}_1})\kett{y}.
\end{eqnarray}
But since $I\otimes \ket{\tilde{x}_1}\bra{\tilde{x}_1}\geq 0$ we see that
\begin{equation}
  \braa{y}Z^{(k+1)}\kett{y}\geq\braa{y}Z^{(k)}\kett{y}
\end{equation}
for all $\kett{y}\in\HH_S\otimes\HH_C^*$.  We see that the update to $Y^{(k)}$ moved one negative eigenvalue to 0 while no new negative eigenvalues can be created.  Thus the procedure will require no more than $m$ iterations where $m$ is the number of negative eigenvalues for $Z^{(0)}$.

\subsection{Initial dual points}

Having established a procedure to generate a dual feasible point given an arbitrary intial point $Y^{(0)}$, we will now present initialization options.  While we can start with any Hermitian operator in $\LL(\HH_C^*)$ including $0$, we do not recommend such an unstructured choice as each iteration is imperfect.  Each iteration adds $|x|/|\lambda_1|^2$ to the dual function value.  If $|\lambda_1|$ is not close to 1, the iteration is not efficient.  We will use more educated initializations to begin closer to feasibility, thus minimizing the number of iterations and improving the bounding properties of the resulting dual feasible point.

We have already presented one method for initialization with the SVD dual point.  In most cases we've seen, this point is already feasible and in fact is a relatively loose bound.  Its advantage lies in its easy computation, but other choices provide better bounding properties.  We would prefer an initial $\Yi$ such that $\Zi$ is non-positive with eigenvalues very close to 0.  If this is the case, we will require only small perturbations (and thus a small dual function value) to achieve a positive semidefinite $Z^{(k)}$.

Consider an initial $Y^{(0)}$ of the form given in (\ref{eq:dual feasible form}).  We choose an initial $\Yi$ in the same way that was used in the proof of Theorem \ref{thm:PauliChannel}:
\begin{equation}
  w_q=\lambda_{\max}((\CC)_{qq}).
\end{equation}
This is very simple to calculate, though it will not generally be dual feasible.  This is the logical choice when we begin with the EigQER recovery, as the only useful information we have is the projective syndrome measurement.  This initialization often iterates to a better bound than the SVD dual point and requires no further information than the partition $\{\SSS_q\}$ provided by any of the sub-optimal QER methods from Chapter \ref{chap:NearOptQER}.  It has one drawback, however, in that $\Zi$ almost certainly has eigenvalues much greater than 0.  For the $\ket{v_i}$ associated with the largest eigenvalue of $(\CC)_{qq}$, $\bra{v_i}\Zi\ket{v_i}=0$.  However, unless $(\CC)_{qq}$ has only one distinct eigenvalue there will be vectors $\kett{x}\in\SSS_q$ such that $\braa{x}\Zi\kett{x}\geq0$, and perhaps quite large, relatively.  Such vectors indicate portions of the Hilbert space where $\Yi$ is already greater than the optimal dual feasible point.  While this likely cannot be avoided in the iterations, it seems wasteful to begin at such a point if not necessary.

We have an alternative choice for $\Yi$ arising from the block SDP QER algorithms of Sec.~\ref{sec:BlockSDP}.  These algorithms already provide information useful for generating a dual feasible point.  When solving the SDP on a subspace $\SSS_q$ one can simultaneously generate the optimal dual function value $Y_q^\star\in\LL(\SSS_q^*)$.  This can be computed just as in Sec.~\ref{sec:optimality_eqs}.  Given such optimal subspace dual points, define the block diagonal operator
\begin{equation}
  \Yi=\begin{bmatrix}
    Y_0^\star &&&\\
    & \ddots &&\\
    && Y_q^\star &\\
    &&&\ddots
  \end{bmatrix}
\end{equation}
as the initial point.  We know that $I\otimes Y_q^\star -(\CC)_{qq}\geq 0$, so there will be $\kett{x}$ for which $\braa{x}\Zi\kett{x}\geq0$.  However, since $Y_q^\star$ is optimal within $\LL(\SSS_q^*)$, we know that we are not being overly wasteful with the initialization.

\subsection{Iterated block dual}

Let's consider the computational burden of the iterated dual bound.  At each iteration we must compute the smallest eigenvalue and associated eigenvector of $Z^{(k)}$, a $2^{n+k}\times 2^{n+k}$ Hermitian matrix.  (We can accomplish this by looking for the largest eigenvalue of $\eta I-Z^{(k)}$ where $\eta\geq1$ is an arbitrary offset to ensure positivity.)  This must be repeated at most $2^{n+k}$ times to ensure dual feasibility, though there may be significantly fewer iterations if the $Z^{(0)}$ is nearly positive semidefinite already.  As mentioned in Sec.~\ref{sec:EigQER}, this can be accomplished in $\mathcal{O}(2^{2(n+k)})$ flops by the power method.  This is very costly if we must repeat the iteration many times.

The block diagonal structure of the initial points suggests a slightly modified alternative procedure with some computational advantage. Consider the optimal dual points $Y_i$ and $Y_j$ in $\LL(\SSS_i^*)$ and $\LL(\SSS_j^*)$.  We can use the same iterative procedure as before to compute a dual feasible $Y_{ij}\in\LL(\SSS_i^*\oplus\SSS_j^*)$ requiring only $\mathcal{O}(2^{2k}(d_i+d_j)^2)$ flops per iteration with a maximum of  $2^{k}(d_i+d_j)$ iterations.  We can generate a dual feasible point on the whole space $\LL(\HH_C^*)$ by successively combining subspace blocks.  Eventually we will have to iterate over the full space, but we will have done most of the work in the smaller blocks, and the full $2^{n+k}\times 2^{n+k}$ eigen decomposition will require few iterations.

In the examples we have processed, the iterated block dual procedure created nearly identical bounds (often within $10^{-5}$ of each other and never more than $10^{-4}$) as the original algorithm.  The computational burden is reduced by approximately $20\%$.

\subsection{Examples}

\begin{figure}
    \includegraphics[width=\columnwidth]{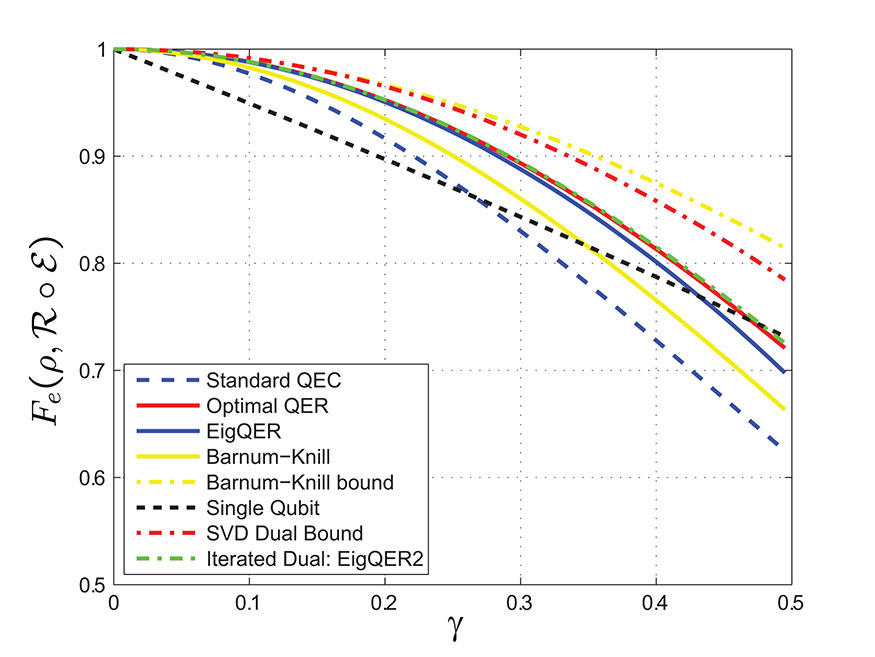}
  \caption[Dual bound comparison for the amplitude damping channel and the five qubit code.]{Dual bound comparison for the amplitude damping channel and the five qubit code.  The iterated dual initialized with the Block EigQER algorithm with $M=2$ is essentially indistinguishable from the optimal recovery performance, thus producing a very tight bound.  Included for comparison are the EigQER performance, the SVD dual bound, and both a channel-adapted recovery and associated bound derived by Barnum and Knill in \cite{BarKni:02}.}\label{fig:AmpDamp5_dual}
\end{figure}

We provide several examples to demonstrate the utility of the iterated dual bound.  At the same time, we we illustrate the near optimality of the algorithms from Chapter \ref{chap:NearOptQER}.  In Fig.~\ref{fig:AmpDamp5_dual}, we show several bounds for channel-adapted QER for the amplitude damping channel and the five qubit code.  In this case, we know the optimal performance and can see that the iterated dual bound,  beginning with the BlockEigQER with $M=2$, is quite tight.  This is in contrast to the SVD dual bound, which was also shown in Fig.~\ref{fig:EigDUAL 5}.  We have included in Fig~\ref{fig:AmpDamp5_dual} the numerical channel-adapted recovery and performance bound from \cite{BarKni:02}.  We see that this bound is looser than even the SVD dual bound for this example.

\begin{figure}
    \includegraphics[width=\columnwidth]{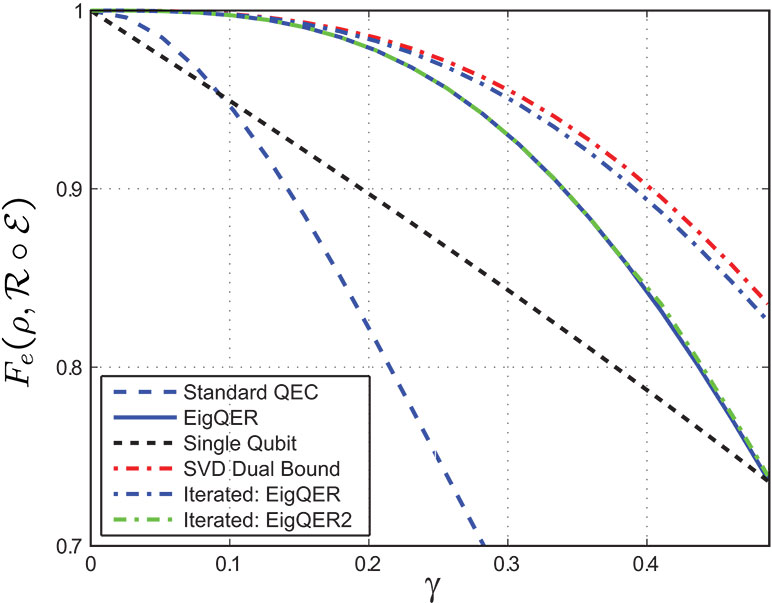}
  \caption[Dual bound comparison for the amplitude damping channel and the nine qubit Shor code.]{Dual bound comparison for the amplitude damping channel and the nine qubit Shor code.  The iterated dual bound initialized with the BlockEigQER recovery with $M=2$ produces a bound that is tight to the EigQER recovery operation.  This demonstrates that the EigQER recovery operation is essentially optimal in this case.  Notice that the iterated bound initialized with the EigQER recovery operation does not generate a tight bound.}\label{fig:AmpDamp9_dual}
\end{figure}

Figure \ref{fig:AmpDamp9_dual} shows several dual bounds for the amplitude damping channel and the nine qubit Shor code.  While we cannot compute the optimum directly, we see that the EigQER performance curve and the iterated bound derived from BlockEigQER with $M=2$ are essentially equivalent.  We can conclude that EigQER operation is essentially in this case.  While not shown, iterations for BlockEigQER with $M=4$ and $M=8$ achieved essentially the same bound.  Note that neither the SVD dual bound nor the iterated bound beginning with the EigQER recovery operation are tight, illustrating the importance of a good initialization for the dual iterations.

\begin{figure}
    \includegraphics[width=\columnwidth]{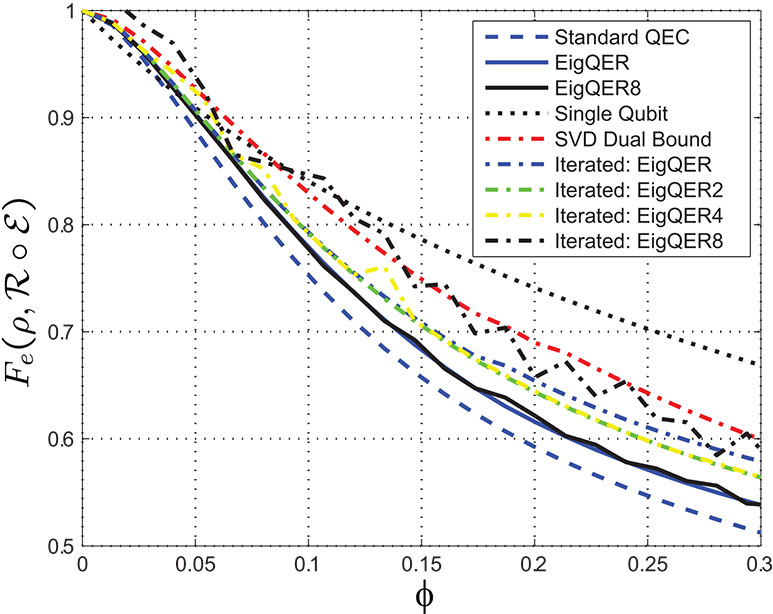}
\caption[Dual bound comparison for the pure state rotation channel with $\theta=5\pi/12$ and the seven qubit Steane code.]{Dual bound comparison for the pure state rotation channel with $\theta=5\pi/12$ and the seven qubit Steane code.  Note that the iterated bounds are generally, though not universally, better than the SVD dual bound.  We also see that the shorter block lengths for the BlockEigQER algorithm generally produce a tighter bound, despite slightly poorer recovery performance.}\label{fig:PureState7_dual}
\end{figure}

Our final example is the pure state rotation channel with $\theta=5\pi/12$ and the seven qubit Steane code.  In Fig.~\ref{fig:PureState7_dual}, we can distinguish between several initialization methods for the dual iterative bound.  We see that none of the recovery operations approach the bound performance for large $\phi$, though the performance is relatively tight as the noise level drops ($\phi\rightarrow 0)$.  Notice that in general the iterative bounds are better than the SVD dual bound, however there are points, especially for the BlockEigQER algorithm with $M=8$, where the iterated bound is poor.  It is interesting to note that the longer block lengths (larger $M$) usually generate better recovery performance (which can be seen with slight improvement even in this case) yet often produce poorer bounds.  Anecdotal experience suggests that the best iterative starting point is the BlockEigQER recovery operation with $M=2$.

Finally, we should point out the gap for large $\phi$ between the recovery performance and the dual bounds.  Absent a better recovery operation or a smaller performance bound, we have no way to know whether the bound or the recovery is further removed from the optimal.  However, this region is below the baseline performance for a single unencoded qubit, and thus is not of serious concern.

\section{Summary}

The bounds presented in this chapter justify describing the recovery operations of Chapter \ref{chap:NearOptQER} as `near-optimal.'  We have demonstrated several numerical methods to upper bound the channel-adapted QER performance using the dual function.  In this way we can certify the convergence of the constrained recovery operations EigQER, BlockEigQER, and OrderQER.  In the cases we have considered, the bounds suggest that the structured recovery operations do not suffer serious performance losses compared to the optimal.  Examples of bounds and recovery performance for all of the considered examples are included in Appendix \ref{chap:App Figures}.

\chapter{High Rate Channel-Adapted QEC for Amplitude Damping}\label{chap:ThreeQubitCode}

The primary assertion of this dissertation is that one can improve both the performance and the efficiency of quantum error correction by adapting QEC procedures to the physical noise process.  To this point, we have developed and interpreted mathematical and algorithmic tools with general application.  That is to say, given any model for the noise process and an appropriately short code we can apply optimal (Chapter \ref{chap:OptQER}) and structured near-optimal (Chapter \ref{chap:NearOptQER}) algorithms to provide channel-adapted encoding and recovery operations.

It is important to note that the aforementioned tools are not, in themselves, complete solutions to the problem of channel-adapted QEC.  When designing an error correction procedure, there is more to consider than whether an encoding or a recovery is physically legitimate.  This motivated our exploration of near-optimal recovery operations, where we imposed a projective syndrome measurement constraint on recovery operations.  Even given such a constraint, to implement channel-adapted QEC efficiently we need to design encoding and decoding procedures with sufficiently simple  structure to allow efficient implementation.  Furthermore, while the optimization routines focus on the entanglement fidelity and ensemble average fidelity due to their linearity, we should still like to understand the minimum fidelity, or worst case performance.

To explore these issues in greater depth, we must depart from the construction of general tools and consider channel-adapted QEC for a specific channel model.  We examine the amplitude damping channel, introduced in Sec.~\ref{sec:AmpDampChannel} and used as a primary exampe throughout the dissertation.  The amplitude damping channel is a logical choice for several reasons.  First of all, it has a useful physical interpretation: it models the decay from an excited state to the ground state for a qubit.  Second, amplitude damping cannot be written with scaled Pauli matrices as the operator elements; thus Theorem \ref{thm:PauliChannel} does not apply.  Finally, due to its structure, the amplitude damping channel can still be analyzed with the stabilizer formalism, greatly aiding analysis.

We begin with a qualitative understanding of the $[4,1]$ code and its optimal channel-adapted recovery operation.  We first interpret the recovery in terms of the code words and then in terms of the code stabilizers.  We see that we can understand both the dominant errors and the recovery operation in terms of stabilizer operations.  The stabilizer interpretation permits a simple generalization for higher rate amplitude damping codes and recovery operations.  In particular, we define two classes of amplitude damping-adapted error correcting codes that can be derived and understood with a simple stabilizer structure.

\section{Qualitative Analysis of Channel-Adapted QER for Approximate [4,1] Code}\label{sec:qualitative analysis 4,1}

\begin{table}[bt]
\begin{center}
  \begin{tabular}{|c|c|}
    \hline
    $R_1$ & $\ket{0_L}(\alpha \bra{0000} +\beta\bra{1111})+\ket{1_L}(\frac{1}{\sqrt{2}}\bra{0011}+\frac{1}{\sqrt{2}}\bra{1100})$\\
    $R_2$ & $\ket{0_L}(\beta \bra{0000} -\alpha\bra{1111})+\ket{1_L}(\frac{1}{\sqrt{2}}\bra{0011}-\frac{1}{\sqrt{2}}\bra{1100})$\\
    $R_3$ & $\ket{0_L}\bra{0111}+\ket{1_L}\bra{0100}$\\
    $R_4$ & $\ket{0_L}\bra{1011}+\ket{1_L}\bra{1000}$\\
    $R_5$ & $\ket{0_L}\bra{1101}+\ket{1_L}\bra{0001}$\\
    $R_6$ & $\ket{0_L}\bra{1110}+\ket{1_L}\bra{0010}$\\
    $R_7$ & $\ket{0_L}\bra{1001}$\\
    $R_8$ & $\ket{0_L}\bra{1010}$\\
    $R_9$ & $\ket{0_L}\bra{0101}$\\
    $R_{10} $& $\ket{0_L}\bra{0110}$\\
    \hline
  \end{tabular}
  \end{center}
  \caption[Optimal QER operator elements for the 4 qubit code.]{Optimal QER operator elements for the [4,1] code.  Operators $R_1$ and $R_2$ correspond to the ``no dampings'' term $E_0^{\otimes 5}$ where $\alpha$ and $\beta$ depend on $\gamma$.  $R_3-R_6$ correct first order dampings.  $R_7-R_{10}$ partially correct some second order dampings, though as only $\ket{0_L}$ is returned in these cases superposition is not preserved.}\label{tab:4 qubit recovery}
\end{table}

Let's consider the optimal channel-adapted recovery for the [4,1] `approximate' code of \cite{LeuNieChuYam:97}.  Described in Sec.~\ref{sec:4 qubit code}, this is an example of a channel-adapted code, designed specifically for the amplitude damping channel rather than arbitrary qubit errors.  Its initial publication demonstrated the utility of channel-adaptation (though without using such a term) for duplicating the performance of standard quantum codes with both a shorter block length and while achieving a higher rate.  In \cite{LeuNieChuYam:97}, the authors proposed a recovery (decoding) circuit and demonstrated its strong performance in minimum fidelity.

It is interesting to note that the recovery operation (described in quantum circuit form in Fig.~\ref{fig:Approx QEC Recovery}) is not a projective syndrome measurement followed by a unitary rotation as is standard for generic codes; yet the optimal recovery \emph{does} conform to such a structure.  Recall that the logical codewords are given by
\begin{eqnarray}\label{eq:4qubitcode_v2}
\ket{0_L}&=& \frac{1}{\sqrt{2}}(\ket{0000}+\ket{1111})\\
\ket{1_L}&=& \frac{1}{\sqrt{2}}(\ket{0011}+\ket{1100})\label{eq:4qubitcode1_v2}.
\end{eqnarray}
The optimal recovery operation is given in Table \ref{tab:4 qubit recovery}.  We will analyze each of the operator elements in turn.  For clarity of presentation, we begin with first and second order damping errors and then we turn our attention to the recovery from the `no damping' term.

\subsection{Recovery from first and second order damping errors}

Recall that the amplitude damping channel on a single qubit has operator elements
\begin{equation}\label{eq:ampdamp v2}
E_0=\left [ \begin{array}{ccc} 1 & 0 \\ 0 &\sqrt{1-\gamma} \end{array} \right ]\hspace{.5 cm} \textrm{and} \hspace{.5 cm}
E_1=\left [ \begin{array}{ccc} 0 & \sqrt{\gamma} \\ 0 & 0 \end{array} \right ],
\end{equation}
neither of which is a scaled unitary operator.  Let us denote a first order damping error as $E_1^{(k)}$, which consists of the qubit operator $E_1$ on the $k^{th}$ qubit and the identity elsewhere.  Consider now the effect of $E_1^{(1)}$ on the codewords of the $[4,1]$ code:
\begin{eqnarray}
  E_1\otimes I^{\otimes 3} \ket{0_L}=\sqrt{\gamma}\ket{0111},\\
  E_1\otimes I^{\otimes 3} \ket{1_L}=\sqrt{\gamma}\ket{0100}.
\end{eqnarray}
We see that the code subspace is perturbed onto an orthogonal subspace spanned by $\{\ket{0111},\ket{0100}\}$.  $R_3$ projects onto this syndrome subspace and recovers appropriately into the logical codewords.  Recovery operators $R_4$, $R_5$, and $R_6$ similarly correct damping errors on the second, third, and fourth qubits.  Notice that the first order damping errors move the information into mutually orthogonal subspaces.  It is therefore not hard to see that the set of errors $\{I^{\otimes 4},E_1^{(k)}\}_{k=1}^4$ satisfy the error correcting conditions for the $[4,1]$ code.  (That the $[4,1]$ code satisfies the error correcting conditions for damping errors was pointed out in \cite{Got:97}.)

Consider now the subspace spanned by $\{\ket{1010},\ket{0101},\ket{0110},\ket{1001}\}$.  By examining the logical codewords in (\ref{eq:4qubitcode_v2}) and (\ref{eq:4qubitcode1_v2}), we see that this subspace can only be reached by multiple damping errors.  Unfortunately, in such a case we lose the logical superpositions as only $\ket{0_L}$ is perturbed into this subspace.  Consider, for example the two damping error $E_1^{(1)}E_1^{(3)}$.  We see that
\begin{eqnarray}
  E_1^{(1)}E_1^{(3)}\ket{0_L}&=&\gamma\ket{0101},\\
  E_1^{(1)}E_1^{(3)}\ket{1_L}&=&0.
\end{eqnarray}
While we cannot fully recover from such an error, we recognize that these higher order errors occur with probability $\gamma^2$.  Furthermore, we see that operator elements $R_7-R_{10}$ do recover the $\ket{0_L}$ portion of the input information.   This contributes a small amount to the overall entanglement fidelity, though would obviously not help the minimum fidelity case.  Indeed, $R_7-R_{10}$ do not contribute to maintaining the fidelity of an input $\ket{1_L}$ state.

We should also note that only a subset of all second order dampings are partially correctable as above.  We reach the syndrome subspaces from $R_7-R_{10}$ only when a qubit from the first pair and a qubit from the second pair is damped, allowing the $\ket{0_L}$ state to be recovered.  If both the first and second qubits (or both the third and fourth qubits) are damped, the resulting states are no longer orthogonal to the code subspace.  In fact, these are the only errors that will cause a logical bit flip, recovering $\ket{0_L}$ as $\ket{1_L}$ and vice versa.

\subsection{Recovery from the distortion of the `no damping' case}\label{sec:4qubit no damping}

We turn now to the recovery operators $R_1$ and $R_2$.  Together these project onto the syndrome subspace with basis vectors $\{\ket{0000},\ket{1111},\ket{1100},\ket{0011}\}$ which includes the entire code subspace.  We just saw that $I^{\otimes 4}$ together with single qubit dampings are correctable, but $\E_{a}^{\otimes 4}$ does not have an operator element proportional to $I^{\otimes 4}$.  Instead, the `no dampings' term is given by $E_0^{\otimes 4}$ which depends on the damping parameter $\gamma$.  Indeed, consider the effect of the no damping term on the logical code words:
\begin{eqnarray}
  E_0^{\otimes 4}\ket{0_L}&=&\frac{1}{\sqrt{2}}(\ket{0000}+(1-\gamma)^2\ket{1111})\\
  E_0^{\otimes 4}\ket{1_L}&=&\frac{1-\gamma}{\sqrt{2}}(\ket{1100}+\ket{0011}).
\end{eqnarray}

A standard recovery operation projects onto the code subspace.  Consider the effect of such a recovery on an arbitrary input state $a\ket{0_L}+b\ket{1_L}$. The resulting (un-normalized) state is
\begin{equation}
  a(1-\gamma+\frac{\gamma^2}{2})\ket{0_L}+b(1-\gamma)\ket{1_L}.
\end{equation}
The extra term $\frac{\gamma^2}{2}$ distorts the state from the original input.  While this distortion is small as $\gamma\rightarrow 0$, both the original recovery operation of Fig.~\ref{fig:Approx QEC Recovery} proposed in \cite{LeuNieChuYam:97} and the optimal recovery seek to reduce this distortion by use of a $\gamma$-dependent operation.  We analyze the optimal recovery operation for this term and compare its efficacy with the simpler projection.

We see that $R_1$ projects onto a perturbed version of the codespace with basis vectors $\{(\alpha\ket{0000}+\beta\ket{1111}),(\frac{1}{\sqrt{2}}\ket{0011}+\frac{1}{\sqrt{2}}\ket{1100})\}$ where $\alpha$ and $\beta$ are chosen to maximize the entanglement fidelity.  We can use any of the numerical techniques of Chapters \ref{chap:OptQER} and \ref{chap:NearOptQER} to compute good values for $\alpha$ and $\beta$, but we would like an intuitive understanding as well.  $\alpha$ and $\beta$ (where $|\beta|=\sqrt{1-|\alpha|^2}$) adjust the syndrome measurement $P_1$ so that it is no longer $\ket{0_L}\bra{0_L}+\ket{1_L}\bra{1_L}$, the projector onto the code subspace.  If we choose them so that $\bra{0_L}P_1\ket{0_L}=\bra{1_L}P_1\ket{1_L}$ then we will perfectly recover the original state when syndrome $P_1$ is detected for the no damping case.  If syndrome $P_2$ is detected, the no damping state will be distorted, but for small $\gamma$, the second syndrome is a relatively rare occurrence.  It could even be used as a classical indicator for a greater level of distortion.

\begin{figure}
  \begin{center}
    \includegraphics[width=\columnwidth]{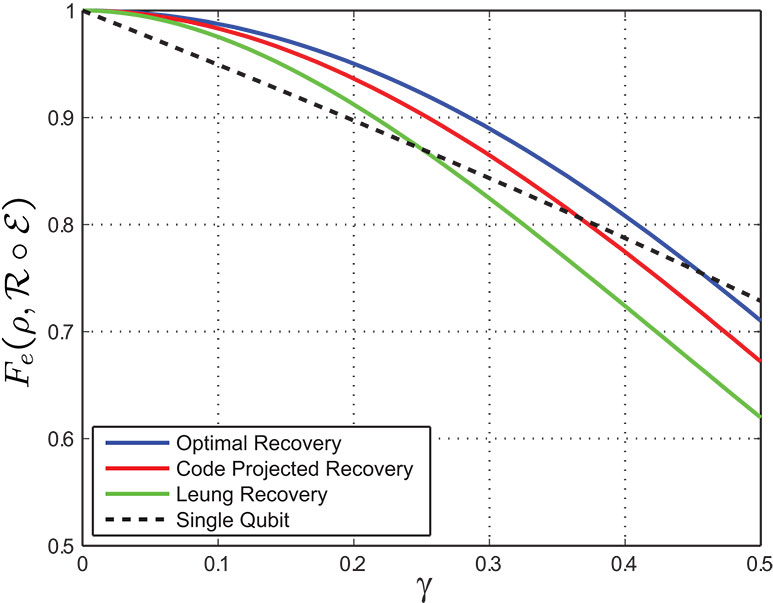}
    \caption[Optimal vs.~code projection recovery operations for the four qubit code.]{Optimal vs.~code projection recovery operations for the [4,1] code.  We compare the entanglement fidelity for the optimal recovery operation and the recovery that includes a projection onto the code subspace.  For comparison, we also include the original recovery operation proposed in \cite{LeuNieChuYam:97} and the baseline performance of a single qubit.  While the optimal recovery outperforms the code projector recovery, the performance gain is likely small compared to the cost of implementing the optimal.}\label{fig:AmpDamp4_stab_rec}
  \end{center}
\end{figure}

We can see in Fig.~\ref{fig:AmpDamp4_stab_rec} that the benefit of the optimal recovery operation is small, especially as $\gamma\rightarrow 0$, though not negligible.  Furthermore, the standard projection onto the code space is a simple operation while the optimal recovery is both $\gamma$-dependent and relatively complex to implement.  For this reason, it is likely preferable to implement the more straightforward code projection, which still reaps most of the benefits of channel-adaptation.

\section{Amplitude Damping Errors in the Stabilizer Formalism}

The stabilizer formalism provides an extremely useful and compact description for quantum error correcting codes.  As we laid out in Sec.~\ref{sec:stabilizers}, code descriptions, syndrome measurements, and recovery operations can be understood by considering the $n-k$ generators of an $[n,k]$ stabilizer code.  In standard practice, Pauli group errors are considered and if $\{X_i,Y_i,Z_i\}_{i=1}^n$ errors can be corrected, we know we can correct an arbitrary error on one of the qubits since the Pauli operators are a basis for single qubit operators.

Let's consider the $[4,1]$ code in terms of its stabilizer group $G=\langle XXXX,ZZII,$\\$IIZZ\rangle$.  We can choose the logical Pauli operators $\bar{X}=XXII$ and $\bar{Z}=ZIZI$ to specify the codewords in (\ref{eq:4qubitcode_v2}) and (\ref{eq:4qubitcode1_v2}).  We saw in Sec.~\ref{sec:qualitative analysis 4,1} that $E_1^{(i)}$ damping errors together with $I^{\otimes4}$ are correctable errors.  Since each of these errors is a linear combination of Pauli group members:
\begin{equation}
  E_1^{(i)}=\frac{\sqrt{\gamma}}{2}(X_i+iY_i),
\end{equation}
we might presume that $\{I,X_i,Y_i\}_{i=1}^4$ are a set of correctable operations and the desired recovery follows the standard stabilizer syndrome measurement structure.  This is not the case.  Consider that the operator $X_1X_2$ (or equivalently $XXII$) is in the normalizer $N(G)$ of the code stabilizer, and thus $\{X_1,X_2\}$ are not a correctable set of errors.

How, then, can the $[4,1]$ code correct errors of the form $X_i+iY_i$?  Instead of projecting onto the stabilizer subspaces and correcting $X_i$ and $Y_i$ separately, we take advantage of the fact that the errors happen in superposition and project accordingly.  As we saw, $X_i+iY_i$ and $X_j+iY_j$ project into orthogonal subspaces when $i\neq j$ and we can recover accordingly.  In fact, the correct syndrome structures can also be described in terms of stabilizers; understanding these syndromes enables design and analysis of other amplitude damping codes.

Let $G=\langle g_1,\ldots,g_{n-k}\rangle$ be the generators for an $[n,k]$ stabilizer code.  We wish to define the generators for the subspace resulting from a damping error $X_i+iY_i$ on the $i^{th}$ qubit.  First, we should note that we can always write the generators of $G$ so that at most one generator commutes with $X_i$ and anti-commutes with $Y_i$ (corresponding to a generator with an $X$ on the $i^{th}$ qubit), at most one generator that anti-commutes with both $X_i$ and $Y_i$ (corresponding to a generator with an $Z$ on the $i^{th}$ qubit), and all other generators commute with both operators.  Let $\ket{\psi}\in C(G)$ be an arbitrary state in the subspace stabilized by $G$.  If $g\in G$ such that $[g,X_i]=[g,Y_i]=0$, then
\begin{equation}
  (X_i+iY_i)\ket{\psi}=(X_i+iY_i)g\ket{\psi}=g(X_i+iY_i)\ket{\psi}.
\end{equation}
From this we see that the $i^{th}$ damped subspace is stabilized by the commuting generators of $G$.  Now consider an element of $G$ that anti-commutes with $X_i$ and $Y_i$.  Then
\begin{equation}
  (X_i+iY_i)\ket{\psi}=(X_i+iY_i)g\ket{\psi}=-g(X_i+iY_i)\ket{\psi},
\end{equation}
so $-g$ is a stabilizer of the $i^{th}$ damped subspace.  Finally, consider a $g$ which commutes with $X_i$ but anti-commutes with $Y_i$:
\begin{equation}
  (X_i+iY_i)\ket{\psi}=(X_i+iY_i)g\ket{\psi}=g(X_i-iY_i)\ket{\psi}.
\end{equation}
We see that neither $g$ nor $-g$ is a stabilizer for the subspace.  It is, however, not hard to see that $Z_i$ is a generator:
\begin{equation}
  Z_i(X_i+iY_i)\ket{\psi}=(iY_i-i^2X_i)\ket{\psi}=(X_i+iY_i)\ket{\psi}.
\end{equation}
In this manner, given any code stabilizer $G$, we can construct the stabilizer for each of the damped subspaces.

\begin{table}
\begin{center}
\begin{tabular}{c}
    $1^{st}$ subspace\\
    \hline
  \begin{tabular}{c@{}c@{}c@{}c@{}c}
    -&$Z$&$Z$&$I$&$I$\\
    &$I$&$I$&$Z$&$Z$\\
    &$Z$&$I$&$I$&$I$
  \end{tabular}
  \end{tabular}
  \hspace{20pt}
  \begin{tabular}{c}
    $2^{nd}$ subspace\\
    \hline
   \begin{tabular}{c@{}c@{}c@{}c@{}c}
   -&$Z$&$Z$&$I$&$I$\\
    &$I$&$I$&$Z$&$Z$\\
    &$I$&$Z$&$I$&$I$
  \end{tabular}
  \end{tabular}
  \hspace{20pt}
  \begin{tabular}{c}
    $3^{rd}$ subspace\\
    \hline
  \begin{tabular}{c@{}c@{}c@{}c@{}c}
    &$Z$&$Z$&$I$&$I$\\
    -&$I$&$I$&$Z$&$Z$\\
    &$I$&$I$&$Z$&$I$
  \end{tabular}
  \end{tabular}
  \hspace{20pt}
  \begin{tabular}{c}
    $4^{th}$ subspace\\
    \hline
  \begin{tabular}{c@{}c@{}c@{}c@{}c}
    &$Z$&$Z$&$I$&$I$\\
    -&$I$&$I$&$Z$&$Z$\\
    &$I$&$I$&$I$&$Z$
  \end{tabular}
  \end{tabular}
  \end{center}
  \caption[Stabilizers for each of the damped subspaces of the four qubit code.]{Stabilizers for each of the damped subspaces of the $[4,1]$ code.}\label{tab:4 qubit damped subspaces}
\end{table}

Consider now the stabilizer description of each of the damped subspaces for the $[4,1]$ code.  These are given in Table \ref{tab:4 qubit damped subspaces}.  Recall that two stabilizer subspaces are orthogonal if and only if there is an element $g$ that stabilizes one subspace while $-g$ stabilizes the other.  It is easy to see that each of these subspaces is orthogonal to the code subspace, as either $-ZZII$ or $-IIZZ$ is included.  It is equally easy to see that the first and second subspaces are orthogonal to the third and fourth.  To see that the first and second subspaces are orthogonal, note that $-IZII$ stabilizes the first subspace, while $IZII$ stabilizes the second.  Equivalently, $-IIZI$ stabilizes the fourth subspace, thus making it orthogonal to the third.

We can now understand the optimal recovery operation in terms of the code stabilizers.  Consider measuring $ZZII$ and $IIZZ$.  If the result is $(+1,+1)$ then we conclude that no damping has occurred and perform the non-stabilizer operations of $R_1$ and $R_2$ to minimize distortion.  If we measure $(-1,+1)$ we know that either the first or the second qubit was damped.  We can distinguish by measuring $ZIII$, with $+1$ indicating a damping on the first qubit and $-1$ a damping on the second.  If our first syndrome is $(+1,-1)$, we can distinguish between dampings on the third and fourth by measuring $IIZI$.
If our first syndrome yields $(-1,-1)$ we conclude that multiple dampings occurred.  We could simply return an error, or we can do the partial corrections of $R_7-R_{10}$ by further measuring both $ZIII$ and $IIZI$.  It is worth pointing out a feature of the stabilizer analysis highlighted by this multiple dampings case.  Each of the damping subspaces from Table \ref{tab:4 qubit damped subspaces} has three stabilizers and thus encodes a 2 dimensional subspace.  Consider applying $E_1^{(1)}$ to the third damped subspace, equivalent to damping errors on qubits 1 and 3.  Note that there is no generator with an $X$ in the first qubit; the resulting subspace is stabilized by
\begin{equation}
  \langle -ZZII,-IIZZ,IIZI,ZIII \rangle.
\end{equation}
As this has four independent generators, the resulting subspace has dimension 1.  We saw this in the previous section, where for multiple dampings the recovery operation does not preserve logical superpositions but collapses to the $\ket{0_L}$ state.

Stabilizer descriptions for amplitude damping-adapted codes are quite advantageous.  Just as in the case of standard quantum codes, the compact description facilitates analysis and aids design.  While the recovery operations for the amplitude damping codes are not quite as neatly described as the standard stabilizer recovery, the stabilizer formalism facilitates the description.  Furthermore, by considering stabilizer descriptions of the $[4,1]$ code and its recovery operation, we may design other channel-adapted amplitude damping codes.  We will rely on stabilizers throughout the remainder of the chapter.

\section{Evidence for a [3,1] Amplitude Damping Code}

In the previous section, we saw that the $[4,1]$ code, despite its original label of `approximate,' perfectly corrects for the first order damping errors $\{I^{\otimes 4}, E_1^{(k)}\}$.  While correcting these errors is not the traditional choice for quantum error correcting codes, this set accurately represents the amplitude damping channel to first order in $\gamma$.  Given this fact, it is reasonable to look for other codes for which first order damping errors satisfy the error correction conditions.

For an $[n,k]$ code, there are $n$ first order damping errors.  One way to satisfy the error correcting conditions is if each of these rotates the code subspace into mutually orthogonal subspaces of dimension $2^k$ without (significantly) distorting the code subspace.  We saw this to be the case with the $[4,1]$ code.  For such an encoding to exist, $\HH_C$ must have dimension $d_C\geq 2^k(n+1)$ as each error, plus the $I^{\otimes 4}$ operator, must result in an orthogonal subspace.  This is satisfied for the $[4,1]$ code as $d_C=16\geq 2(4+1)=10$.  This inequality holds with equality for a $[3,1]$ code: $d_C=8=2(3+1)$, suggesting the existence of a good $[3,1]$ amplitude damping code.

This degrees of freedom argument is actually quite intuitive: we know of good $[3,1]$ codes for both the bit flip and phase flip channels.  (See \cite{NieChu:B00}$\S$10.1.)  These channels are similar to our treatment of the amplitude damping channel as there are only $n$ first order errors for an $[n,k]$ code.  As they both have good $[3,1]$ codes, this further suggests the existence of a $[3,1]$ amplitude damping code.

As we mentioned in Sec.~\ref{sec:opt encoding}, several authors \cite{ReiWer:05,KosLid:06,FleShoWin:J07a} have suggested iterative methods to determine channel-adapted quantum codes.  Given an initial choice for an encoding isometry $U_C$, we can determine the optimal recovery operation $\R$.  If we then hold $\R$ fixed, we can determine the optimal encoding $U_C$.  In this way we iteratively approach an encoding/recovery pair that is locally optimal.

\begin{figure}
  \begin{center}
    \includegraphics[width=\columnwidth]{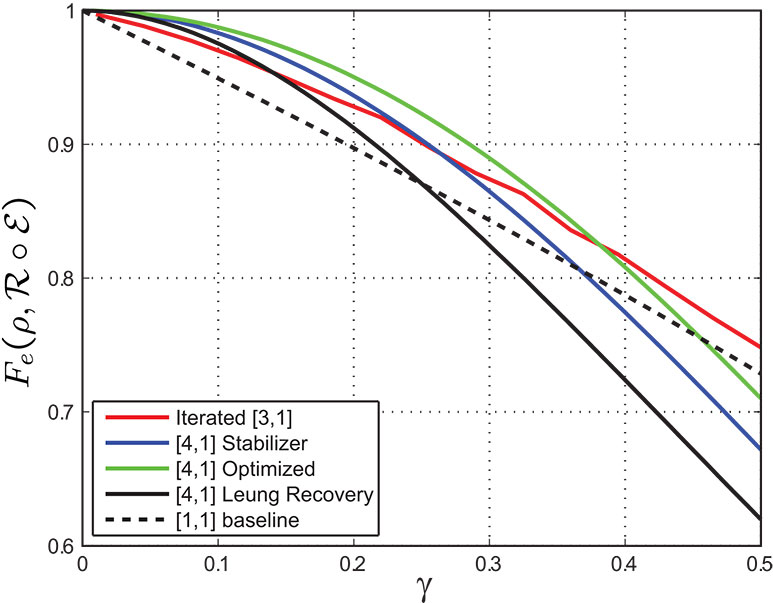}
    \caption[Performance of iterated three qubit amplitude damping code.]{Performance of iterated $[3,1]$ amplitude damping code.  This code is determined via iterative optimization of a random encoding.  For comparative purposes, we include various recovery performances for the $[4,1]$ code.}\label{fig:31 iterated}
  \end{center}
\end{figure}

Figure \ref{fig:31 iterated} shows the entanglement fidelity performance for $[3,1]$ codes determined by optimizing random encodings for the amplitude damping channel.  We see that there are $[3,1]$ codes that are better than the baseline performance, though for small $\gamma$ the $[4,1]$ code has higher fidelity.  It is interesting to see that for larger $\gamma$, $[3,1]$ codes continue to perform above the baseline even when the $[4,1]$ does not.  This arises as the $\{E_1^{(k)}\}$ are no longer representative of the first order errors.  At this point, the optimization procedure tunes to the correct representation of the channel, while the $[4,1]$ codes do not as the encoding is not properly channel-adapted.

The numerically obtained $[3,1]$ codes are difficult to both analyze and utilize.  First of all, the iterated optimization problem is not convex and has many local optima.  These optima have nearly identical performance, though no obvious similar structure.  We know that, due to the symmetry of the channel, swapping any of the qubits of the code will not change its performance.  Given all of the local solutions, there appear to be other sources of ambiguity as well.  We also note that numerically obtained quantum codes are challenging to implement.  With no clear structure, both the encoding and recovery quantum circuits may require a large number of gates.  Finally, we should note that the iterated codes do not satisfy the quantum error correcting conditions for first order dampings, though the damped subspaces are approximately orthogonal.

To alleviate these issues, we would like to find a good $[3,1]$ stabilizer code.   Unfortunately, none exist which perfectly correct $\{E_1^{(k)}\}$ errors.  A $[3,1]$ code is stabilized by a group $G$ with two independent generators.  We saw the effect of damping errors on stabilizer subspaces in the previous section.  For each damped subspace to be mutually orthogonal and rank 2, the generators must satisfy several requirements.  First of all, for every qubit, there must be a $g\in G$ such that $\{g,X_i\}=\{g,Y_i\}=0$.  In words, there must be a $g$ with a $Z$ on the $i^{th}$ qubit, for $i=1,2,3$.  If this is not the case, then the damped subspace will not be orthogonal to the code subspace.  Second, for every qubit, we need a $g$ with either an $X$ or a $Y$ on that qubit.  If this is not the case, then the damped subspace will have dimension 1, since the $i^{th}$ damped subspace is always stabilized by $Z_i$.  To satisfy these requirements, the two generators of $G$ must anti-commute at each qubit; since there are three qubits, this means the generators anti-commute which is a contradiction.

We can illustrate the impossibility of a good $[3,1]$ amplitude damping stabilizer code by example.  As there are a relatively small number of possibilities for $[3,1]$ stabilizer code, it is a simple matter to compute the optimal recovery operation for each encoding, given the amplitude damping channel.  From such an exercise (for small $\gamma$), we determine that the best $[3,1]$ stabilizer code has the stabilizer group $\langle XYZ, ZZI \rangle$.  The first and second damped subspaces are stabilized by $\langle -ZZI,ZII\rangle$ and $\langle -ZZI, IZI\rangle$.  It is not hard to see that these are mutually orthogonal subspaces and are also orthogonal to the code subspace.  When we look at the third damped subspace, we see that it is stabilized by $\langle -XYZ,ZZI,IIZ\rangle$.  As this has three generators, the stabilized subspace has only one dimension.  If were to utilize this encoding scheme, a damping on the third qubit would not be correctable as only half of the logical space could be preserved.  From a minimum fidelity standpoint, such an encoding would be worse than no error correction procedure at all.

\section{Generalization of the [4,1] Code for Higher Rates}\label{sec:generalized 41}

While we are unable to generate a $[3,1]$ stabilizer code for the amplitude damping channel, it is still interesting to consider good channel-adapted codes of longer block lengths with the same or higher rate.  Fortunately, the stabilizer analysis for the $[4,1]$ code provides a ready means to generalize for higher rate code.  Consider the three codes given in Table \ref{tab:six-two and eight-three} (A).  Each of these is an obvious extension of the $[4,1]$ code, but with a higher rate.  Indeed the general structure can be extended as far as desired generating an $[2(M+1),M]$ code for all positive integers $M$.  We can thus generate a code with rate arbitrarily close to $1/2$.

While the codes presented in Table \ref{tab:six-two and eight-three} (A) have an obvious pattern related to the $[4,1]$ code, we will find it more convenient to consider the stabilizer in standard form as given in Table \ref{tab:six-two and eight-three} (B).  The standard form, including the choice of $\bar{X}_i$ and $\bar{Z}_i$, provides a systematic means to write the encoding circuit.  The change is achieved through a reordering of the qubits which, due to the symmetry of the channel, has no effect on the error correction properties.

Let's consider the form of the $M+2$ stabilizer group generators.  Just as with the $[4,1]$ code, the first generator has an $X$ on every qubit.  The physical qubits are grouped into $M+1$ pairs; for each pair $(i,j)$ there is a generator $Z_iZ_j$.

The structure of the stabilizers makes it easy to see that $\{I^{\otimes 2(M+1)},E_1^{(k)}\}_{k=1}^{2(M+1)}$ satisfy the error correcting conditions for the $[2(M+1),M]$ code.  To see this, we will show that the damped subspaces are mutually orthogonal, and orthogonal to the code subspace.  Consider a damping on the $i^{th}$ qubit, where $i$ and $j$ are a pair.  The resulting state is stabilized by $Z_i$, $-Z_iZ_j,$ and the remaining $Z$-pair generators.  We will call this the $i^{th}$ damped subspace.  This subspace is clearly orthogonal to the code subspace, due to the presence of the $-Z_iZ_j$ stabilizer. For the same reason, the $i^{th}$ damped subspace is clearly orthogonal to the $k^{th}$ damped subspace for $k\neq j$.  Finally, the $i^{th}$ and $j^{th}$ damped subspaces are orthogonal as we see that $Z_i$ stabilizes the $i^{th}$ and $-Z_i$ stabilizes the $j^{th}$.

\begin{figure}[tb]
  \centerline{
    \Qcircuit @C=1.3em @R=.7em {
    \lstick{\ket{0}} & \qw & \qw & \qw & \qw & \qw & \gate{H} & \ctrl{17} & \qw \\
    \lstick{\ket{0}} & \qw & \qw & \qw & \qw & \qw & \qw & \targ & \qw \\
    & & &  & & & & &\\
    & & & \vdots & & & & &\\
    & & &  & & & & &\\
    \lstick{\ket{0}} & \qw & \qw & \qw & \qw & \qw & \qw & \targ & \qw \\
    \lstick{\ket{0}} & \qw & \qw & \qw & \targ & \qw & \qw & \targ & \qw \\
    & & &  & & & & &\\
    & & & \vdots & & & & &\\
    & & &  & & & & &\\
    \lstick{\ket{0}} & \qw & \targ & \qw & \qw & \qw & \qw & \targ & \qw \\
    \lstick{\ket{0}} & \targ & \qw & \qw & \qw & \qw & \qw & \targ & \qw \\
    \lstick{k_1} & \ctrl{-1} & \qw & \qw & \qw & \qw & \qw & \targ & \qw \\
    \lstick{k_2} & \qw & \ctrl{-3} & \qw & \qw & \qw & \qw & \targ & \qw \\
    & & &  & & & & &\\
    & & & \vdots & & & & &\\
    & & &  & & & & &\\
    \lstick{k_M} & \qw & \qw & \qw & \ctrl{-11} & \qw & \qw & \targ & \qw
    }
  }
  \caption{Circuit to encode the arbitrary state of $M$ qubits given in qubits $k_1\cdots k_M$ into $2(M+1)$ physical qubits.  This is the $[2(M+1),M]$ code in standard form.}\label{fig:generalized_leung_encoding}
\end{figure}
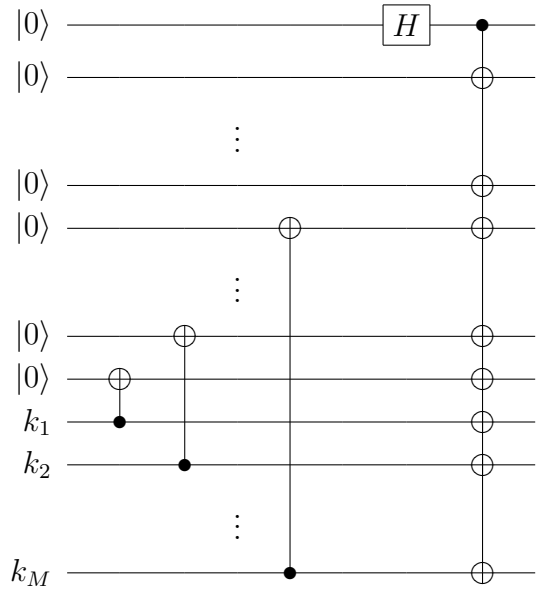

By writing the $[2(M+1),M]$ codes in the standard form, it is easy to generate an encoding circuit.  The circuit to encode the arbitrary state $\ket{\psi}$ in the $M$ qubits $k_1\cdots k_M$ is given in Fig.~\ref{fig:generalized_leung_encoding}.  The encoding circuit requires $3M+1$ CNOT operations and one Hadamard gate.

\clearpage
Let's write out the logical codewords  of the $[6,2]$ code given the choice of $\bar{Z}_i$ in Table \ref{tab:six-two and eight-three}:
\begin{eqnarray}
  \ket{00_L}&=&\frac{1}{\sqrt{2}}(\ket{000000}+\ket{111111})\\
  \ket{01_L}&=&\frac{1}{\sqrt{2}}(\ket{001001}+\ket{110110})\\
  \ket{10_L}&=&\frac{1}{\sqrt{2}}(\ket{000110}+\ket{111001})\\
  \ket{11_L}&=&\frac{1}{\sqrt{2}}(\ket{110000}+\ket{001111}).
\end{eqnarray}
Each codeword is the equal superposition of two basis states.  We can see by inspection that the damped subspaces are mutually orthogonal:  $E_1^{(k)}$ will eliminate one of the two basis states from each codeword and the resulting basis states do not overlap.

\subsection{[2(M+1),M] Syndrome measurement}

We begin the recovery by first measuring the $Z$-pair stabilizers.  A $-1$ result on the $(i,j)$-pair stabilizer indicates a damping of either the $i^{th}$ or $j^{th}$ qubit.  This holds true even if multiple $Z$-pair stabilizers measure $-1$. Such a result indicates multiple damped qubits.  Once we have identified the qubit pair, we perform an additional stabilizer measurement to determine which of the qubits was damped.  As an example, if the $(i,j)$-pair was damped, we measure $Z_i$, with a $+1$ result indicating a damping on the $i^{th}$ qubit and a $-1$ indicating a damping on the $j^{th}$ qubit.  We perform this measurement for all pairs which measure $-1$.



\begin{table}
\begin{center}
\begin{tabular}{ccc} 

  \begin{tabular}{c}
    $[6,2]$ code\\
    \hline
    \begin{tabular}{c@{}c@{}c@{}c@{}c@{}c}
    $X$&$X$&$X$&$X$&$X$&$X$\\
    $Z$&$Z$&$I$&$I$&$I$&$I$\\
    $I$&$I$&$Z$&$Z$&$I$&$I$\\
    $I$&$I$&$I$&$I$&$Z$&$Z$
    \end{tabular}
  \end{tabular}
  &
  \begin{tabular}{c}
    $[8,3]$ code\\
    \hline
    \begin{tabular}{c@{}c@{}c@{}c@{}c@{}c@{}c@{}c}
    $X$&$X$&$X$&$X$&$X$&$X$&$X$&$X$\\
    $Z$&$Z$&$I$&$I$&$I$&$I$&$I$&$I$\\
    $I$&$I$&$Z$&$Z$&$I$&$I$&$I$&$I$\\
    $I$&$I$&$I$&$I$&$Z$&$Z$&$I$&$I$\\
    $I$&$I$&$I$&$I$&$I$&$I$&$Z$&$Z$
    \end{tabular}
  \end{tabular}
  &
  \begin{tabular}{c}
    $[10,4]$ code\\
    \hline
    \begin{tabular}{c@{}c@{}c@{}c@{}c@{}c@{}c@{}c@{}c@{}c}
    $X$&$X$&$X$&$X$&$X$&$X$&$X$&$X$&$X$&$X$\\
    $Z$&$Z$&$I$&$I$&$I$&$I$&$I$&$I$&$I$&$I$\\
    $I$&$I$&$Z$&$Z$&$I$&$I$&$I$&$I$&$I$&$I$\\
    $I$&$I$&$I$&$I$&$Z$&$Z$&$I$&$I$&$I$&$I$\\
    $I$&$I$&$I$&$I$&$I$&$I$&$Z$&$Z$&$I$&$I$\\
    $I$&$I$&$I$&$I$&$I$&$I$&$I$&$I$&$Z$&$Z$
    \end{tabular}
  \end{tabular}
  \\ 
  & (A) & \\
  \hline
  \begin{tabular}{c}
    $[6,2]$ standard form\\
    \hline
    \begin{tabular}{c@{}c@{}c@{}c@{}c@{}c}
    $X$&$X$&$X$&$X$&$X$&$X$\\
    $Z$&$Z$&$I$&$I$&$I$&$I$\\
    $I$&$I$&$Z$&$I$&$I$&$Z$\\
    $I$&$I$&$I$&$Z$&$Z$&$I$
    \end{tabular}\\
    \begin{tabular}{c@{ = }c@{}c@{}c@{}c@{}c@{}c}
    $\bar{X_1}$&$I$&$I$&$I$&$X$&$X$&$I$\\
    $\bar{X_2}$&$I$&$I$&$X$&$I$&$I$&$X$\\
    $\bar{Z_1}$&$Z$&$I$&$I$&$I$&$Z$&$I$\\
    $\bar{Z_2}$&$Z$&$I$&$I$&$I$&$I$&$Z$
    \end{tabular}
 \end{tabular}
  &
  \begin{tabular}{c}
    $[8,3]$ standard form\\
    \hline
    \begin{tabular}{c@{}c@{}c@{}c@{}c@{}c@{}c@{}c}
    $X$&$X$&$X$&$X$&$X$&$X$&$X$&$X$\\
    $Z$&$Z$&$I$&$I$&$I$&$I$&$I$&$I$\\
    $I$&$I$&$Z$&$I$&$I$&$I$&$I$&$Z$\\
    $I$&$I$&$I$&$Z$&$I$&$I$&$Z$&$I$\\
    $I$&$I$&$I$&$I$&$Z$&$Z$&$I$&$I$\\
    \end{tabular}\\
    \begin{tabular}{c@{ = }c@{}c@{}c@{}c@{}c@{}c@{}c@{}c}
    $\bar{X_1}$&$I$&$I$&$I$&$I$&$X$&$X$&$I$&$I$\\
    $\bar{X_2}$&$I$&$I$&$I$&$X$&$I$&$I$&$X$&$I$\\
    $\bar{X_3}$&$I$&$I$&$X$&$I$&$I$&$I$&$I$&$X$\\
    $\bar{Z_1}$&$Z$&$I$&$I$&$I$&$I$&$Z$&$I$&$I$\\
    $\bar{Z_2}$&$Z$&$I$&$I$&$I$&$I$&$I$&$Z$&$I$\\
    $\bar{Z_3}$&$Z$&$I$&$I$&$I$&$I$&$I$&$I$&$Z$
    \end{tabular}
  \end{tabular}
  &
    \begin{tabular}{c}
    $[10,4]$ standard form\\
    \hline
    \begin{tabular}{c@{}c@{}c@{}c@{}c@{}c@{}c@{}c@{}c@{}c}
    $X$&$X$&$X$&$X$&$X$&$X$&$X$&$X$&$X$&$X$\\
    $Z$&$Z$&$I$&$I$&$I$&$I$&$I$&$I$&$I$&$I$\\
    $I$&$I$&$Z$&$I$&$I$&$I$&$I$&$I$&$I$&$Z$\\
    $I$&$I$&$I$&$Z$&$I$&$I$&$I$&$I$&$Z$&$I$\\
    $I$&$I$&$I$&$I$&$Z$&$I$&$I$&$Z$&$I$&$I$\\
    $I$&$I$&$I$&$I$&$I$&$Z$&$Z$&$I$&$I$&$I$\\
    \end{tabular}\\
    \begin{tabular}{c@{ = }c@{}c@{}c@{}c@{}c@{}c@{}c@{}c@{}c@{}c}
    $\bar{X_1}$&$I$&$I$&$I$&$I$&$I$&$X$&$X$&$I$&$I$&$I$\\
    $\bar{X_2}$&$I$&$I$&$I$&$I$&$X$&$I$&$I$&$X$&$I$&$I$\\
    $\bar{X_3}$&$I$&$I$&$I$&$X$&$I$&$I$&$I$&$I$&$X$&$I$\\
    $\bar{X_4}$&$I$&$I$&$X$&$I$&$I$&$I$&$I$&$I$&$I$&$X$\\
    $\bar{Z_1}$&$Z$&$I$&$I$&$I$&$I$&$I$&$Z$&$I$&$I$&$I$\\
    $\bar{Z_2}$&$Z$&$I$&$I$&$I$&$I$&$I$&$I$&$Z$&$I$&$I$\\
    $\bar{Z_3}$&$Z$&$I$&$I$&$I$&$I$&$I$&$I$&$I$&$Z$&$I$\\
    $\bar{Z_4}$&$Z$&$I$&$I$&$I$&$I$&$I$&$I$&$I$&$I$&$Z$
    \end{tabular}
  \end{tabular}
  \\
  & (B) &
\end{tabular}
  \end{center}
  \caption[Stabilizers for six, eight, and ten qubit amplitude damping codes.]{Stabilizers for $[6,2]$, $[8,3]$, and $[10,4]$ qubit amplitude damping codes. In (A), these are written in a way to illustrate the connection to the $[4,1]$ code.  In (B), we present the code in the standard form, which we achieve merely by swapping the code qubits and choosing the logical operators systematically.  The standard form provides a convenient description for generating quantum circuits for encoding.}\label{tab:six-two and eight-three}
\end{table}

If multiple stabilizers yield a $-1$ measurement then we have multiple damped qubits.  As before, this reduces by half the dimension of the subspace and we cannot preserve all logical superpositions.  For an example, examine the stabilizers for the $[6,2]$ code when both the first and fifth qubits are damped:
\begin{equation}
    \langle -ZZIIII,
    IIZIIZ,
    -IIIZZI,
    ZIIIII,
    IIIIZI\rangle.
\end{equation}
This subspace has 5 stabilizers and thus has 2 dimensions.  Furthermore, combining the last two stabilizers, we can see that $ZIIIZI=\bar{Z_1}$ stabilizes the subspace,  indicating that the remaining logical information is spanned by $\{\ket{01_L},\ket{00_L}\}$.  In general, for a $[2(M+1),M]$ code, up to $M+1$ dampings can be partially corrected as long as the dampings occur on distinct qubit pairs.  If $m$ is the number of damped qubits, then the resulting subspace has dimension $2^{M+1-m}$.

If all $Z$-pair measurements for the $[2(M+1),M]$ code return $+1$, we determine that we are in the `no dampings' syndrome and may perform some  further operation to reduce distortion as much as possible.  As in the example of the $[4,1]$ code in Sec.~\ref{sec:4qubit no damping}, we can choose to optimize this recovery with a $\gamma$-dependent recovery or we can apply a stabilizer projective measurement.  In the former case, we may calculate an optimized recovery with a SDP or any of the near-optimal methods of Chapter \ref{chap:NearOptQER}.  If we choose a stabilizer measurement, we simply measure $XXXXXX$ where a $+1$ result is a projection onto the code subspace.  A $-1$ result can be corrected by applying a $IIIIIZ$ operation (in fact a $Z$ on any one of the qubits will suffice).  This can be seen by noting that the $-XXXXXX$ stabilizer changes the logical codewords by replacing the $+$ with a $-$.

\subsection{[2(M+1),M] Stabilizer syndrome recovery operations}

In the previous section, we described syndrome measurements to determine which qubits were damped.  We also explained the extent to which multiple qubit dampings are correctable.  We now present a straightforward set of Clifford group operations to recover from each syndrome.

\begin{figure}
\centerline{
  \begin{tabular}{c}
    \Qcircuit @C=1.3em @R=1em {
    \lstick{\ket{0}} & \targ & \targ & \qw & \qw &  \qw &\qw & \qw & \qw & \qw &  \meter\\
    \lstick{\ket{0}} & \qw & \qw &\targ & \targ &  \qw & \qw & \qw & \qw & \qw &  \meter\\
    \lstick{\ket{0}} & \qw & \qw & \qw & \qw & \targ & \targ & \qw & \qw & \qw &  \meter\\
    & & & & & & & & & & \\
    & & & & & & & \vdots & \\
    & & & & & & & & & & \\
    \lstick{\ket{0}} & \qw & \qw & \qw & \qw & \qw & \qw & \qw & \targ & \targ &  \meter\\
    \lstick{k_1} & \ctrl{-7} & \qw & \qw & \qw & \qw & \qw & \qw & \qw & \qw & \qw \\
    \lstick{k_2}& \qw & \ctrl{-8} & \qw & \qw & \qw & \qw & \qw & \qw & \qw & \qw \\
    \lstick{k_3}& \qw & \qw & \ctrl{-8} & \qw & \qw & \qw & \qw & \qw & \qw & \qw \\
    \lstick{k_4}& \qw & \qw & \qw & \qw & \ctrl{-8} & \qw & \qw & \qw & \qw & \qw \\
    & & & & & & & & & & \\
    & & & & & & & \vdots & \\
    & & & & & & & & & & \\
    \lstick{k_{M+1}}& \qw & \qw & \qw & \qw & \qw & \qw & \qw & \ctrl{-8} & \qw & \qw \\
    \lstick{k_{M+2}}& \qw & \qw & \qw & \qw & \qw & \qw & \qw & \qw & \ctrl{-9} & \qw \\
    & & & & & & & & & & \\
    & & & & & & & \vdots \\
    & & & & & & & & & & \\
    \lstick{k_{2M+1}}& \qw & \qw & \qw & \qw & \qw & \ctrl{-17} & \qw & \qw & \qw & \qw \\
    \lstick{k_{2M+2}}& \qw & \qw & \qw & \ctrl{-19} & \qw & \qw & \qw & \qw & \qw & \qw
    }
    \\
    \\
    (A)\\
    \\
    \begin{tabular}{c@{\hspace{2cm}}c}
    \Qcircuit @C=1.3em @R=1em {
    \lstick{\ket{0}} & \gate{H} & \ctrl{6} & \gate{H} & \meter\\
    \lstick{k_1} & \qw & \targ & \qw & \qw \\
    \lstick{k_1} & \qw & \targ & \qw & \qw \\
    & &&&\\
    & \vdots &&&\\
    &&&&\\
    \lstick{k_{2M+2}} & \qw & \targ & \qw &
    }&
    \Qcircuit @C=1.3em @R=1em {
    \lstick{\ket{0}} & \qw &\targ & \meter\\
    &&&\\
    &\vdots &&\\
    &&&\\
    \lstick{k_i} & \qw &\ctrl{-4} &\qw
    }\\
    &\\
    (B)&(C)
    \end{tabular}
    \end{tabular}
    }
  \caption{Syndrome measurement circuits for the $[2(M+1),M]$ code.  Circuit (A) measures each of the $Z$-pair stabilizers. If all of the measurements in (A) are $+1$, we are in the `no damping' syndrome and we perform the syndrome measurement in (B).  If the $(i,j)$-pair stabilizer measures $-1$, we perform the syndrome measurement in (C).}\label{fig:62syndrome}
\end{figure}

\begin{figure}
  \centerline{
    \Qcircuit @C=1.3em @R=.7em {
     & \gate{H} & \ctrl{5} & \gate{X} &\qw\\
     & \qw & \targ & \qw & \qw\\
     & \qw & \targ & \gate{X} & \qw\\
     & \qw & \targ & \qw & \qw\\
     & \qw & \targ & \qw & \qw\\
     & \qw & \targ & \qw & \qw\\
    }
  }
  \caption{Syndrome recovery circuit for the [6,2] code with the first and third qubits damped.}\label{fig:62syndrom recovery}
\end{figure}
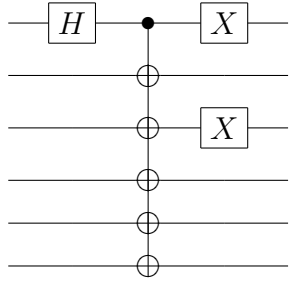

Consider a syndrome measurement in which we determine that $m$ qubits $i_1,\ldots,i_m$ were damped, where $m\leq M+1$.  We recover from this syndrome via the following three steps:
\begin{enumerate}
  \item{Apply a Hadamard gate $H_{i_1}$ on the $i_1$ qubit.}
  \item{With qubit $i_1$ as the control, apply a CNOT gate to every other qubit.}
  \item{Flip every damped qubit: $X_{i_1}\cdots X_{i_m}$.}
\end{enumerate}
The procedure is illustrated as a quantum circuit for a two-damping syndrome and the $[6,2]$ code in Fig.~\ref{fig:62syndrom recovery}.

To see that this is the correct syndrome recovery for the $[2(M+1),M]$ code, we need to examine the effect of the three gate operations on the damped subspace stabilizers.  In the syndrome where $i_1,\ldots,i_m$ are damped, we have three categories of generators for the resulting stabilizer group: $-Z$-pair stabilizers for the damped pairs, $+Z$-pair stabilizers for the non-damped pairs, and $Z_{i_1},\ldots,Z_{i_m}$ for each damped qubit.  We need to see the effect of the recovery gate operations on each of these generators.  Fortunately, we can demonstrate all of the relevant cases with the example of the $[6,2]$ code with the first and fifth qubits damped:
\begin{equation}
  \begin{tabular}{c@{}c@{}c@{}c@{}c@{}c}
    -Z&Z&I&I&I&I\\
    I&I&Z&I&I&Z\\
    -I&I&I&Z&Z&I\\
    Z&I&I&I&I&I\\
    I&I&I&I&Z&I
  \end{tabular}
  \rightarrow^{H_1}
  \begin{tabular}{c@{}c@{}c@{}c@{}c@{}c}
    -X&Z&I&I&I&I\\
    I&I&Z&I&I&Z\\
    -I&I&I&Z&Z&I\\
    X&I&I&I&I&I\\
    I&I&I&I&Z&I
  \end{tabular}
  \rightarrow^{\textrm{CNOT}_1\textrm{'s}}
  \begin{tabular}{c@{}c@{}c@{}c@{}c@{}c}
    -Y&Y&X&X&X&X\\
    I&I&Z&I&I&Z\\
    -I&I&I&Z&Z&I\\
    X&X&X&X&X&X\\
    Z&I&I&I&Z&I
  \end{tabular}
  \rightarrow^{X_1X_5}
  \begin{tabular}{c@{}c@{}c@{}c@{}c@{}c}
    Y&Y&X&X&X&X\\
    I&I&Z&I&I&Z\\
    I&I&I&Z&Z&I\\
    X&X&X&X&X&X\\
    Z&I&I&I&Z&I
  \end{tabular}
  =
  \begin{tabular}{c@{}c@{}c@{}c@{}c@{}c}
    Z&Z&I&I&I&I\\
    I&I&Z&I&I&Z\\
    I&I&I&Z&Z&I\\
    X&X&X&X&X&X\\
    Z&I&I&I&Z&I
  \end{tabular}.
\end{equation}
The final two sets of stabilizers are equivalent since $ZZIIII$ is the product of $XXXXXX$ and $YYXXXX$.  The first four generators of the resulting group are the code stabilizer.  The last generator is $\bar{Z}_1$ which, as we saw before, indicates that the recovered information is spanned by $\{\ket{00_L},\ket{01_L}\}$ while the other two dimensions of information have been lost.

While we have shown that the syndrome recovery operation returns the information to the code subspace, it remains to demonstrate that the information is correctly decoded.  We can demonstrate this by considering the syndrome recovery operation on each of the $\bar{Z}_i$ of the code.  By showing that each of these is correctly preserved, we conclude that the syndrome recovery operation is correct.

We have chosen the $\bar{Z}_i$ so that each has exactly two qubit locations with a $Z$ while the rest are $I$.  There are, therefore, five cases of interest.  In case 1, neither of the damped qubits corresponds to a location with a $Z$.  In case 2, the first damped qubit ($i_1$) corresponds to a location with a $Z$.  In case 3, one of the $Z$ locations corresponds to a damped qubit, but it is not $i_1$.  In case 4, both of the $Z$ locations correspond to a damped qubit, but neither is $i_1$.
Finally, case 5 is when both $Z$ locations correspond to damped qubits and one is $i_1$.

Without loss of generality, we can see the effect of each case by considering an example using $ZIIIZI$ and appropriately selected damped qubits.  Consider case 1:
\begin{equation}
  ZIIIZI\rightarrow^{H_{i_1}}ZIIIZI\rightarrow^{\textrm{CNOT}_{i_1}\textrm{'s}}ZIIIZI\rightarrow^{X_{i_1}\cdots X_{i_m}} ZIIIZI.
\end{equation}
Case 2:
\begin{equation}
  -ZIIIZI\rightarrow^{H_{i_1}}-XIIIZI\rightarrow^{\textrm{CNOT}_{i_1}\textrm{'s}}-YXXXYX\rightarrow^{X_{i_1}\cdots X_{i_m}} YXXXYX.
\end{equation}
Notice that this last is equivalent to $ZIIIZI$ as $XXXXXX$ is in the stabilizer.\\
Case 3:
\begin{equation}
  -ZIIIZI\rightarrow^{H_{i_1}}-ZIIIZI\rightarrow^{\textrm{CNOT}_{i_1}\textrm{'s}}-ZIIIZI\rightarrow^{X_{i_1}\cdots X_{i_m}} ZIIIZI.
\end{equation}
Case 4:
\begin{equation}
  ZIIIZI\rightarrow^{H_{i_1}}ZIIIZI\rightarrow^{\textrm{CNOT}_{i_1}\textrm{'s}}ZIIIZI\rightarrow^{X_{i_1}\cdots X_{i_m}} ZIIIZI.
\end{equation}
Case 5:
\begin{equation}
  ZIIIZI\rightarrow^{H_{i_1}}XIIIZI\rightarrow^{\textrm{CNOT}_{i_1}\textrm{'s}}YXXXYX\rightarrow^{X_{i_1}\cdots X_{i_m}} YXXXYX.
\end{equation}

We see that in all cases, the recovery procedure correctly preserves the geometry of the encoded information, even in the case of multiple qubit dampings.  It is worth emphasizing, however, that when multiple qubits are damped at least half of the information dimensions are lost.

\subsection{Performance comparison}

\begin{figure}
  \begin{center}
  \begin{tabular}{c}
  \includegraphics[width=.8\columnwidth]{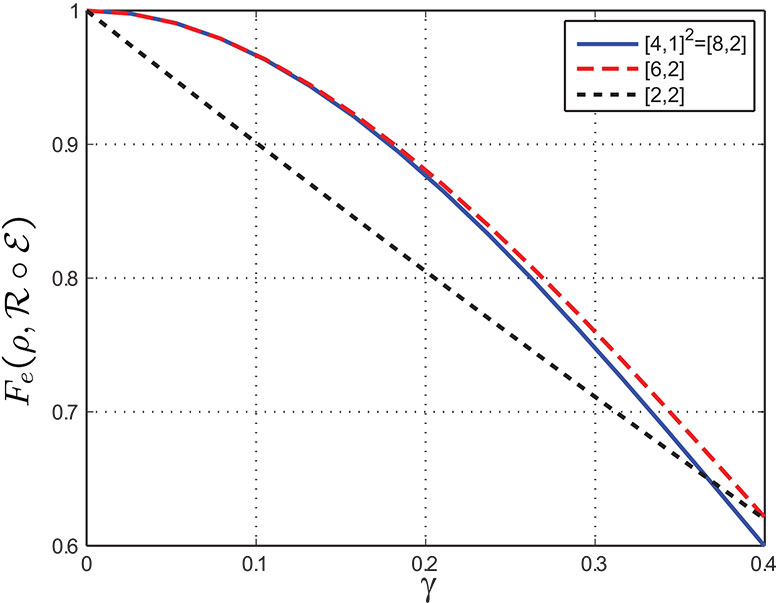}\\
  (A)\\
  \includegraphics[width=.8\columnwidth]{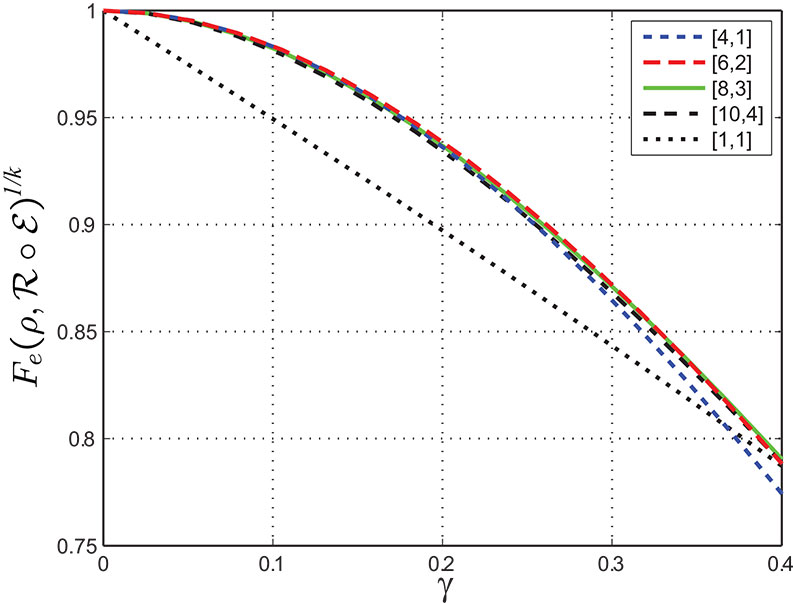}\\
  (B)
  \end{tabular}
    \caption[Performance comparison for generalized amplitude damping codes.]{Performance comparison of generalized amplitude damping codes.  In (A) we compare the $[6,2]$ code with the $[4,1]$ repeated twice.  In (B), we compare the $[4,1]$, $[6,2]$, $[8,3]$ and $[10,4]$ codes.  The entanglement fidelity has been normalized as $1/k$ where $k$ is the number of encoded qubits.  Notice that despite the increasing rates, the normalized entanglement fidelity maintains high performance.}\label{fig:Generalized_Leung}
  \end{center}
\end{figure}

It is useful to compare the performance of each of the $[2(M+1),M]$ codes in terms of the damping parameter $\gamma$.  Consider a comparison between the $[4,1]$ code and the $[6,2]$ code.  To make a valid comparison, we need to establish a common baseline.  We do this by considering the encoding of two qubits with the $[4,1]$ code. For the completely mixed state $\rho=I/2$, this is the equivalent of squaring the single qubit entanglement fidelity:
\begin{equation}
  \bar{F}_e(\rho\otimes\rho,\R\circ\E\otimes\R\circ\E)=\bar{F}_e(\rho,\R\circ\E)^2.
\end{equation}
This comparison is given in Fig.~\ref{fig:Generalized_Leung} (A).  To compare multiple codes, it is more straightforward to normalize each to a single qubit baseline.  This can be done by computing $\bar{F}_e^{(1/k)}$ for an $[n,k]$ code.  The normalized performance for the $[4,1]$, $[6,2]$, $[8,3]$ and $[10,4]$ codes is given in Fig.~\ref{fig:Generalized_Leung} (B).

It is very interesting to note how comparably these codes maintain the fidelity even as the code rate increases.  This is particularly striking when noting that each code can still perfectly correct only a single damping error.  Thus, the $[4,1]^{\otimes 4}$ can correct 4 dampings (as long as they occur on separate blocks) while the $[10,4]$ code can only perfectly correct 1.  Yet we see that the normalized performance is quite comparable.

\begin{figure}
  \begin{center}
    \includegraphics[width=\columnwidth]{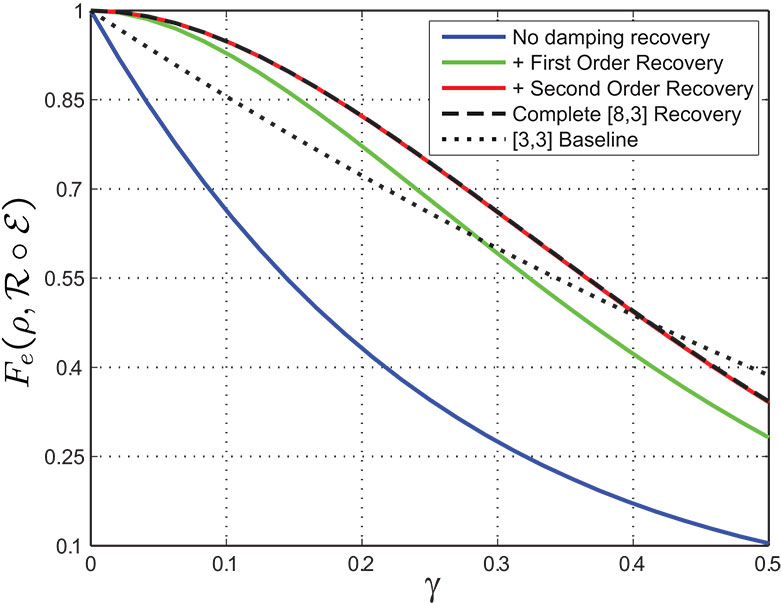}
    \caption[Fidelity contributions for each order error of the eight qubit amplitude damping code.]{Fidelity contributions for each order error of the $[8,3]$ amplitude damping code.  We see that the no damping, first, and second order recovery syndromes contribute to the entanglement fidelity of the recovery operation.}\label{fig:83fidelity_contributions}
  \end{center}
\end{figure}

\clearpage
We take a closer look at the performance of the $[8,3]$ code in Fig.~\ref{fig:83fidelity_contributions}.  We see that, while most of the entanglement fidelity is supplied by correcting no damping and $E_1^{(i)}$ terms, a not insignificant performance benefit arises by partially correcting second order damping errors.  In the case of the $[4,1]$ recovery, we concluded that such contributions improved the entanglement fidelity, but not the minimum fidelity as $\ket{1_L}$ was never preserved by such a recovery.  This is not the case for the higher rates.  Two damping errors eliminate half of the logical space, but different combinations of damping errors will divide the logical space differently.  For example, an damping error on the fifth and sixth qubits means the resulting space is stabilized by $\bar{Z}_1\bar{Z}_2$ thus eliminating logical states $\ket{01x_L}$ and $\ket{10x_L}$ (where $x$ indicates either $0$ or $1$).  On the other hand, a damping on the fifth and seventh qubits results in a space stabilized by $\bar{Z}_1\bar{Z}_3$ eliminating logical states $\ket{0x1_L}$ and $\ket{1x0_L}$.  Thus, correcting second order damping errors still contributes to minimum fidelity performance.

\begin{table}
  \begin{center}
    \begin{tabular}{c}
        Gottesman $[8,3]$ code\\
        \hline
        \begin{tabular}{c@{}c@{}c@{}c@{}c@{}c@{}c@{}c}
        $X$&$X$&$X$&$X$&$X$&$X$&$X$&$X$\\
        $Z$&$Z$&$Z$&$Z$&$Z$&$Z$&$Z$&$Z$\\
        $I$&$X$&$I$&$X$&$Y$&$Z$&$Y$&$Z$\\
        $I$&$X$&$Z$&$Y$&$I$&$X$&$Z$&$Y$\\
        $I$&$Y$&$X$&$Z$&$X$&$Z$&$I$&$Y$
        \end{tabular}
    \end{tabular}
  \end{center}
  \caption{Stabilizers for the [8,3] code due to Gottesman\cite{Got:97}.}\label{tab:Gottesman 83}
\end{table}
Given their identical rates, it is reasonable to compare the $[8,3]$ amplitude damping code presented here with the generic $[8,3]$ stabilizer code due to Gottesman\cite{Got:97}.  The stabilizers for this code are presented in Table \ref{tab:Gottesman 83}.  This code can correct an arbitrary single qubit error, and thus can correct all first order amplitude damping errors, as well as the less probable $Z$ errors.  These are corrected with 25 stabilizer syndrome measurements (Pauli operators on each of the 8 qubits as well as the identity).  This leaves an additional 7 degrees of freedom to correct for higher order errors.  While typically these are not specified, since we know the channel of interest is the amplitude damping channel, we can do a small amount of channel-adaptation by selecting appropriate recovery operations for these syndromes.  Since $X$ and $Y$ errors are the most common, we choose operators with 2 $X$'s or 2 $Y$'s (or one of each).

\begin{figure}
  \begin{center}
    \includegraphics[width=\columnwidth]{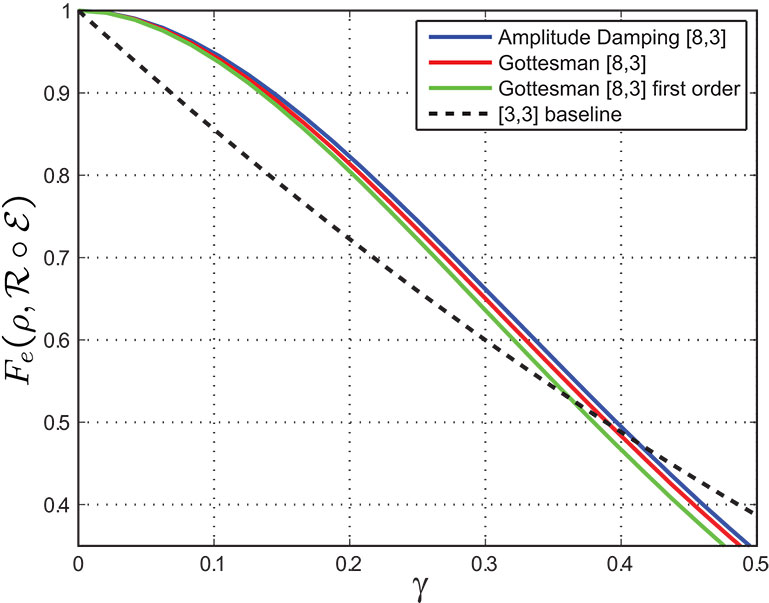}
    \caption[Comparison of the amplitude damping rate $3/8$ code and the generic rate $3/8$ code due to Gottesman.]{Comparison of the amplitude damping $[8,3]$ code and the generic rate $[8,3]$ code due to Gottesman.  We include both the Gottesman recovery where no attention is paid to second order recoveries, as well as a recovery where second order syndromes are chosen to adapt to the amplitude damping channel.}\label{fig:83Gottesman_compare}
  \end{center}
\end{figure}

The comparison between the rate $3/8$ codes is given Fig.~\ref{fig:83Gottesman_compare}.  Here we see that the channel-adapted $[8,3]$ code outperforms the generic Gottesman code, but the effect is minor.  The attention to higher order syndromes is seen to improve the performance of the $[8,3]$ code modestly.  It should be pointed out that both recovery operations can be accomplished with Clifford group operations, and neither is dependent on $\gamma$.

\section{Linear Amplitude Damping Channel Codes}

The channel-adapted codes of the previous section have similar corrective properties to the $[4,1]$ code: $\{I,E_1^{(i)}\}$ are correctable errors while $\{X_i,Y_i\}$ are not.  It is actually quite simple to design channel-adapted codes that correct both $X_i$ and $Y_i$ errors and thus can correct $\{I,E_1^{(i)}\}$ as well. Consider the $[7,3]$ code presented in Table \ref{table:seven three}.  The first three stabilizers can be readily identified as the classical $[7,4]$ classical Hamming code parity check matrix (replacing 0 with $I$ and 1 with $Z$).  They are also three of the six stabilizers for the Steane code.  Measuring these three stabilizers, an $X_i$ will result in a unique three bit measurement syndrome $(M_1,M_2,M_3)$.  (In fact, a nice property of the Hamming code is that the syndrome, replacing $+1$ with 0 and $-1$ with 1, is just the binary representation of $i$, the qubit that sustained the error.)  Unfortunately, a $Y_i$ error will yield the same syndrome as $X_i$.  We add the $XXXXXXX$ generator to distinguish the two, resulting in 14 orthogonal error syndromes for the $\{X_i,Y_i\}_{i=1}^7$.

\begin{table}
  \begin{center}
    \begin{tabular}{c}
      $[7,3]$ linear code\\
      \hline
    \begin{tabular}{c@{}c@{}c@{}c@{}c@{}c@{}c}
      $I$&$I$&$I$&$Z$&$Z$&$Z$&$Z$\\
      $I$&$Z$&$Z$&$I$&$I$&$Z$&$Z$\\
      $Z$&$I$&$Z$&$I$&$Z$&$I$&$Z$\\
      $X$&$X$&$X$&$X$&$X$&$X$&$X$
      \end{tabular}
    \end{tabular}
  \end{center}
  \caption[Amplitude damping channel-adapted seven qubit linear code.]{Amplitude damping channel-adapted $[7,3]$ linear code. Looking at the first three generators, this is clearly based on the classical Hamming code. The fourth generator differentiates between $X$ and $Y$ syndromes.}\label{table:seven three}
\end{table}

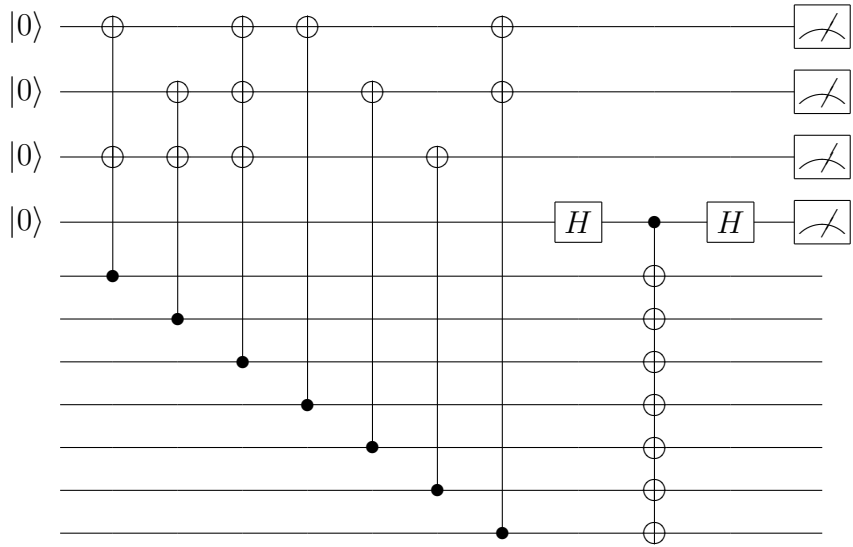
\begin{figure}[tb]
  \centerline{
    \Qcircuit @C=1.3em @R=.7em {
    \lstick{\ket{0}} & \targ & \qw & \targ & \targ & \qw & \qw & \targ & \qw & \qw & \qw &\meter\\
    \lstick{\ket{0}} & \qw & \targ & \targ & \qw & \targ & \qw & \targ & \qw & \qw & \qw &\meter\\
    \lstick{\ket{0}} & \targ & \targ & \targ & \qw & \qw & \targ & \qw & \qw & \qw & \qw &\meter\\
    \lstick{\ket{0}} & \qw & \qw & \qw & \qw & \qw & \qw & \qw & \gate{H} & \ctrl{7} & \gate{H} &\meter\\
    & \ctrl{-4} & \qw & \qw & \qw & \qw & \qw & \qw & \qw & \targ & \qw&\qw\\
    & \qw & \ctrl{-4} & \qw & \qw & \qw & \qw & \qw & \qw & \targ &\qw&\qw\\
    & \qw & \qw & \ctrl{-6} & \qw & \qw & \qw & \qw & \qw & \targ &\qw&\qw\\
    & \qw & \qw & \qw &\ctrl{-7} &  \qw & \qw & \qw & \qw & \targ &\qw&\qw\\
    & \qw & \qw & \qw & \qw &\ctrl{-7} &  \qw & \qw & \qw & \targ &\qw&\qw\\
    & \qw & \qw & \qw & \qw & \qw &\ctrl{-7} &  \qw & \qw & \targ &\qw&\qw\\
    & \qw & \qw & \qw & \qw & \qw & \qw &\ctrl{-10} &  \qw & \targ &\qw&\qw
    }
  }
  \caption{Syndrome measurement circuit for the $[7,3]$ amplitude damping code.}\label{fig:73syndrome}
\end{figure}

\begin{figure}
  \begin{center}
    \includegraphics[width=\columnwidth]{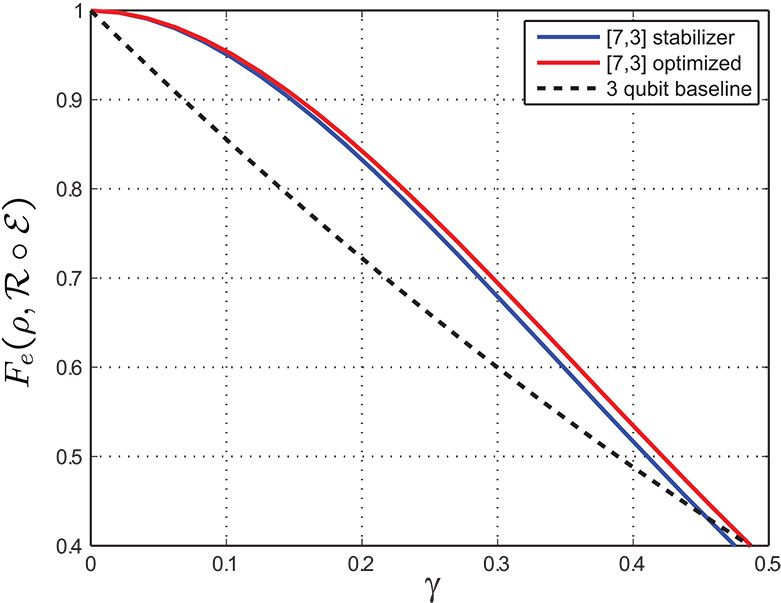}
    \caption[Optimal vs.~code projection recovery operations for the seven qubit linear amplitude damping code.]{Optimal vs.~code projection recovery operations for the [7,3] code.  We compare the entanglement fidelity for the optimal recovery operation and the recovery that includes a projection onto the code subspace.  For comparison, we also include the baseline performance of three unencoded qubits.  While the optimal recovery outperforms the code projector recovery, the performance benefit is likely small compared to the cost of implementing the optimal.}\label{fig:73optimized}
  \end{center}
\end{figure}

As in previous examples, we have a choice of recovery operations for the `no dampings' syndrome.  We can minimize the `no damping' distortion as was done in previous cases by computing the optimal or EigQER recovery within this subspace.  This will result in a $\gamma$-dependent recovery operation.  Alternatively, we can simply measure $XXXXXXX$ with a $+1$ projecting onto the code subpsace and a $-1$ requiring a correction of $Z_i$.  We compare these recovery operations in Fig.~\ref{fig:73optimized}.

\begin{figure}
  \begin{center}
    \includegraphics[width=\columnwidth]{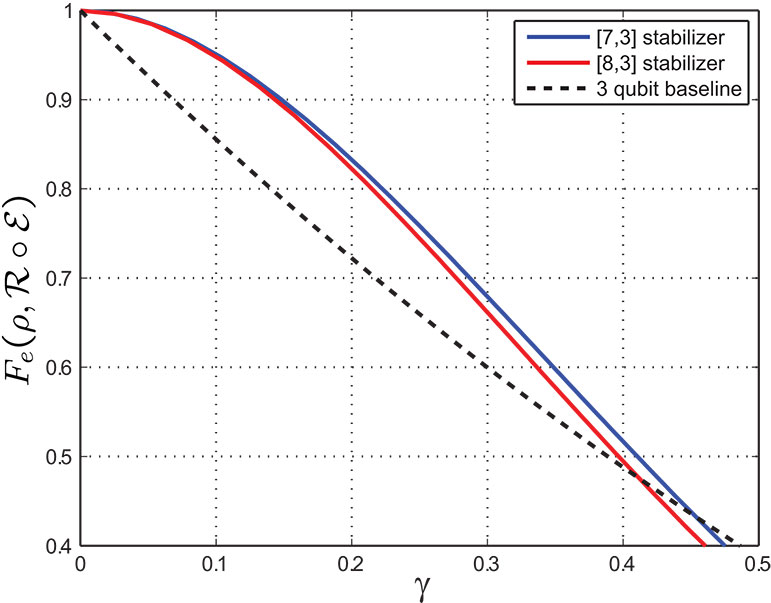}
    \caption[Comparison of the seven and eight qubit amplitude damping codes.]{Comparison of the $[7,3]$ and $[8,4]$ qubit amplitude damping codes.  We see that the $[7,3]$ performance is slightly better, despite the higher rate.}\label{fig:AmpDamp87}
  \end{center}
\end{figure}

We see in Fig.~\ref{fig:AmpDamp87} that the $[7,3]$ code slightly outperforms the $[8,3]$ code of Sec.~\ref{sec:generalized 41}.  The $[7,3]$ code perfectly corrects first order dampings and does not correct any second order dampings while the $[8,3]$ code partially corrects for higher order dampings.  The performance advantage of the $[7,3]$ code arises from the decreased length: the probability of a higher order damping error decreases as only seven physical qubits are needed.

Given its structure, it is logical to compare the $[7,3]$ amplitude damping code to the $[7,1]$ Steane code, as both are derived from the classical Hamming code.  We saw in Fig.~\ref{fig:AmpDamp579_eigQER} of Chapter \ref{chap:NearOptQER} that the Steane code is not particularly well adaptable to amplitude damping errors; despite its extra redundancy, the channel-adapted $[5,1]$ code significantly outperforms the channel-adapted Steane code.  This is particularly unfortunate as the Steane code can be implemented with such efficiency, with particular value for fault tolerant quantum computing.  The $[7,3]$ code provides a useful compromise position.

\begin{figure}
  \begin{center}
    \includegraphics[width=\columnwidth]{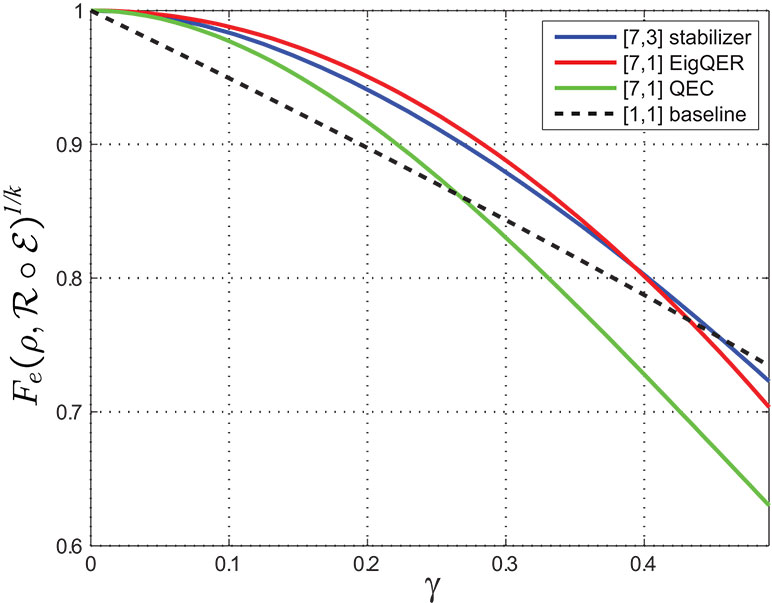}
    \caption[Comparison of the seven qubit amplitude damping code and the seven qubit Steane code.]{Comparison of the $[7,1]$ Steane code and the $[7,3]$ amplitude damping code, normalized by ${}^{1/k}$.    We see that the $[7,3]$ performance is very similar to the EigQER optimized recovery for the Steane code.}\label{fig:AmpDamp73 vs 71}
  \end{center}
\end{figure}

We see in Fig.~\ref{fig:AmpDamp73 vs 71} the performance comparison for $[7,3]$ code and the $[7,1]$ code (with and without channel-adapted recovery).   It comes as no surprise that the $[7,3]$ code outperforms the $[7,1]$ with standard stabilizer recovery: each perfectly corrects the first order damping errors, but the $[7,3]$ code has done so while preserving the times as much information.  It is interesting to see how close the $[7,3]$ performance is to the channel-adapted $[7,1]$.  As we saw in Fig.~\ref{fig:OrderQER}, the channel-adapted $[7,1]$ at least partially corrects some second-order damping errors while the $[7,3]$ does not.  This is mitigated by the higher rate of the $[7,3]$ code as again, three times as much information is preserved.

The $[7,3]$ code is not the only high rate linear code for amplitude damping errors.  Consider any classical linear code that can correct 1 error for which codewords have even parity.  We can convert this code to a quantum amplitude damping code in the same way as the $[7,3]$ code.  If $H$ is the parity check matrix for an $[n,k]$ classical linear code, then each row can be made a quantum code stabilizer replacing 1's with $Z$ and 0's with $I$.  To distinguish $X_i$ and $Y_i$ errors we include $X^{\otimes n}$ as a generator.  Since the classical code has even parity, we know that this generator commutes with the others.  This construction yields a $[n,k-1]$ quantum amplitude damping code that corrects for single amplitude damping errors.

The $[7,3]$ code we have presented here follows the structure proposed in \cite{Got:97} for amplitude damping codes; namely the code is a combination of an $X$-error correcting code and a $Z$-error detecting code.  It is not immediately clear how to generalize to $t$ error correcting linear codes.  Instead of a single generator to distinguish $X_i$ and $Y_i$ errors, we require an extra $t$ generators as we must distinguish $X_i$ and $Y_i$ for each corrected damping.

\section{Amplitude Damping Errors and the Shor Code}

We now turn our attention to the $[9,1]$ Shor code and its performance with a recovery operation channel-adapted to amplitude damping errors.  We saw through the EigQER results in Fig.~\ref{fig:AmpDamp579_eigQER} that the Shor code provides remarkably good protection from the amplitude damping channel.  We also saw in Fig.~\ref{fig:AmpDamp9_dual} that the recovery operation generated by EigQER is essentially optimal.  Thus, in this case, the optimal channel-adapted recovery operation can be described as a projective syndrome measurement followed by a unitary operation.   Given this intuitive structure, we can analyze the amplitude damping channel-adapted recovery operation.

We begin by noting that first order errors $\{E_1^{(k)}\}$ are perfectly correctable.  This comes as no surprise, since the Shor code can correct an arbitrary single qubit operation.  What may be surprising is that second order errors $\{E_1^{(j)}E_1^{(k)}\}$ are also perfectly correctable.  This was pointed out in \cite{Got:96} and can be seen through the same kind of stabilizer analysis of damped subspaces as we employed for the $[2(M+1),M]$ codes.

\begin{table}
  \begin{tabular}{ccc}
     \begin{tabular}{c}
        Qubit 1 damped\\
        \hline
        \begin{tabular}{c@{}c@{}c@{}c@{}c@{}c@{}c@{}c@{}c@{}c}
            -&$Z$&$Z$&$ I$&$ I$&$ I$&$ I$&$ I$&$ I$&$ I$\\
            &$I $&$Z$&$ Z$&$ I$&$ I$&$ I$&$ I$&$ I$&$ I$\\
            &$I$&$ I$&$ I$&$ Z$&$ Z$&$ I$&$ I$&$ I$&$ I$\\
            &$I$&$ I$&$ I$&$ I$&$ Z$&$ Z$&$ I$&$ I$&$ I$\\
            &$I$&$ I$&$ I$&$ I$&$ I$&$ I$&$ Z$&$ Z$&$ I$\\
            &$I$&$ I$&$ I$&$ I$&$ I$&$ I$&$ I$&$ Z$&$ Z$\\
            &$I$&$ I$&$ I$&$ X$&$ X$&$ X$&$ X$&$ X$&$ X$\\
            &$Z$&$I$&$I$&$I$&$I$&$I$&$I$&$I$&$I$
        \end{tabular}
     \end{tabular}
     &
     \begin{tabular}{c}
        Qubits 2 \& 3 damped\\
        \hline
        \begin{tabular}{c@{}c@{}c@{}c@{}c@{}c@{}c@{}c@{}c@{}c}
            -&$Z$&$Z$&$ I$&$ I$&$ I$&$ I$&$ I$&$ I$&$ I$\\
            &$I $&$Z$&$ Z$&$ I$&$ I$&$ I$&$ I$&$ I$&$ I$\\
            &$I$&$ I$&$ I$&$ Z$&$ Z$&$ I$&$ I$&$ I$&$ I$\\
            &$I$&$ I$&$ I$&$ I$&$ Z$&$ Z$&$ I$&$ I$&$ I$\\
            &$I$&$ I$&$ I$&$ I$&$ I$&$ I$&$ Z$&$ Z$&$ I$\\
            &$I$&$ I$&$ I$&$ I$&$ I$&$ I$&$ I$&$ Z$&$ Z$\\
            &$I$&$ I$&$ I$&$ X$&$ X$&$ X$&$ X$&$ X$&$ X$\\
            -&$Z$&$I$&$I$&$I$&$I$&$I$&$I$&$I$&$I$
        \end{tabular}
     \end{tabular}
     &
     \begin{tabular}{c}
        Qubits 1 \& 7 damped\\
        \hline
        \begin{tabular}{c@{}c@{}c@{}c@{}c@{}c@{}c@{}c@{}c@{}c}
            -&$Z$&$Z$&$ I$&$ I$&$ I$&$ I$&$ I$&$ I$&$ I$\\
            &$I $&$Z$&$ Z$&$ I$&$ I$&$ I$&$ I$&$ I$&$ I$\\
            &$I$&$ I$&$ I$&$ Z$&$ Z$&$ I$&$ I$&$ I$&$ I$\\
            &$I$&$ I$&$ I$&$ I$&$ Z$&$ Z$&$ I$&$ I$&$ I$\\
            -&$I$&$ I$&$ I$&$ I$&$ I$&$ I$&$ Z$&$ Z$&$ I$\\
            &$I$&$ I$&$ I$&$ I$&$ I$&$ I$&$ I$&$ Z$&$ Z$\\
            &$Z$&$I$&$I$&$I$&$I$&$I$&$I$&$I$&$I$\\
            &$I$&$I$&$I$&$I$&$I$&$I$&$Z$&$I$&$I$
        \end{tabular}
     \end{tabular}
  \end{tabular}
\caption{Stabilizers for several damped subspace syndromes for the Shor code.}\label{tab:Shor subspaces}
\end{table}

We see in Table \ref{tab:Shor subspaces} a few representative syndrome subspaces for damping errors on the Shor code.  From these subspaces, we surmise that the first step in making a syndrome measurement is to measure the first 6 code stabilizers (each of which has a pair of $Z$'s).  Depending on those outcomes, we can make a further stabilizer measurement.

As an example, consider when the first stabilizer returns a $-1$ and the rest return $+1$.  In that case, we can conclude that either the first qubit was damped, or both the second and third qubits were damped.  These can be distinguished by measuring $Z_1$ with a $+1$ indicating qubit one and a $-1$ indicating both qubits two and three.

It is interesting to note that in this case, we will have only measured 7 stabilizers, and thus need one further measurement to achieve a 2 dimensional subspace.  A natural choice would be to measure $IIIXXXXXX$; alternatively, this is an opportunity for a $\gamma$-dependent measurement instead.  As before, such an operation can improve performance at the cost of circuit complexity.  In most of this chapter, we have leaned toward the simpler operation, concluding that $\gamma$-dependent operations provide some performance benefit but not enough to justify the added complexity.  We will see that in the case of the Shor code, the performance gain may be sufficiently large to warrant a $\gamma$-dependent operation.

Before this consideration, let's turn to another syndrome for multiple qubit dampings.
The Shor code is divided into three blocks of three qubits each;
we already examined an example where two qubits on the same block are both damped.
The third subspace in Table \ref{tab:Shor subspaces} is an example of two qubits damped from different blocks; in this case the first and fifth stabilizers are both measured to be $-1$.  While the most likely cause of this syndrome is a two-qubit damping, we can further measure $Z_1$ and $Z_7$ to correct for a three or four-qubit damping occurrence.

\begin{figure}
  \includegraphics[width=\columnwidth]{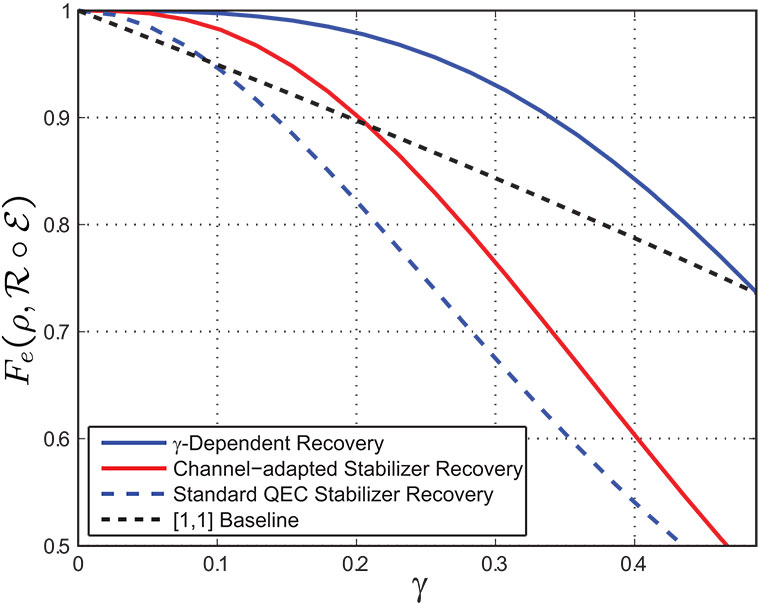}
  \caption{Channel-adapted stabilizer recovery vs.~$\gamma$-dependent recovery for the Shor code and the amplitude damping channel}\label{fig:Shor_recoveries}
\end{figure}

The preceding discussion of stabilizer subspaces provides two options for a channel-adapted recovery operation.  Both begin with a projective syndrome measurement of the first 6 code generators.  At that point, we may either make a set of stabilizer measurements to project onto the damped subspaces, or we may make a $\gamma$-dependent syndrome recovery to minimize this distortion.  It turns out that the best $\gamma$-dependent syndrome recovery has equivalent performance to the EigQER recovery operation and is therefore essentially optimal.  The stabilizer recovery, while simple to implement with Clifford group operations, has significantly weaker performance.  We compare the two recovery operations for various values of $\gamma$ in Fig.~\ref{fig:Shor_recoveries}.

How should we understand the extensive performance gains for the $\gamma$-dependent recovery?
As we saw in Sec.~\ref{sec:4qubit no damping}, we may reduce the distortion introduced by $E_0$ even though it cannot be perfectly corrected.  The $\gamma$-dependent operation arises when we have a remaining degree of freedom after determining the syndrome.  For the $[2(M+1),M]$ codes and the $[7,3]$ code, we only have such freedom in the `no damping' syndrome; in all of the damping syndromes, the syndrome measurement requires a full set of stabilizer measurements.  We saw that for the Shor code first order dampings require only 7 stabilizer measurements to determine the syndrome, leaving one extra degree of freedom.  We also have an extra degree of freedom when two qubits from the same block are damped.  These constitute all of the first and some of the second order syndromes, each of which can be optimized to minimize $E_0$ distortion.

\section{Summary}

We have developed several quantum error correcting codes channel-adapted for the amplitude damping channel.  All of the encodings can be compactly described in the stabilizer formalism.  While optimized $\gamma$-dependent recovery operations are possible, a much simpler recovery operation using only stabilizer measurements and Clifford group operations achieves nearly equivalent performance.
The channel-adapted codes have much higher rates (with short block lengths) than generic quantum codes.

\chapter{Conclusion}\label{chap:Conclusion}

Quantum error correction is an essential consideration for any quantum information processing system, whether it be a quantum computer, quantum memory device, or a quantum communication system.  The standard approach is to design procedures capable of correcting arbitrary single qubit errors.  Indeed, the first fundamental breakthrough in quantum error correction was the discovery that correcting Pauli errors $\{I,X_i,Y_i,Z_i\}$ on the $i^{th}$ qubit is equivalent to correcting an arbitrary error on the $i^{th}$ qubit.

We began the introduction with a quote from \cite{NieChu:B00} about quantum error correction: ``Many authors have what appears to be a suspicious fondness for the depolarizing channel...''  The authors' point, explained in the ensuing paragraphs, is the principle of arbitrary qubit error correction.  The depolarizing channel has operator elements $\{\sqrt{1-3p}I,\sqrt{p}X,\sqrt{p}Y,\sqrt{p}Z\}$, and by perfectly correcting these errors, we have a procedure for arbitrary errors.  While we do not dispute this point, we suggest that such procedures are in some sense `tuned' to the depolarizing channel.

QEC systems will exist in connection with a physical device which has been carefully designed to limit the effects of noise.  Furthermore, it is quite reasonable to expect errors to adhere to a specific form; the errors to be corrected are not arbitrary but have a more defined structure.  What we have shown in this dissertation is that we can `tune' or channel-adapt QEC to the structure of the noise.\clearpage

In essence, this tuning illustrates an important principle for QEC that is sometimes obscured by the term `arbitrary error:'  all QEC schemes are approximate.  For any noise model, the QEC system protects against the most likely errors.  For standard, generic QEC, the highest order errors are presumed to be single qubit errors; standard one-error correcting stabilizer codes protect against any arbitrary single qubit error and fail when more than qubit is in error.  The channel is approximated by single qubit errors and protected accordingly.  The approximate codes for the amplitude damping channel make an alternate assumption about the most likely errors.  By assuming greater structure for the likely errors, we may design more efficient QEC procedures.  In both cases, the QEC will be effective only inasmuch as the physical noise matches the assumed model.

To conclude this dissertation, we summarize the results of each of the preceding chapters.  We discuss our original contributions and how they connect to the existing body of knowledge.  We then lay out remaining open questions and the natural extensions of channel-adapted QEC. We end with s few comments on the broader impact of this work on the quantum computing community.

\section{Chapter Summaries}

We showed in Chapter \ref{chap:OptQER} how to cast channel-adapted quantum error correction as a solvable convex optimization problem.  We evaluated QEC performance with the average entanglement fidelity, which measures how well the procedure preserves quantum information.  Since the average entanglement fidelity is linear in the channel and CPTP maps correspond to positive semidefinite operators, we can determine the optimal channel-adapted recovery operation using a semidefinite program (SDP).  We demonstrated that channel-adapted recoveries improve performance over generic recoveries for certain channels.  By example, both the amplitude damping channel and the pure state rotation channel allow improved performance via channel-adaptation.  We also proved the form of the optimal recovery operation for channels described by scaled Pauli operators and a stabilizer encoding.  The optimal recovery measures each generator and the most likely Pauli operation is the syndrome recovery;  in many cases this corresponds to the generic stabilizer code recovery operation.

The optimal channel-adapted recovery operation has several drawbacks that we addressed in Chapter \ref{chap:NearOptQER}.  First of all, computation of the optimal solution grows exponentially and quickly becomes impractical for codes with block lengths greater than five.  Second of all, while the optimal recovery is physically legitimate, it is desirable for implementation and interpretation purposes to further constrain recovery operations.  We presented several methods to design channel-adapted recovery operations constrained to begin with a projective syndrome measurement.  These algorithms approach the optimal recovery operation and help us to understand the mechanism by which channel-adaptation improves fidelity.

Chapter \ref{chap:DualBounds} derived performance bounds on the sub-optimal recovery operations of Chapter \ref{chap:NearOptQER}.  Each algorithm constructs a dual feasible point that upper bounds the average entanglement fidelity.  Using these methods, we showed that recovery operations constrained to projective syndrome measurements approach the optimal recovery, especially as noise levels approach 0.

In Chapter \ref{chap:ThreeQubitCode} we focused our attentions on the amplitude damping channel and determined several classes of channel-adapted encodings and recoveries.  We began with an analysis of the optimal recovery operation of the channel-adapted encoding of \cite{LeuNieChuYam:97}.  Concluding that this $[4,1]$ code perfectly corrects qubit damping errors, we presented two classes of stabilizer codes with stabilizer recoveries that also correct qubit damping errors.  These codes have higher rates than the $[4,1]$ yet have comparable performance.  We presented quantum circuit descriptions of the encodings, syndrome measurements, and syndrome recovery operations.  These have simple forms with Clifford group operations.

\section{Contributions}

Researchers have known that channel-adapted QEC could yield more efficient procedures for some time.  The subject, however, had few examples and few defined tools, whether computational or analytical.  We first pointed out the utility of the semidefinite program for optimal channel-adapted QEC in \cite{FleShoWin:J07a}. (It was contemporaneously understood though unpublished by \cite{ReiWer:05} and \cite{KosLid:06}.  Furthermore, the SDP was used in \cite{YamHarTsu:05} as a sub-optimal method using minimum fidelity as the performance measure.)

The structured near-optimal  recovery operations of Chapter \ref{chap:NearOptQER} are a unique contribution to the field.    These are the first demonstration of good recovery operations constrained for easier implementation.  That they approach the optimal performance despite the added constraints provides evidence that channel-adapted techniques have utility in reasonable physical circumstances.  These near-optimal techniques can be used to numerically analyze longer codes, one of the serious criticisms of the SDP and other optimal methods.

The Lagrange dual function is a standard method for bounding optimization problems.  The bounds in Chapter \ref{chap:DualBounds}, however, are the first use of such bounding techniques for channel-adapted QEC.  (\cite{BarKni:02} presented a channel-adapted QEC bound which is not based in Lagrange duality and which was looser than our numerical bounds in every example.)  We presented numerical techniques for the construction of apparently tight  bounds for every circumstance in which the near-optimal recovery techniques are computationally tractable.

The channel-adapted amplitude damping codes of Chapter \ref{chap:ThreeQubitCode} are a significant generalization of the approximate code of \cite{LeuNieChuYam:97}.  Indeed, the numerical tools presented earlier in the dissertation led to a better recovery operation for the $[4,1]$ code which could be understood with the stabilizer formalism.  The stabilizer formalism allows a straightforward generalization to higher rate amplitude damping codes.  We generalized the $[4,1]$ code to $[2(M+1),M]$ codes with simple quantum circuits for encoding and recovery. We furthermore showed how an even-parity, one error correcting classical linear code can be converted into a channel-adapted amplitude damping code.

\section{Open Questions}

\subsection{Channel-adapted fault tolerant quantum computing}\label{sec:FTQC}

Quantum error correction, as we have treated it, operates under a benign set of circumstances.  In essence, the assumption is of a noisy communications channel between perfect quantum computers.  This model is quite useful as it illustrates the principles of QEC, but it has limited physical reality.  Perhaps its only direct physical value is in the construction of quantum memory, where we can presume that the decoherence is much more likely in the storage mechanism than in the retrieval device.  QEC is, however, an essential first step toward understanding the more generally applicable fault tolerant quantum computing (FTQC).

The natural extension of channel-adapted quantum error correction is, therefore, channel-adapted FTQC.  The two operate under the same principles, but for FTQC, the model is more stringent.  Rather than perfect quantum computers on either side of a noisy channel, we now must consider a quantum computer in which each component is subject to noise.  For fault tolerant operation, we apply an encoding and recovery procedure that insulates our logical computation from the decoherence of each gate.

The basic premise of channel-adaptivity is still sound for FTQC; it is still reasonable to suppose that errors will arrive with some known structure.  If we adapt our fault tolerant procedure to this structure, we will likely be able to improve performance and/or reduce overhead.

The most useful results in FTQC have been the demonstration of error thresholds.  Gates are assumed to fail with some probability $p$.  The threshold theorems have shown that if $p$ falls below some threshold then there are FT techniques that will yield computation of arbitrarily high fidelity.  The various threshold theorems provide important research quests for both theorists and experimentalists. Theorists seek FT constructions with ever higher thresholds; FT thresholds provide experimentalists with device fidelity goals.

FT thresholds are obtained through the concept of concatenated QEC codes and logical operations.  Consider a quantum circuit designed to carry out some quantum algorithm where every gate is presumed to be perfect; this is the ideal logical operation.  As we must implement this operation with imperfect gates, we replace the logical qubits with $n$ physical qubits, using some QEC code.  Each gate is replaced with a set of gates to perform the same function on the encoded logical qubits.  We also include error correcting blocks that detect and recover from error syndromes; presumably we add such a block after each of the logical gates.  Just as is done in the simpler QEC framework, we protect our computation from noise by encoding redundancy into our system.

One may be troubled by the idea of adding noisy gates in order to protect from noise.  After all, each error correction block is also made up of imperfect quantum gates.  The remarkable results of FTQC thresholds indicate that such constructions can indeed protect from noise.  The intuitive picture (ignoring the technical details) is that as long as the physical gates are reasonably good ($p$ is below the threshold), the error correction blocks result in logical operations that are less noisy than the underlying physical operations.

Concatenated codes repeat the process.  We can encode each of the logical qubits (now consisting of $n$ physical qubits) with yet another quantum error correcting code and corresponding logical operations.  As before, this procedure improves the fidelity of each logical operation.  In essence, this second level of error correction targets the higher order errors ignored by the first layer of the code.  The procedure can be repeated until the computation has reached the desired fidelity.

Channel-adaptivity as a likely role at the lowest level of a concatenated code.  By measuring a physical noise process, we may choose an encoding and recovery procedure that more efficiently targets the first order errors.  We could choose to determine the structure of the higher order errors passed along to the second level of the concatenated code; for example, it is not hard to characterize the errors inherent in the $[2(M+1),M]$ code after recovery.  It is more likely, however, that the higher levels of error correction will be more effective as generic QEC.  While we will likely have a good model for the physical noise process, such a model will never be perfect and may characterize the noise only to the first order.  In such a case it would be foolish to adapt QEC too completely to such an estimate.  Instead, the arbitrary error nature of generic QEC can be used.

The error correction blocks in a fault tolerant system have two basic requirements.  First, they must not allow errors to propagate.  If an error occurs on one qubit, it must not be allowed to corrupt other qubits.  We must construct a universal set of quantum gates such that errors do not propagate.  Second, the error correction blocks must be implemented in a relatively simple way.  Stated simply, the more complicated the quantum operation, the more it is subject to noise and error.  Thus, a FT procedure that is itself highly complex will not prove very effective at shielding a computation from noise.

The results of Chapter \ref{chap:ThreeQubitCode} are quite encouraging for channel-adapted FTQC, at least when considering the amplitude damping channel.  Consider a noise model in which the output of every quantum gate is subject to the amplitude damping channel.  Can we construct a fault-tolerant computational model using the amplitude damping codes of Chapter \ref{chap:ThreeQubitCode}?  Does this result in a good FT threshold?  How much improvement do we see from each level of concatenation?  Do channel-adapted methods improve the computational fidelity with less overhead?  These are the natural extensions of the channel-adapted QEC presented in this dissertation.

\subsection{Channel-adaptation with physical models}

In the examples throughout this dissertation, we have examined two channel models.  The first, the amplitude damping channel, is easily motivated by physical noise processes.  The second, the pure state rotation channel, is not.  Both channels demonstrate the performance gains that can be achieved through channel-adaptation.  These are, however, merely examples to demonstrate the principle of channel-adapted QEC; for practical use, we must channel-adapt to a physically observed noise process.

There is an obvious connection to quantum process tomography.  Given some physical device, whether a quantum memory element or a computational gate, one can measure the overall quantum operation.  The departure of the observed operation from the ideal can be considered the physical noise in the operation which must be corrected.

For a given noise process, how necessary is the arbitrary error model that underlies generic QEC?  Is there sufficient structure to the noise to justify channel-adapted techniques?  How well does the amplitude damping channel describe observed physical noise?  Is there a physical application for the channel-adapted amplitude damping codes?  Will a hybrid concatenation of a channel-adapted and a generic code provide good protection from a physical quantum channel?

\section{Broader Impact}

The answer to these open questions will determine the impact of this dissertation in the greater quantum computing community.  It is our belief that we have demonstrated a significant inefficiency within the generic QEC construction that can and should be exploited.  Generic QEC procedures have been monumentally important in advancing the field of quantum computing: without the principles QEC and FTQC, we would have little hope of building sufficiently noise-free quantum computers for any interesting applications.  As an engineering principle, however, the inefficiency of generic QEC procedures may limit its utility.

We have presented valuable tools to understand the impact of channel-adapted QEC.   As we approach the construction of larger-scale quantum information devices, these tools will aid in developing efficient error correction and fault tolerant procedures.  In this way, we are striving to connect the theoretical results of quantum error correction with the observed realities of  quantum devices.

\appendix
\chapter{QER Figures and Tables}\label{chap:App Figures}

This appendix includes a more comprehensive collection of figures and tables describing the performance of each channel-adapted QER algorithm.  Plots include the performance curves for the optimal QER operation (where available), the EigQER algorithm, the BlockEigQER algorithm with $m=2,4,8$, and the OrderQER algorithm (where available).  We present the svd dual performance bound and the lowest iterated performance bound.  We also include the channel-adapted recovery operation of Barnum and Knill\cite{BarKni:02}, with the associated performance bound.  These figures are presented for the five, seven, and nine qubit codes and for the amplitude damping channel and the pure state rotation channel with $\theta=5\pi/12$ and $\pi/4$.  Each scenario also includes a table with a quadratic polynomial fit for small noise values.

\begin{figure}[h]
  \begin{center}
    \includegraphics[width=\columnwidth]{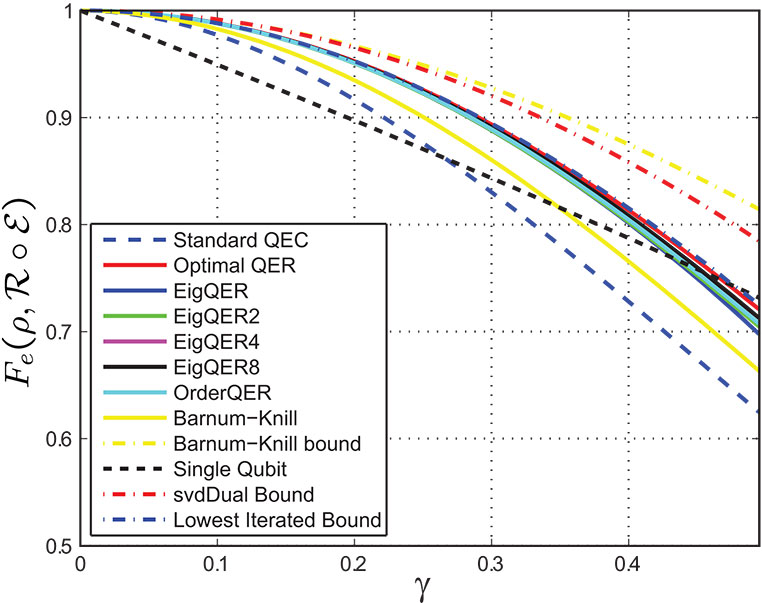}
  \end{center}
  \caption{QER performance for the amplitude damping channel and the five qubit code.}
\end{figure}

\begin{table}[h]
\begin{center}
  \begin{tabular}{|l|r|r|r|}
  \hline
  & $\gamma^0$ & $\gamma$ & $\gamma^2$\\
  \hline
  Optimal QER & 1 & 0 & -1.19\\
  Generic QEC & 1 & -.01 & -2.21\\
  EigQER & 1 & 0 & -1.22\\
  EigQER2 & 1 & 0 & -1.22\\
  EigQER4 & 1 & 0 & -1.21\\
  EigQER8 & 1 & 0 & -1.20\\
  OrderQER & 1 & 0 & -1.22\\
  Lowest Bound & 1 & 0 & -1.19\\
    \hline
  \end{tabular}
  \end{center}
  \caption{Asymptotic QER performance for the amplitude damping channel and the five qubit code.}
\end{table}

\begin{figure}[h]
  \begin{center}
    \includegraphics[width=\columnwidth]{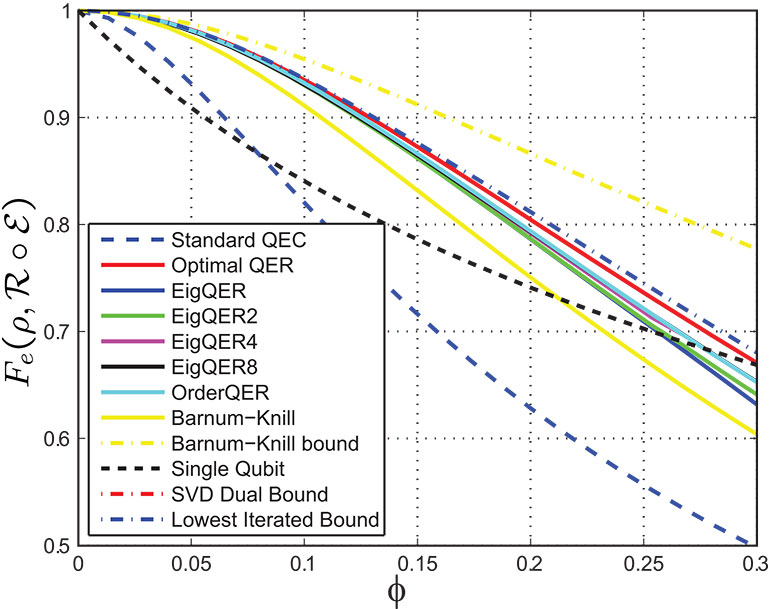}
  \end{center}
  \caption{QER performance for the pure state rotation channel with $\theta=5\pi/12$ and the five qubit code.}
\end{figure}

\begin{table}[h]
\begin{center}
  \begin{tabular}{|l|r|r|r|}
  \hline
  & $\phi^0$ & $\phi$ & $\phi^2$\\
  \hline
  Optimal QER & 1 & -.13 & -5.27\\
  Generic QEC & 1.01 & -1.24 & -6.02\\
  EigQER & 1 & -.13 & -5.73\\
  EigQER2 & 1 & -.13 & -5.76\\
  EigQER4 & 1 & -.13 & -5.64\\
  EigQER8 & 1 & -.14 & -5.45\\
  OrderQER & 1 & -.12 & -5.61\\
  Lowest Bound & 1 & 0 & -1.19\\
    \hline
  \end{tabular}
  \end{center}
  \caption{Asymptotic QER performance for the pure state rotation channel with $\theta=5\pi/12$ and the five qubit code.}
\end{table}

\begin{figure}[h]
  \begin{center}
    \includegraphics[width=\columnwidth]{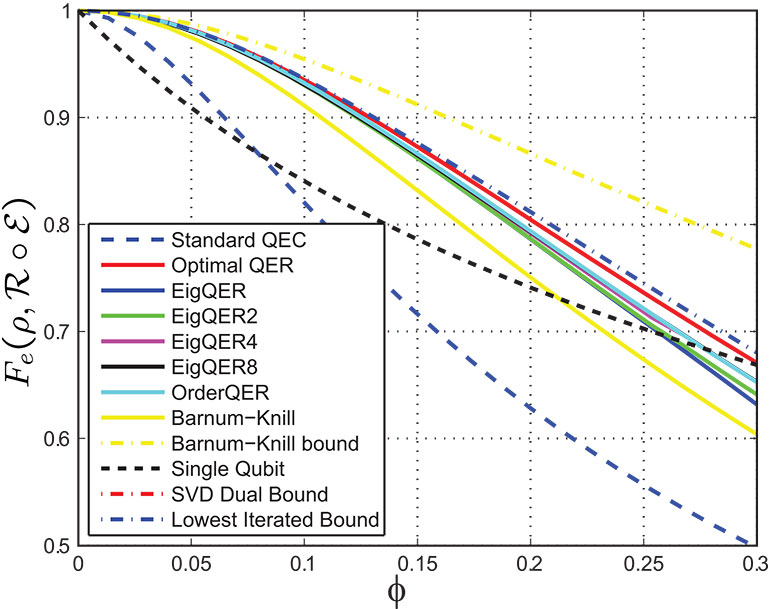}
  \end{center}
  \caption{QER performance for the pure state rotation channel with $\theta=\pi/4$ and the five qubit code.}
\end{figure}

\begin{table}[h]
\begin{center}
  \begin{tabular}{|l|r|r|r|}
  \hline
  & $\phi^0$ & $\phi$ & $\phi^2$\\
  \hline
  Optimal QER & 1 & -.06 & -8.20\\
  Generic QEC & 1.01 & -.24 & -12.92\\
  EigQER & 1 & -.05 & -8.66\\
  EigQER2 & 1 & -.05 & -8.66\\
  EigQER4 & 1 & -.05 & -8.55\\
  EigQER8 & 1 & -.06 & -8.45\\
  OrderQER & 1 & -.05 & -8.56\\
  Lowest Bound & 1 & -.06 & -8.09\\
    \hline
  \end{tabular}
  \end{center}
  \caption{Asymptotic QER performance for the pure state rotation channel with $\theta=\pi/4$ and the five qubit code.}
\end{table}

\begin{figure}[h]
  \begin{center}
    \includegraphics[width=\columnwidth]{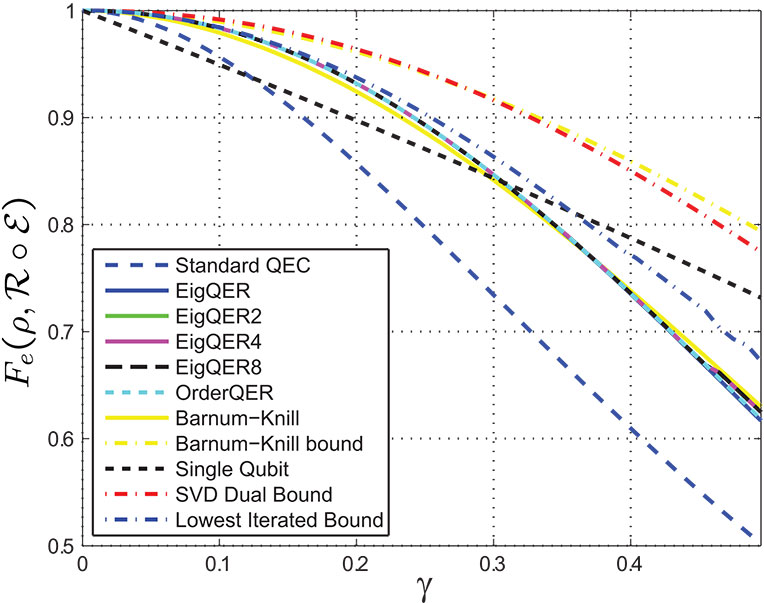}
  \end{center}
  \caption{QER performance for the amplitude damping channel and the seven qubit Steane code.}
\end{figure}

\begin{table}[h]
\begin{center}
  \begin{tabular}{|l|r|r|r|}
  \hline
  & $\gamma^0$ & $\gamma$ & $\gamma^2$\\
  \hline
  Generic QEC & 1 & -.04 & -4.00\\
  EigQER & 1 & 0 & -1.68\\
  EigQER2 & 1 & 0 & -1.68\\
  EigQER4 & 1 & 0 & -1.68\\
  EigQER8 & 1 & 0 & -1.68\\
  OrderQER & 1 & 0 & -1.68\\
  Lowest Bound & 1 & 0 & -1.59\\
    \hline
  \end{tabular}
  \end{center}
  \caption{Asymptotic QER performance for the amplitude damping channel and the seven qubit Steane code.}
\end{table}

\begin{figure}[h]
  \begin{center}
    \includegraphics[width=\columnwidth]{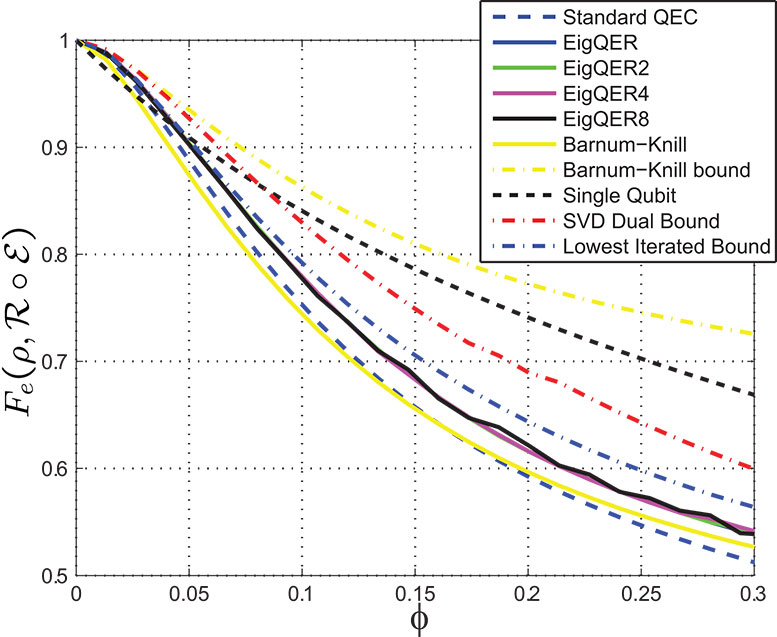}
  \end{center}
  \caption{QER performance for the pure state rotation channel with $\theta=5\pi/12$ and the seven qubit Steane code.}
\end{figure}

\begin{table}[h]
\begin{center}
  \begin{tabular}{|l|r|r|r|}
  \hline
  & $\phi^0$ & $\phi$ & $\phi^2$\\
  \hline
  Generic QEC & 1.01 & -2.47 & -.87\\
  EigQER & 1.01 & -2.04 & -2.45\\
  EigQER2 & 1.01 & -2.04 & -2.45\\
  EigQER4 & 1.01 & -2.07 & -2.19\\
  EigQER8 & 1.01 & -2.07 & -2.17\\
  Lowest Bound & 1.01 & -2.01 & -1.43\\
    \hline
  \end{tabular}
  \end{center}
  \caption{Asymptotic QER performance for the pure state rotation channel with $\theta=5\pi/12$ and the seven qubit Steane code.}
\end{table}

\begin{figure}[h]
  \begin{center}
    \includegraphics[width=\columnwidth]{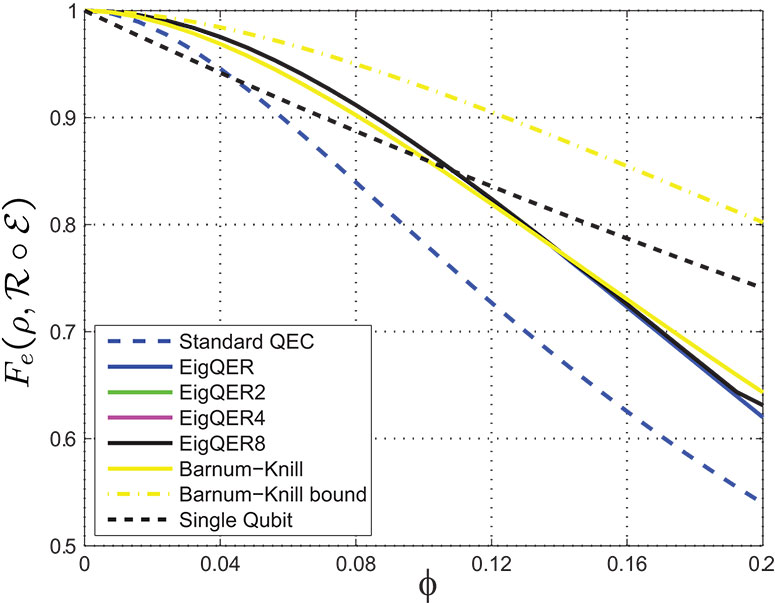}
  \end{center}
  \caption{QER performance for the pure state rotation channel with $\theta=\pi/4$ and the seven qubit Steane code.}
\end{figure}

\begin{table}[h]
\begin{center}
  \begin{tabular}{|l|r|r|r|}
  \hline
  & $\phi^0$ & $\phi$ & $\phi^2$\\
  \hline
  Generic QEC & 1 & -.6787 & -18.09\\
  EigQER & 1.01 & -.12 & -12.62\\
  EigQER2 & 1.01 & -.12 & -12.62\\
  EigQER4 & 1.01 & -.12 & -12.62\\
  EigQER8 & 1.01 & -.12 & -12.62\\
    \hline
  \end{tabular}
  \end{center}
  \caption{Asymptotic QER performance for the pure state rotation channel with $\theta=\pi/4$ and the seven qubit Steane code.}
\end{table}

\bibliography{bibfiles/quantum}
\bibliographystyle{plain}

\end{document}